\documentclass[%
reprint,
amsmath,amssymb,
aps,
]{revtex4-2}

\usepackage{newtxtext,newtxmath}

\usepackage[T1]{fontenc}
\usepackage{ae,aecompl}

\usepackage{graphicx}	
\usepackage{amsmath}

\begin{document}


\title{
       Energy self-extraction of a Kerr black hole through its frame-dragged force-free magnetosphere }


\author{Isao Okamoto$^{1,2}$}
\email[]{okamoto@nao.ac.jp}


\affiliation{
	$^{1}$ National Astronomical Observatory, 2-21-1 Oosawa, Mitaka-shi, Tokyo 181-8588, Japan \\ 
	$^{2}$ Institute of Black Hole Mining, 114-6 Ochikawa, Hino-shi, Tokyo 191-0034, Japan}

\author{Yoogeun Song$^{3,4}$}

\email[]{ygsong1004@gmail.com}
	
\affiliation{
	$^{3}$ Korea Astronomy and Space Science Institute, 776 Daedeok-daero, Yuseong-gu, Daejeon 34055, Korea\\
	$^{4}$ University of Science and Technology, 217 Gajeong-ro, Yuseong-gu, Daejeon 34113, Korea
}


\date{\today}


\newcommand{\dpsty}{\displaystyle}

\newcommand{\AOo}{A_{0 \omega}} 
\newcommand{\Om}{\Omega}	\newcommand{\ellH}{\ell_{\rm H}}
\newcommand{\om}{\omega}   \newcommand{\omN}{\omega_{\rm N}}  
\newcommand{\OLOmNb}{ {\bar{\Omega}_{\rm F}} } 
\newcommand{\vpN}{\vp_{\rm N}} \newcommand{\Dlvp}{\Delta\vp}
\newcommand{\omoL}{\omega_{\rm oL}}  \newcommand{\omiL}{\omega_{\rm iL}} 
\newcommand{\omoF}{\omega_{\rm oF}}  \newcommand{\omiF}{\omega_{\rm iF}} 
\newcommand{\OmFm}{\Omega_{{\rm F}\omega}}
\newcommand{\OmFmOL}{(\OmFm)_{\rm oL}}  \newcommand{\OmFmIL}{(\OmFm)_{\rm iL}} 
\newcommand{\eps}{\epsilon}	

\newcommand{\epsGTE}{\epsilon_{\rm EX}}		
\newcommand{\vre}{\varrho_{\rm e}} 
\newcommand{\Psic}{\Psi_{\rm c}}   \newcommand{\Psib}{\bar{\Psi}}
\newcommand{\Psiz}{\Psi_0} \newcommand{\Psio}{\Psi_1}  \newcommand{\Psit}{\Psi_2}
\newcommand{\OmF}{\Omega_{\rm F}}  \newcommand{\OmH}{\Omega_{\rm H}}
\newcommand{\OmNS}{\Omega_{\rm NS}}   \newcommand{\SNS}{S$_{\rm NS}$}
\newcommand{\OLom}{\overline{\om}}

\newcommand{\SOL}{S$_{\rm oL}$}   \newcommand{\SIL}{S$_{\rm iL}$}
\newcommand{\SOF}{S$_{\rm oF}$}   \newcommand{\SIF}{S$_{\rm iF}$}
\newcommand{\Sns}{S$_{\rm NS}$}
\newcommand{\SH}{S$_{\rm H}$}   \newcommand{\SSN}{S$_{\rm N}$}	 \newcommand{\Sinf}{S$_{\infty}$} 
\newcommand{\SZAM}{S$_{\rm ZAMD}$}

\newcommand{\calE}{{\cal E}} 
\newcommand{\calEout}{{\cal E}_{\rm (out)}} \newcommand{\calEin}{{\cal E}_{\rm (in)}}  
\newcommand{\calEns}{{\cal E}_{\rm NS}}		\newcommand{\calEBH}{{\cal E}_{\rm BH}}
\newcommand{\calD}{\cal D}
\newcommand{\calDout}{{\cal D}_{\rm (out)}} \newcommand{\calDin}{{\cal D}_{\rm (in)}} 
\newcommand{\calCout}{{\cal C}_{\rm (out)}}  \newcommand{\calCin}{{\cal C}_{\rm (in)}} 

\newcommand{\calP}{{\cal P}}	\newcommand{\calPE}{{\cal P}_{\rm E}}  \newcommand{\calPJ}{\calP_J} 
\newcommand{\calC}{{\cal C}}  
\newcommand{\calR}{{\cal R}}  \newcommand{\calRffinf}{\calR_{{\rm ff}\infty}}  
\newcommand{\calI}{{\cal I}}  
\newcommand{\vccalI}{\mbox{\boldmath ${\cal I}$} }  
\newcommand{\calIffinf}{{\cal J}_{{\rm ff}\infty}}	 \newcommand{\calIffH}{{\cal J}_{\rm ffH}}          
\newcommand{\vpL}{\vp_{\rm L}}
\newcommand{\SL}{S$_{\rm L}$}

\newcommand{\calPEout}{\calP_{\rm E, (out)}}  \newcommand{\calPEinU}{\calPE^{\rm (in)}}  \newcommand{\calPEin}{\calP_{\rm E,(in)}}  
\newcommand{\calPJout}{\calP_{\rm J, (out)}}  \newcommand{\calPJinU}{\calPJ^{\rm (in)}}  \newcommand{\calPJin}{\calP_{\rm J,(in)}}  
\newcommand{\calRinf}{{\cal R}_{\infty}}
\newcommand{\calS}{{\cal S}}   \newcommand{\calSJ}{{\cal S}_{J}} 
\newcommand{\SffH}{\calS_{\rm ffH}}    \newcommand{\Th}{T_{\rm H}}			
\newcommand{\Sffinf}{\calS_{{\rm ff}\infty}}   \newcommand{\SN}{\calS_{\rm N}}	 
\newcommand{\SG}{{\cal G}_{\rm N}}  
\newcommand{\GN}{{\cal G}_{\rm N}}  
\newcommand{\SNG}{{\calS}_{\rm N.or.G}}  
\newcommand{\Szamd}{S$_{\rm ZAMD}$}  

\newcommand{\epsGTEb}{\bar{\epsilon}_{\rm EX}}		
\newcommand{\epsE}{{\varepsilon}_{\rm E}}
\newcommand{\epsJ}{\varepsilon_{\rm J}}	
\newcommand{\SepsE}{S_{\eps_{\rm E}=0}}
\newcommand{\omepsE}{\om_{\eps_{\rm E}=0}}

\newcommand{\OmFb}{\bar{\Omega}_{\rm F}}    \newcommand{\omNb}{\bar{\om}_{\rm N}}   
\newcommand{\SNb}{\bar{\S}_{\rm N}}

\newcommand{\rH}{r_{\rm H}}

\newcommand{\Srhozero}{S$_{\varrho_{e}=0}$}  \newcommand{\ellN}{\ell_{\rm N}}
\newcommand{\Dl}{\Delta}  \newcommand{\Dlom}{\Delta\om}  \newcommand{\Dlell}{\Delta\ell}  \newcommand{\al}{\alpha}  \newcommand{\vp}{\varpi}  

\newcommand{\lb}[1]{\label{eq:#1}} 		\newcommand{\rf}[1]{\ref{eq:#1}}
\newcommand{\lbs}[1]{\label{sec:#1}} 	\newcommand{\rfs}[1]{\ref{sec:#1}}
\newcommand{\lbf}[1]{\label{ff:#1}} \newcommand{\rff}[1]{\ref{ff:#1}}
\newcommand{\lbt}[1]{\label{tbl:#1}} \newcommand{\rft}[1]{\ref{tbl:#1}}

\newcommand{\beeq}{\begin{equation}} \newcommand{\eneq}{\end{equation}}

\newcommand{\benu}{\begin{enumerate}}   \newcommand{\enen}{\end{enumerate}} 
\newcommand{\vF}{v_{\rm F}}    
\newcommand{\Iot}{I_{\overline{12}}}  \newcommand{\Ito}{I_{\overline{21}}}    
\newcommand{\Iout}{I_{\rm (out)}}    \newcommand{\Iin}{I_{\rm (in)}}	 
\newcommand{\IoutU}{I^{\rm (out)}}    \newcommand{\IinU}{I^{\rm (in)}}
\newcommand{\Igap}{I_{\rm gap}} 
\newcommand{\uvt}{\mbox{\boldmath $t$}}   \newcommand{\uvp}{\mbox{\boldmath $p$}}  \newcommand{\uvn}{\mbox{\boldmath $n$} }
\newcommand{\vcnb}{\mbox{\boldmath $\nabla$}}

\newcommand{\vcj}{\mbox{\boldmath $j$}}  
\newcommand{\vcjp}{\mbox{\boldmath $j$}_{\rm p}}    \newcommand{\jt}{j_{\rm t}}
\newcommand{\jvl}{j_{\perp}} 	\newcommand{\jpl}{j_{\parallel}} 
\newcommand{\Evl}{E_{\perp}} 	\newcommand{\Epl}{E_{\parallel}} 
\newcommand{\LPdr}[2]{\left( \frac{d#1}{d#2}\right)}
\newcommand{\dr}[2]{\frac{d#1}{d#2}} 

\newcommand{\PlDr}[2]{\frac{\partial#1}{\partial#2}}   \newcommand{\LPPlDr}[2]{\left(\frac{\partial#1}{\partial#2}\right)}
\newcommand{\Ftpldr}[2]{\left(\frac{\partial#1}{\partial#2}\right)}
\newcommand{\LPfrac}[2]{\left(\frac{#1}{#2}\right)}
\newcommand{\DN}[1]{[#1]_{\rm N}}   \newcommand{\DG}[1]{[#1]_{\rm G}} 
\newcommand{\PN}[1]{(#1)_{\rm N}}    \newcommand{\PG}[1]{(#1)_{\rm G}}
\newcommand{\Pout}[1]{(#1)_{\rm (out)}}	\newcommand{\Pin}[1]{(#1)^{\rm (in)}}
\newcommand{\PffH}[1]{(#1)_{\rm ffH}}	\newcommand{\Pffinf}[1]{(#1)_{{\rm ff}\infty}}
\newcommand{\lmb}{\lambda}   \newcommand{\Lmb}{\Lambda}
\newcommand{\sg}{\sigma}	\newcommand{\kp}{\kappa}

\newcommand{\vcv}{\mbox{\boldmath $v$}}   \newcommand{\vcm}{\mbox{\boldmath $m$}}	\newcommand{\vcp}{\mbox{\boldmath $p$}}
\newcommand{\vcbt}{\mbox{\boldmath $\beta$}} 
\newcommand{\vcx}{\mbox{\boldmath $x$}}  \newcommand{\vcg}{\mbox{\boldmath $g$}}  \newcommand{\vcH}{\mbox{\boldmath $H$}} 
\newcommand{\vcA}{\mbox{\boldmath $A$}}   \newcommand{\vcell}{\mbox{\boldmath $\ell$}}
\newcommand{\vcvp}{\mbox{\boldmath $v$}_{\rm p}}  \newcommand{\vvt}{v_{\rm t}}
\newcommand{\vcB}{\mbox{\boldmath $B$}}  \newcommand{\vcE}{\mbox{\boldmath $E$}}
\newcommand{\vcBp}{\mbox{\boldmath $B$}_{\rm p}}    \newcommand{\vcBt}{\mbox{\boldmath $B$}_{\rm t}}
\newcommand{\Bp}{B_{\rm p}} \newcommand{\Bt}{B_{\rm t}}
\newcommand{\vcEp}{\mbox{\boldmath $E$}_{\rm p}}	\newcommand{\OLEp}{\overline{\vcEp}}
\newcommand{\vcS}{\mbox{\boldmath $S$}}
\newcommand{\vcSJ}{\mbox{\boldmath $S$}_{\rm J}}  
\newcommand{\vcSJin}{\mbox{\boldmath $S$}_{\rm J,(in)}}  \newcommand{\vcSJout}{\mbox{\boldmath $S$}_{\rm J,(out)}}  
\newcommand{\vcSJinU}{\mbox{\boldmath $S$}_{\rm J}^{\rm (in)}}    \newcommand{\vcSJoutU}{\mbox{\boldmath $S$}_{\rm J}^{\rm (out)}}
\newcommand{\vcSE}{\mbox{\boldmath $S$}_{\rm E}} 
\newcommand{\vcSEout}{\mbox{\boldmath $S$}_{\rm E,(out)}}	
\newcommand{\vcSEin}{\mbox{\boldmath $S$}_{\rm E,(in)}}  \newcommand{\vcSEinU}{\mbox{\boldmath $S$}_{\rm E}^{\rm (in)} }
\newcommand{\vcSsd}{\mbox{\boldmath $S$}_{\rm SD}}  
\newcommand{\vcSEM}{\mbox{\boldmath $S$}_{\rm EM}} 		

\newcommand{\vcSEMout}{\mbox{\boldmath $S$}_{\rm EM,(out)}}	\newcommand{\vcSEMin}{\mbox{\boldmath $S$}_{\rm EM,(in)}}

\newcommand{\vcSsdin}{\vcS_{\rm SD,(in)}}	\newcommand{\vcSsdout}{\vcS_{\rm SD,(out)}}
\newcommand{\vcSsdinU}{\vcS_{\rm SD}^{(\rm in)}}	

\newcommand{\Mrot}{\mbox{\boldmath $M$}_{\rm rot}}	\newcommand{\Mirr}{\mbox{\boldmath $M$}_{\rm irr}}  

\newcommand{\OLOmFm}{\overline{\Omega}_{{\rm F}\omega}}   
\newcommand{\OLvF}{\overline{v}_{\rm F}} \newcommand{\OLvarepsJ}{\overline{\varepsilon}_{J}}    
\newcommand{\OLOmFmout}{\overline{\Omega}_{{\rm F}\omega},{\rm out}} 
\newcommand{\OLvcSsd}{\overline{\vcS}_{\rm SD}}
\newcommand{\OLvcSsdin}{\overline{\vcS}_{\rm SD,(in)}}		
\newcommand{\OLvcSsdout}{\overline{\vcS}_{\rm SD,(out)}}   \newcommand{\OLvcSsdinU}{\overline{\vcS}_{\rm SD}^{(\rm in)}}
\newcommand{\OLvcSEM}{\overline{\vcS}_{\rm EM}}   \newcommand{\OLvcSEMout}{\overline{\vcS}_{\rm EM,(out)}}
\newcommand{\OLvcSEMinU}{\overline{\vcS}_{\rm EM}^{\rm (in)} }    
\newcommand{\OLvcSEMin}{\overline{\vcS}_{\rm EM,(in)}}   
\newcommand{\OLvcSJinU}{\overline{\vcSJ}^{in}}    
\newcommand{\OLvcSJout}{\vcSJout}

\newcommand{\CS}{C$_{\rm S}$} 
\newcommand{\CP}{C$_{\rm P}$}
\newcommand{\RN}{R$_{\rm N}$}  \newcommand{\RP}{R$_{\rm P}$}

\newcommand{\vcSEMinU}{\vcSEM^{\rm (in)}}

\newcommand{\barom}{\bar{\om}}
\newcommand{\vtt}{v_{\rm t}}  

\newcommand{\zamp}{ZAM-P} 	\newcommand{\zam}{zero-angular-momentum} 		\newcommand{\zamo}{zero-angular-momentum-observer} 
\newcommand{\SgapI}{S$_{\rm G(in)} $}  \newcommand{\SgapO}{S$_{\rm G(out)}$}
\newcommand{\calRffH}{{\cal R}_{{\rm ffH}}} 

\newcommand{\rSch}{r_{\rm Sch}}

\newcommand{\ggel}{ \mathrel{
\raise1.1ex\hbox{$\scriptstyle >$}
\mkern-10mu\raise.4ex\hbox{$\scriptstyle =$}
\mkern-10mu\lower0.3ex\hbox{$\scriptstyle <$} }}
\newcommand{\lleg}{ \mathrel{
\raise1.1ex\hbox{$\scriptstyle <$}
\mkern-10mu\raise.4ex\hbox{$\scriptstyle =$}
\mkern-10mu\lower0.3ex\hbox{$\scriptstyle >$} }}
\newcommand{\ggo}{ \mathrel{\raise.3ex\hbox{$>$}\mkern-14mu\lower0.6ex\hbox{$\sim$}} }
\newcommand{\lo}{\mathrel{\raise.3ex\hbox{$<$}\mkern-14mu\lower0.6ex\hbox{$\sim$}} }

\newcommand{\SiF}{S$_{\rm iF}$}	\newcommand{\SoF}{S$_{\rm oF}$}
\newcommand{\TTH}{T_{\rm H}}  \newcommand{\SSH}{S_{\rm H}} \newcommand{\RH}{\calR_{\rm H}}  
\newcommand{\EH}{E_{\rm H}} \newcommand{\BH}{B_{\rm H}}

\newcommand{\xN}{x_{\rm N}}	\newcommand{\xM}{x_{\rm M}}	\newcommand{\xE}{x_{\rm E}}	
\newcommand{\xoL}{x_{\rm oL}}    \newcommand{\xiL}{x_{\rm iL}}

\newcommand{\SSM}{{S$_{\rm M}$}}  \newcommand{\SSMc}{{S$_{\rm Mc}$}}  \newcommand{\SE}{S$_{\rm E}$} 
\newcommand{\omM}{\om_{\rm M}}
\newcommand {\hc}{h_{\rm c}}

\newcommand{\PFS}{pseudo-flat space}
\newcommand{\RTDy}{rotational-tangential discontinuity}  
\newcommand{\RTDs}{rotational-tangential discontinuities}

\newcommand{\FDo}{FD $\omega$-}

\newcommand{\SNN}{S$_{\rm N}$}


\begin{abstract} 
{ It is shown that when only the condition} $0<\OmF<\OmH$ 
is satisfied, the Kerr black hole frame-drags its surrounding force-free magnetosphere with the field-line-angular-velocity (FLAV) $\OmF$, where $\OmH$ is the horizon angular-velocity. Then, the zero-angular-momentum-observers (ZAMOs) circulating with the frame-dragging-angular-velocity $\om$ will see that the `null surface' \SSN\ where $\omN=\OmF$ always exists.  
They will see that the outer domain $\calDout$ outside \SSN\ is prograde-rotating
with $\OmFm>0$, whereas the inner domain $\calDin$ inside is retrograde-rotating 
with $\OmFm<0$, where $\OmFm=\OmF-\om$ denotes the ZAMO-FLAV.  \emph{This surface} \SSN\ must be the magneto-centrifugal divider of the force-free magnetosphere, with a kind of plasma-shed on it. 
Subsequently, the force-free and freezing-in conditions break down on \SSN, thereby allowing the particle-current sources to be set up on \SNN. \emph{This surface} also is the ZAM-surface \SZAM, on which no flow of angular momentum nor electric current can cross.  Because the electric field $\vcEp$ reverses sign on \SNN, the Poynting flux reverses direction from outward to inward on \SSN.  
  An electromagnetic self-extraction of energy will be possible only through the frame-dragged magnetosphere, with the inner domain $\calDin$ nested between the horizon and \emph{this surface} \SSN, in order to comply with the first and second laws of thermodynamics. 

\end{abstract}


\maketitle


\section{\label{sec:level1}Introduction}   
\setcounter{equation}{0}
\numberwithin{equation}{section}

More than four decades have passed since the pioneering paper by \citet{bla77} 
was published on the electromagnetic extraction of energy from Kerr black holes (BHs).  This controversial task however still remains a big challenge at the latest new frontier of black hole astrophysics in modern classical physics (\citet{tho17}).  The purpose of this paper is to venture challenging this formidable task. 

Fundamental concepts and expressions as well as the basic formulation of general-relativity, thermodynamics, and electrodynamics, necessary for elucidating self-extraction of energy from Kerr holes, have fortunately been given in a nearly complete form by \cite{bla77,tho17,mac82,zna77,zna78,phi83a,tho86} 
already four decades ago. 

Extending the `single-pulsar model' \cite{bla77,bla79}, Macdonald \& Thorne \citep{mac82} and Thorne, Price \& Macdonald \citep{tho86} regarded the `horizon battery' as explicitly existent in the event horizon.  Phinney \citep{phi83a,phi83b} 
were the first that tried to develop a comprehensive model for `BH-driven hydromagnetic flows' or jets for active galactic nuclei (AGNs), making use of the pulsar wind theory (e.g., \citep{oka78,ken83}). 

Since then, it was thought that a `magnetized' Kerr hole would possess not only a battery but also an internal resistance $Z_{\rm H}$ on the horizon, as seen in `a little table on BH circuit theory for engineers' \cite[Fig 3]{phi83a}. The image in the 1980s looks like the magnetosphere consisting of {\em double} wind structures with a \emph{negligible} violation of the force-free condition for particle production and a {\em single} series circuit with a battery on the horizon and two resistances on the horizon and infinity (referred to as the `single-pulsar model') (see \citet{pun89, pun90}). Based on these pioneering works, we remodel the pulsar-type force-free magnetosphere by making full use of the $3+1$-formulation of black hole electrodynamics \cite{mac82} and the Membrane Paradigm \cite{tho86}.   

 In order to perform this formidable task of self-extraction of energy from Kerr holes, referring to some important descriptions on general-relativistic and thermodynamic aspects of energy extraction by \cite{bla77, zna77,zna78, mac82,tho86}, we need to  unequivocally unify  the pulsar electrodynamics and BH thermodynamics into BH gravito-thermo-electrodynamics (GTED). 

We introduce some central critical premises; one of them is that the large-scale poloidal magnetic field $\vcBp$ trapped in some way (i.e., frame-dragged) by the hole extends from near the horizon \SH\ to the infinity surface \Sinf, with the field-line-angular-velocity (FLAV) $\OmF=$~constant (i.e., Ferraro's law of iso-rotation holds). When the distant observers see that the iso-rotation holds for $\OmF$, { the zero-angular-momentum-observers (ZAMOs) will see that the FLAV that they measure violates it. }
Then, when $0<\OmF<\OmH$, the ZAMO-FLAV $\OmFm=\OmF-\om$ changes sign on the `null' surface \SSN\ with $\omN=\OmF$ \cite{bla77}. Therefore, the force-free magnetosphere is inexorably divided into the outer semi-classical (SC) and inner general-relativistic (GR) domains by \SSN\  (and $\OmFm= 0$). In turn, when the SC domain $\calDout$ progradely rotates ($\OmFm>0$), the GR domain $\calDin$ retrogradely rotates ($\OmFm<0$). The electric field $\vcEp$ changes direction as well. When the poloidal field $\vcBp$ with $\OmF(\Psi)$ is continuous across \SSN, these conditions as a whole subsequently give rise to the breakdown of the force-free and freezing-in conditions on \SSN\ between the two light surfaces. One of the important questions is \emph{how do we determine $\omN=\OmF$ in terms of the hole's angular velocity (AV) $\OmH$.} 

Through coupling with `potential-gardient' $\OmF$, the frame-dragging-angular-velocity (FDAV) $\om$ acquires a function of `gravito-electric potential gradient,' indispensable for managing `self-extraction' of energy, fulfilling the first and second laws of thermodynamics. It indeed is the FDAV $\om$ that combines BH  thermodynamics and pulsar electrodynamics.  
Despite the hindering presence of the event horizon, it therefore is the existence of \emph{this surface} \SSN\ that enables the Kerr hole to manipulate its magnetosphere, to launch the Poynting flux both out- and in-wards from the underlying particle-current sources hidden under \SSN\ in the force-free limit. 

It will be instructive to remind that a Kerr hole itself is by nature not an electrodynamic object, but basically a thermodynamic object, being fated to obey the four laws of thermodynamics \citep{tho86, OK90, OK91,KO91}. { The electromagnetic process of extraction of energy therefore operates \emph{only} under severe control of the first and second laws, because the former defines the efficiency of extraction, and the latter poses an important restriction on the efficiency \cite{bla77}.} The efficiency has a physical meaning only when \emph{this surface} \SSN\ exists, really resulting from unification of (BH) thermodynamics and (pulsar) electrodynamics, and it is the frame-dragging  that bridges the event horizon \SH\ between them. 

We extend the `Membrane Paradigm' \cite{tho86} from one membrane to three membranes. The first two `resistive' ones on the infinity and horizon surfaces, $\Sffinf$ and $\SffH$, terminate the outer and inner force-free domains, $\calDout$ and $\calDin$, by particle acceleration on $\Sffinf$ and entropy production on $\SffH$, respectively, and the third one is the `inductive' membrane $\SN$ on \emph{this surface} \SSN, which covers the particle-current sources. 
 
\citet{bla22} recently constructed the models for the ergo-magnetosphere, ejection disc, and magnetopause in M87. They argued that the force-free approximation is justifiable in the vicinity of the BH.  In fact, there is good chemistry between thermodynamics for the Kerr hole with \emph{two} hairs (e.g.,\ entropy $S$ and angular momentum $J$) and electrodynamics for the pulsar magnetosphere with \emph{two} conserved quantities ($\OmF$ and $I$).  We show that the force-free approximation is flexible and robust enough to accept the existence of the `null surface' \SSN\ \cite{bla77,oka92}, where the breakdown of the force-free condition (as well as the freezing-in one) takes place. Nonetheless, \emph{how to incorporate these into the force-free formalism} has unfortunately been a puzzling question over the past few decades, 
despite that \SSN\ always exists even when the hole loses energy \cite{bla77}. 

We show a brief outline of each section in what follows.  
In Sec.\  \ref{FFPM},	
we discuss the `battery and resistance' for a pulsar force-free magnetosphere in terms of the two `conserved' quantities $\OmF(\Psi)$ and $I(\Psi)$.  

We clarify the fundamental properties of thermodynamics and electrodynamics for BH force-free magnetospheres (see Sec.\ \ref{ther-rot}), 
and define the overall efficiency of extraction by $\epsGTEb=\OmFb/\OmH$ (see Sec.\ \ref{BH-FFM}),  
where $\OmFb$ is the weighted mean of $\OmF(\Psi)$ by $I(\Psi)$.  
Sec.\  \ref{FF-Frz-Conditions} 
introduces the force-free and freezing-in conditions, whose breakdown on the null surface \SSN\ is indispensable for extracting energy from a Kerr hole. \emph{This surface} divides the magnetosphere into the two force-free domains: SC domain $\calDout$ and GR domain $\calDin$.

In Sec.\ \ref{2fluxes},   
we show that the `conserved' overall energy flux $\vcSE$ consists of two `non-conserved' fluxes, e.g., the `Electromagnetic-Poynting' flux $\vcSEM$ and  `Spin-Down-energy' flux $\vcSsd$.  
In Sec.\ \ref{I&LSs}, 
the two light surfaces \SOL\ and \SIL\ are defined for the two domains $\calDout$ and $\calDin$ (see Sec.\ \ref{SoL/SiL}), respectively, 
and the two eigenfunctions for $I(\Psi)$, $\Iout$, and $\Iin$, are derived for the respective domains (see Sec.\ \ref{Iout/in}).

It is clarified in Sec.\ \ref{second/restr}     
as to how the second law imposes such a restriction of efficiency as $0\lo \epsGTE \approx\epsGTEb\lo 1$. 
In Sec.\ \ref{EnegR}, 
variations of the energy and angular-momentum densities of the electromagnetic fields  are discussed along each field line (FL) across the null surface \SSN.

In Sec.\  \ref{NullS}, 
we show that 
the ZAMOs will see that a violation of iso-rotation by frame-dragging leads finally to a breakdown of the freezing-in and force-free conditions on \SSN.  \emph{This surface} existing between the two light surfaces defines the `magneto-centrifugal divider' of the force-free magnetosphere by $\OmFm\ggel 0$ into the outer SC domain $\calDout$ with \SOL\ for the outflow and the inner GR domain $\calDin$ with \SIL\ for the inflow, and hence some pair-production mechanism must be at work there \citep{zna78}.  

In Sec.\ \ref{indMem}, 
it is argued that the `inductive membrane' $\SN$ between \SOL\ and \SIL\ 
must be installed with a pair of unipolar induction batteries with electromotive forces (EMFs), $\calEout$ and $\calEin$, driving currents to flow through the circuits $\calCout$ and $\calCin$, in $\calDout$ and $\calDin$, respectively (see FIG.\ \rff{DC-C}). There will be a huge voltage drop $\Dl V$ across \SSN\ between the two EMFs for particle production. 


In Sec.\  \ref{m-mdGap},  
we show how critical is the Constraint $\PN{\vcj}=I(\ell,\Psi)=0$ on \SSN\ due to a  \emph{complete} violation of the force-free condition, because  
\emph{this surface} \SSN\ will be widened to such a gap $\GN$ as filled with zero-angular-momentum-particles (ZAM-particles) pair-produced due to the voltage drop $\Dl V$. 
We use such a simple model for the Gap structure as shown in FIG.\ \rff{GapI}.
Some important properties of the ZAM-particles, magnetization of the Gap $\GN$, and the plasma-shed on \SSN\ are clarified (see Secs.\  \ref{GapStruc}$\sim$~\ref{plasma-shed}).  

In Sec.\ \ref{BC-SN},   
we argue the `boundary condition' for determining the eigenvalue of $\OmF=\omN$ in the steady axisymmetric state. 
 A new evidence will be helpful in understanding an `enigmatic' flow of \emph{positive} angular momentum from the horizon membrane $\SffH$, beyond the inductive membrane $\SN$ covering the Gap $\GN$, to the infinity membrane $\Sffinf$  (see Sec.\ \ref{BCagain}). 
Thus, the ZAM-Gap $\SG$ will allow us to impose the conservation law of angular momentum as the `boundary condition' determining the eigenfunction $\OmF(\Psi)=\omN$; this means that the eigen-magnetosphere with $\omN=\OmF$ is `frame-dragged' by the hole's rotation (see Sec.\ \ref{Feigenv}). 
 
 Sec.\ \ref{TW-P-M}  
attempts to explain that the null surface \SSN\ is a new kind of \RTDy\ (RTD) in the GR setting \cite{lan84,oka15a}. We conjecture that this RTD involving the voltage drop $\Dl V$ between the two EMFs will bring up a new mechanism of pair-particle creation at work on \SSN\ toward $\GN$. As opposed to the `single-pulsar model' based on BH electrodynamics \cite{bla77,bla79,mac82,tho86}, we propose in this paper the `twin-pulsar model' based on gravito-thermo-electrodynamics (or GTED), because there will be two `pulsar-type magnetospheres' coexisting, outer prograde- and inner retrograde-rotating, respectively, with the RTD Gap between them for the supply of electricity and particles.   

Sec.\ \ref{FluxC} 
discusses the `energetics and structure' of the twin-pulsar model as opposed to that of the single-pulsar model (cf.,\ e.g.,\ Ch. IV D in \cite[]{tho86}).  
The last one, Sec.\ \ref{Dis-Con}, is devoted to discussion and conclusions with some remaining issues listed. 

In Appendix \ref{ED3/1F},  
by making full use of the $3+1$ formulation and the Membrane Paradigm \cite{mac82,tho86} with the freezing-in and force-free conditions as well as frame-dragging taken into account,  we examine necessary quantities and relations, such as the two outer and inner light surfaces \SOL\ and \SIL\ and the densities of the electromagnetic energy and angular momentum $\epsE$ and $\epsJ$ shown in Sec.\ \ref{EnegR}. 
Appendix \ref{SNshapAp} shows the place and shape of the null surface \SSN\ in the force-free magnetosphere for the parameter $0\lo h=a/\rH\lo 1$.  


\section{Pulsar electrodynamics: revisited}  \label{FFPM}  
\setcounter{equation}{0}
\numberwithin{equation}{section}  

The force-free pulsar magnetosphere filled with perfectly conductive plasma is considered under the two basic presumptions of the force-free and freezing-in conditions (see Eqs.\ (\rf{ff-a}a,b)).      
The combination of them creates a degenerate state of $\vcE\cdot\vcB=\vcj\cdot\vcE=\vcv\cdot\vcE =0$. We then have $\vcv=\vcj/\vre$ for the `force-free' plasma consisting of charge-separated particles. We presume that the neutron star (NS) is spinning around the `zero-angular-momentum' axis with the AV $\OmNS$. 

The magnetosphere is characterized by the two \emph{conserved} quantities, $\OmF(\Psi)$ and $I(\Psi)$. The AV $\OmF$ possesses two-sidedness as the FLAV and electric potential gradient, i.e.,\ $\OmF=-2\pi c(dA_0/d\Psi)$, and Ferraro's law of iso-rotation holds, i.e., $\OmF$ is constant along each field line (or FL) emanating from the star.  Also, a `current/angular-momentum duality' with respect to $I$ holds in the degenerate state. Note that both quantities $\OmF$ and $I$ cannot be determined within the force-free domains. 

When $\vcBp=-(\uvt\times\vcnb\Psi/2\pi \vp)$ is defined for the poloidal component of the magnetic field $\vcB$, the toroidal component is given by $\Bt=-(2I/\vp c)$, as the swept-back component of $\vcBp$ by inertial loading (e.g.,\ particle acceleration), and the electric field is given by $\vcEp=-(\OmF/2\pi c)\vcnb\Psi$ from the induction equation with the freezing-in condition ($E_{\rm t}\equiv 0$ by axial symmetry). 
The Poynting and angular momentum fluxes become, in terms of $\OmF$ and $I$, 
\beeq 	
\vcSE=\OmF \vcSJ, \ \ \vcSJ=(I/2\pi c)\vcBp , 
\lb{vcSEM/SJ} \eneq  
where the toroidal component of $\vcSE$ (and other fluxes) will be omitted throughout the paper. 

Obviously, there is no necessity of a breakdown of the force-free condition in the pulsar force-free magnetosphere ({except for} a negligible violation of force-freeness for additional particle supply needed), and ``energy and angular momentum from a rotating neutron star can be extracted by the mechanism of Goldreich \& Julian (1969)'' as is discussed in \citep{gj69, oka74}. When a rotating magnetized star like an NS behaves like a unipolar-induction battery,  the force-free conditions  will naturally be \emph{violated} inside the `battery,' 
because there will be no electric field in the matter-dominated interior of the star, i.e., $\vcE=0$, and also there is no-inertial loading $I=0$ on the FLs in the context of magnetospheric theory. Also, a unipolar induction  battery has no internal resistance, i.e., $I=0$ within the battery. Then, the potential gradient $\OmF$ for FLs  emanating from the star will be determined on the NS surface \SNS\ by imposing the `boundary condition,' i.e., 
\beeq
I=\OmF-\OmNS=0. 
\lb{b-cSNS}  \eneq  
In terms of pulsar {\em thermo}-electrodynamics, the whole spin-down energy from the star will be transferred through the resistive membrane $\Sffinf$ to kinetic energy of particles. 
 As usual, we may suppose that the EMF of the NS's unipolar-induction battery is given in terms of potential gradient $\OmF$ by 
\beeq
\calEns=-\frac{1}{2\pi c}\int^{\Psi_2}_{\Psi_1}\OmF(\Psi)d\Psi 
\lb{nsEMF} \eneq	 
\citep[\S 63]{lan84};\citep{oka12b},  
which drives currents along the current-field-streamline $\Psit$ with $\vcjp>0$ and return currents along $\Psio$ with $\vcjp<0$, where $\Psiz<\Psio<\Psic<\Psit<\bar{\Psi}$ and $\vcjp\lleg 0$ for $\Psi\lleg\Psic$ (see Eq\  (\rf{c-c-c})),  where $\bar{\Psi}$ is the last limiting FL satisfying $I(\Psiz)=I(\bar{\Psi})=0$ (see FIG.\ 2 in \cite{okam06} for one example of $I(\Psi)$). The surface return currents flow from $I(\Psit)$ to $I(\Psio)$, crossing FLs between $\Psio$ and $\Psit$ on the resistive membrane $\Sffinf$, and the Ohmic dissipation there formally represents particle acceleration taking place in $\Sffinf$. 

In reality, in order to determine the eigenfunction $I(\Psi)$, one needs to trace a kind of process to terminate the force-free domain by restoring particle inertia so far neglected in the force-free domain \cite{mac82}, which can be expressible by several equivalent ways \citep{okam06}. 
One of them is the `criticality condition' at the fast magneto-sonic surface S$_{\rm F}$ ($\approx$\Sinf) in wind theory, or the infinity resistive membrane $\Sffinf$ with the surface resistivity $\calR=4\pi/c=377$ Ohm in circuit theory, containing a layer from S$_{\rm F}$ at $\ell=\ell_{\rm F}$ to \Sinf\ at $\ell=\ell_{\infty}$. The conversion of field energy to kinetic energy takes place on $\Sffinf$ in the form of the MHD particle acceleration \citep{oka74,oka78}; 
\beeq
I_{\rm NS}(\Psi)=\frac{1}{2}\OmF(\Bp\vp^2)_{{\rm ff}\infty}	
\lb{I/Pul}  
\eneq  
(see Eq.\ (\rf{Iout/in-a})), which is equivalent to the `radiative' condition and Ohm's law for the surface current on $\Sffinf$.

Now, we can regard the toroidal field $\Bt$ as the swept-back component of $\vcBp$ due to inertial loadings on the terminating surface $\Sffinf$ of the force-free domain. Accordingly, the behavior of $I=I(\ell, \Psi)$ from the stellar surface to infinity will be described as follows; 
\begin{eqnarray}
	I(\ell,\Psi) =  \left\{
	\begin{array} {ll} 
		0 & ;\ell \lo \ell_{\rm NS}, \\[1mm] 
		I_{\rm NS}(\Psi) &;\ \ell_{\rm NS} \lo \ell \lo \ell_{\rm F},  \\[1mm]
		\to 0 & ; \ell_{\rm F} \lo \ell \lo \ell_{\infty}    
	\end{array}  \right.    \lb{I/Pul-M}  \end{eqnarray}    
(see Eq.\ (\rf{OL-I}) for a Kerr hole's force-free magnetosphere). 
We do not intend to consider complicated interactions of the force-free pulsar wind with the interstellar media permeated by the general magnetic FLs in this model. We assume simply that $I(\ell,\Psi)$ approaches zero for $\ell\to\infty$ and also $\vp\to\infty$. This presumes that all the Poynting energy will be transferred eventually to the particle kinetic energy.  The condition $I=0$ in $\ell \lo \ell_{\rm NS}$ means that there will be no unipolar induction battery that has an internal resistance. 

The wind theory and circuit theory must be complementary with each other, where $\OmF$ and $I$ take the two sides of the same coin respectively \cite{oka15a}; 
$\OmF$ related closely to the magneto-centrifugal particle acceleration in the former and to an EMF due to the unipolar induction battery on the NS surface in the latter, as related to the source of the Poynting flux at \SNS. Also, $I$ denotes the angular momentum flux as well as the current function. 

We stress that there is no reason nor necessity for the `force-free' condition to break down within the `force-free' pulsar magnetosphere, when a magnetized NS is regarded as behaving like a battery with particle acceleration as an external resistance far from the star, but with no internal resistance. The FLAV $\OmF$ is given by the NS's surface AV $\Om_{\rm NS}$. This is because each FL emanating from \SNS\ is anchored inside the star. 
This means that the efficiency of `extraction of energy' is $\eps_{\rm NS}=\OmF/\OmNS=1$ in the sense that no dissipation of rotational energy inside the NS takes place. This case will thermodynamically be called an `adiabatic extraction.'  

\section{Thermodynamics and electrodynamics}  \label{BHTsub} 
\setcounter{equation}{0}
\numberwithin{equation}{section}

\subsection{Basic thermodynamic properties of the Kerr hole}  \label{ther-rot}  
The no-hair theorem tells us that Kerr holes possess only two hairs.  This indicates that when one chooses the entropy $S$ and the angular momentum $J$, as two `extensive' variables, then all other thermodynamic quantities are expressed as functions of these two. For example, the BH's mass-energy $M$ is expressed in terms of $S$ and $J$, as follows;   
\beeq 
M=\sqrt{(\hbar cS/4\pi kG) + (\pi kcJ^2/\hbar GS)}.  
\lb{massF} \eneq  
As one can in principle utilize a Kerr hole as a Carnot engine \citep{KO91}, 
it may be regarded as a `thermodynamic object,' but not as an electrodynamic one, because the Kerr hole by itself stores no extractable electromagnetic energy. Therefore, the Kerr hole may be regarded as a huge rotating mass of `entropy matter' (see, e.g., \cite{bas90}), 
fundamentally different from the magnetized rotating NS which consists of `normal matter' with magnetic field lines emanating outside. Therefore, its evolutionary behaviors, such as due to extraction of angular momentum, are strictly governed by the four laws of thermodynamics (see, e.g., Ch.III C3 in \cite{tho86} for a succinct summary). 

The mass $M$ of the hole is divided into the `irreducible' and `rotational' masses \citep{tho86}, i.e.,
\begin{subequations}
	\begin{eqnarray}
		M=M_{\rm irr}+M_{\rm rot},  \hspace{0.3cm}~~~~~ \lb{massa} \\  
		M_{\rm irr}=\frac{M}{\sqrt{1+h^2}} =\sqrt{c^4 A_{\rm H}/16\pi G^2} =\sqrt{\hbar cS/4\pi kG}, \hspace{0.3cm} ~~~~~ \\ 
		M_{\rm rot}= M[1- 1/\sqrt{1+h^2}], \hspace{0.3cm}  ~~~~~ 	
	\end{eqnarray}  \lb{mass/irr/red}   \end{subequations}  
where $A_{\rm H}$ is the horizon surface area, and $h$ is defined as the ratio of $a\equiv J/Mc$ to the horizon radius $\rH$, i.e., 
\beeq 
h=\frac{a}{\rH} =\frac{2\pi kJ}{\hbar S}=\frac{2GM\OmH}{c^3}.  
\lb{Khole-h} \eneq  

The Kerr hole's thermo-rotational state is uniquely specified by its entropy $S$ and angular momentum $J$, or its mass-energy $M$ and the spin-parameter $h$ (see Eqs.\ (\rf{massF})--(\rf{Khole-h})). 

The evolutional state of the BH losing energy is specified as the time line of function $h(t)$ for the `outer-horizon in $0\leq h\leq 1$. We see $h=0$ for a Schwarzschild BH and $h=1$ for an extreme Kerr hole \citep{OK90,OK91} (see also Eqs.\ (10.4a,b,...,f) in \cite{oka92}). 

The hole's irreducible mass $M_{\rm irr}$ and surface area $A_{\rm H}$ are functions of $S$ only, but the rotational mass $M_{\rm rot}$ may be a function of $S$ as well as $J$. Therefore, when the hole loses angular momentum by, e.g., an influx of \emph{negative} angular momentum ($dJ<0$), the hole's total mass and rotational mass will decrease ($dM<0$ and $dM_{\rm rot}<0$, respectively), while $dM_{\rm irr}>0$ and $\Th dS>0$ always hold. 

Different from a magnetized NS consisting of `normal matter,' a Kerr hole with the mass function, $M=M(S,J)$ in Eq.\ (\rf{massF}) will be the biggest rotating mass of `entropy matter.' Then, a naive question comes to mind: \emph{how do magnetic field lines manage to thread and survive in `entropy matter' under the horizon?}  If a battery really existed in the horizon, this might indeed necessitate the threading of FLs into the `imperfect conductor' \citep{bla77, zna78} covered by the horizon, i.e., $(\vcBp)_{\rm H}\neq 0$. 

The zeroth law of thermodynamics indicates that two `intensive' variables, $\Th$ (the surface temperature) and $\OmH$, conjugate to $S$ and $J$, respectively, are constant on \SH, e.g., $\om\to\OmH$ for $\al\to 0$.  In passing, the third law indicates that ``by a finite number of operations one cannot reduce the surface temperature to the absolute zero with $h=1$.''  In addition, ``the finite processes of mass accretion with angular momentum cannot accomplish the extreme Kerr state with $h=1$, $\Th=0$ and $\OmH=c^3/2GM$'' (as discussed in \citep{OK91}). Incidentally, the `inner-horizon' thermodynamics can formally be constructed analogously to the `outer-horizon' thermodynamics \citep{OK93, cve18}. 

It is the first and second laws that govern the extraction process of energy, i.e., 
\begin{subequations}  \begin{eqnarray} 
		c^2 dM=\Th dS + \OmH dJ ,  \lb{1st-l} \\  
		\Th dS \geq 0 ,  
	\end{eqnarray} \lb{1-2laws}   \end{subequations}  
where $\Th$ and $\OmH$ are uniquely expressed in terms of $J$ and $S$ from Eq.\ (\rf{massF}) or $M$ and $h$ \citep{OK90}; 
\beeq 
		\Th=c^2 (\partial M/\partial S)_J, 
		\quad     
		\OmH=c^2 (\partial M/\partial J)_S.  
		\lb{2nd-l}   \eneq  

\subsection{The efficiency of the force-free magnetosphere}  \label{BH-FFM}  
We describe the hole's evolutionary change in terms of two variables $M(t)$ and $J(t)$, when the hole's angular momentum and energy is extracted through 
the force-free magnetosphere with \emph{conserved} quantities $\OmF(\Psi)$ and $I(\Psi)$, i.e., $dJ<0$ and $c^2 dM<0$.  
The change in universal time $t$ of the hole's total mass-energy become, from the first law (\rf{1st-l}), 	
	\beeq 
	c^2 (dM/dt)   = \Th(dS/dt)+  \OmH (dJ/dt).
	\lb{c.mass} \eneq  
Since the hole's gravity produces a gravitational redshift of ZAMO clocks, their lapse of proper time $d\tau$ is related to the lapse of global time $dt$ by the lapse function $\al$, i.e., $d/dt=\al$ \cite{mac82}. 

The angular momentum and energy fluxes are given by
\beeq
\vcSE =\OmF(\Psi)\vcSJ, \quad \vcSJ=( I(\Psi)/2\pi\al c)\vcBp,  
\lb{E-AmFlux} \eneq		
which are apparently the same as Eq.\ (\rf{vcSEM/SJ}) for the loss through the pulsar magnetosphere except the redshift factor $\al$. The output power $\calPE$ and the loss rate of angular momentum $\calPJ$ are given by
	\begin{subequations}  \begin{eqnarray}  
	 \calPE= -c^2 \dr{M}{t} = \oint \al\vcSE\cdot d\vcA  = \frac{1}{c}\int^{\bar{\Psi}}_{\Psi_0}\OmF(\Psi) I(\Psi) d\Psi,   \nonumber \\
	 \lb{TotFluxE}  \\		
	\calPJ =-\dr{J}{t} = \oint \al\vcSJ \cdot d\vcA  = \frac{1}{c}\int^{\bar{\Psi}}_{\Psi_0} I(\Psi) d\Psi  \hspace{0.8cm}
	\lb{TotFluxJ} 	
	\end{eqnarray} \lb{TotalFa,b}  \end{subequations} 
(see general expressions (3.89) and (3.90) in \cite{tho86}), where $\vcBp\cdot d\vcA=2\pi d\Psi$ and the integration is done over all open field lines in $\Psi_0\leq\Psi\leq \bar{\Psi}$.  Note that the FDAV $\om$ and the hole's AV $\OmH$ do not explicitly appear.  Here, we define the `overall' potential gradient, calculated from $\OmF(\Psi)$ weighted by $I(\Psi)$, i.e.,  
	\beeq \left.
	\OmFb = \int^{\bar{\Psi}}_{\Psi_0} \OmF (\Psi) I(\Psi) d\Psi \right/ \int^{\bar{\Psi}}_{\Psi_0} I(\Psi) d\Psi=\calPE/\calPJ.   
	\lb{GTEepsb}   \eneq  
Then, from Eqs.\ (\rf{TotFluxE},b) and (\rf{GTEepsb}), the first law in Eq.\ (\rf{1st-l}) splits into the two parts, i.e.,
         \begin{subequations}  \begin{eqnarray}			%
	c^2 dM  =\OmFb dJ ,  \lb{first-la} \\	
	\Th dS=- (\OmH-\OmFb) dJ, \lb{second-ib}         
	\end{eqnarray} \lb{first-II}  \end{subequations}  
because $c^2 dM= -\calPE dt=(\calPE/\calPJ) dJ=\OmFb dJ$, and hence we have Eq.\ (\rf{second-ib}) from the first law. 
Eq.\ (\rf{first-la}) corresponds to $\vcSE=\OmF \vcSJ$ in Eq.\ (\rf{E-AmFlux}), and 
seems to justify the statement that ``the radiation condition $\OmFb>0$ requires energy to flow outwards on all the FLs,'' (as discussed in \cite{bla77}) i.e., $c^2 dM<0$, and hence the direction of the `overall' energy flux $\vcSE=\OmF\vcSJ>0$ does not reverse on any given FL.  

The `overall' efficiency $\epsGTEb$ is defined by the ratio of ``actual energy extracted to maximum extractable energy, when unit angular momentum is removed'' (cf.\ \cite{bla77}), i.e.,\ from Eqs.\ (\rf{1st-l}), (\rf{TotalFa,b}a,b) and (\rf{GTEepsb}),
	\beeq 
	\epsGTEb =\frac{(dM/dJ)}{ (\partial M/\partial J)_{S}}=\frac{\calPE}{\OmH\calPJ} =\frac{\OmFb}{\OmH}. 
	\lb{epsGTE1} \eneq  
	\begin{figure*}
				~~~~~~~~~~~~~~~~~~~~~~~~~
	\includegraphics[width=11cm, height = 7cm, angle=-0]{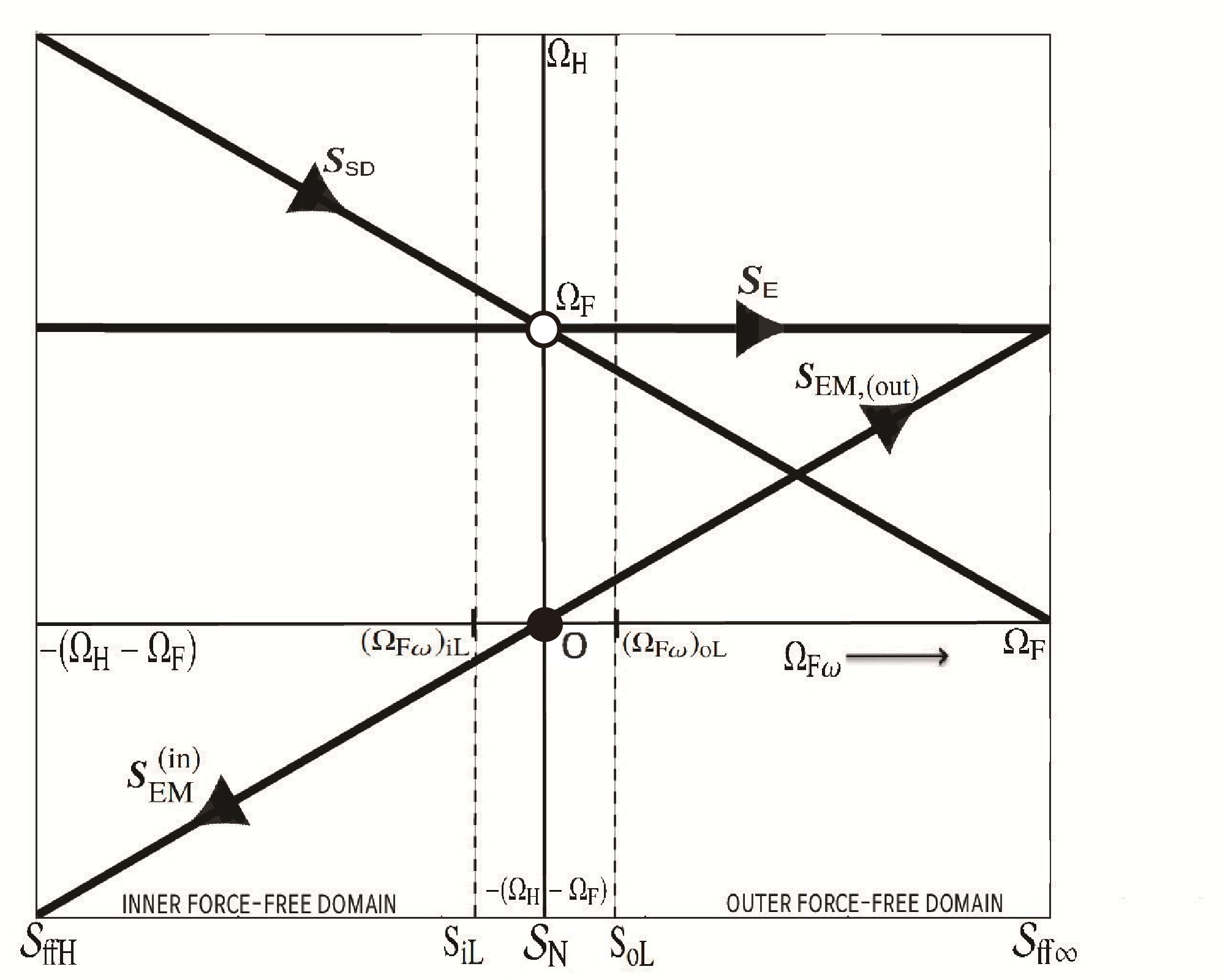} 
	\caption{ The $\OmFm$-dependence of three energy fluxes $\vcSE$, $\vcSEM$, and $\vcSsd$ along each field-current-stream-line (FCSL) (modified from figure 3 in \cite{oka09}).  The point is to `visualize' the effect of frame-dragging, by `coordinatizing' $\OmFm$ as well as $\om$. 	   
	The overall flux of extraction of energy is shown by $\vcSE=\OmF\vcSJ$, and its two components are given by $\vcSEM=\OmFm\vcSJ$ for the Electro-Magnetic Poynting flux and $\vcSsd=\om\vcSJ$ for the frame-dragging Spin-Down energy flux, respectively, where $\vcSJ$ is the angular momentum flux (see Eqs.\ (\rf{Sem})$\sim$(\rf{s.i})).  The ordinate $\vcEp=\OmFm=0$ expresses the null surface \SSN\ ($\om=\OmF$) \cite{oka92}, which always exists, ``when the hole is losing energy electromagnetically'' (as is written in the Caption of Figure 2 of \cite{bla77}). 
	 \emph{This surface} \SSN\ divides the force-free magnetosphere into the two, outer \emph{progradely}-rotating and inner \emph{retrogradely}-rotating, domains, $\calDout$ and $\calDin$, respectively, with the respective light surfaces \SOL\ and \SIL\ (see Sec.\ \ref{SoL/SiL} and Eq.\  (\rf{GapC})).  The three energy fluxes are linked to the three terms of the first law of thermodynamics on the horizon under the resistive membrane $\SffH$, i.e., $c^2(dM/dt)$, $\Th(dS/dt)$, and $\OmH(dJ/dt)$ in Eq.\  (\rf{c.mass}). Note that the hole is actually accepting an influx of \emph{negative} angular momentum from \SSN, which is equivalent to launching \emph{positive} angular momentum $\vcSJ>0$ from the hole. Also, note that there is no reversal of the {\em conserved} flux $\vcSE=\OmF\vcSJ$ (except $\vcSE=\vcSEM=\vcSsd=\vcSJ=0$ on \SSN, due to breakdown of the force-free condition; see Sec.\  \ref{NullS} and FIG.\ \rff{GapI}), while its two component fluxes $\vcSEM$ and $\vcSsd$ are not conserved. Indeed, $\vcSEM$ reverses direction, because the radiation condition for the Poynting flux is given by the sign of $\OmFm\ggel 0$. A pair of batteries as well as particle source must exist under the inductive membrane $\SN$ (see FIGs.\ \rff{DC-C}, \rff{F-WS}).    }
	\lbf{Flux-om} 	\end{figure*}

\newcommand{\IThisS}{\emph{this surface}}			
\subsection{Coupling of unipolar-induction with frame-dragging under the force-free and freezing-in conditions}  \label{FF-Frz-Conditions}   


The necessary and sufficient condition of energy extraction is to adapt to the  first three laws of thermodynamics. To do so, we rebuild a pulsar-type force-free magnetosphere. The first step is to make frame-dragging AV (or FDAV) $\om$ couple with unipolar-induction AV $\OmF$ through a formula given by  
	\beeq
		\vcE=\frac{1}{\al}\left(\vcnb A_0+\frac{\om}{c}\vcnb A_\phi \right)
		\lb{vcEp-A} \eneq 
 (see  Eq.\ (4.7) in \cite{mac82}), where $A_0$ is a scalar potential and $\vcA=(0,0, A_\phi)$ is a vector potential, and we put $A_\phi=\Psi/2\pi$. 
 
 The two fundamental conditions, i.e., the force-free and  freezing-in conditions are assumed in the BH's stationary axisymmetric magnetosphere with $\vcE_{\rm t} \equiv 0$; 
\begin{subequations}
\begin{eqnarray}  
	\vre\vcE+\vcj /c \times \vcB=0, \lb{ff-fz}  \\  
	\vcE+\vcv/c\times\vcB=0.  \lb{fi-c}   
	\end{eqnarray} \lb{ff-a}  \end{subequations}  %
When the FLAV is given by $\OmF=-2\pi c(dA_0/d\Psi)$ in the steady axisymmetric state \cite{mac82}, the electric field measured by the ZAMOs reduces to
	\begin{subequations}   \begin{eqnarray}  
		\vcEp=-\frac{\OmFm}{2\pi \al c}\vcnb\Psi, \lb{EpBt} \\
		\OmFm=\OmF-\om  \lb{OmFm}
		 \end{eqnarray} \lb{vcEp-OmFm}  \end{subequations}   
(see Sec.\ \ref{D-OmF}; Eq.\ (5.3) in \cite{mac82}), where $\OmFm$ is 
the ZAMO-measured FLAV. By the way, the toroidal component of magnetic field is given by
	\beeq
	\vcBt=-\frac{2I}{\vp\al c} \uvt.  
	\lb{vcBt} \eneq	
	
While the FLAV $\OmF$ complies with the iso-rotation, the ZAMO-FLAV $\OmFm$ does not. That is, the ZAMOs will see that the force-free magnetosphere is rotating differentially, because of frame-dragging; $\OmFm$ increases along each FL from the horizon \SH, through the vertical axis on the null surface \SSN\  (see FIG.\ \rff{Flux-om}), to the infinity surface \Sinf, as follows; 
		\beeq
		(\OmFm)_{\rm H}= -(\OmH-\OmF) < (\OmFm)_{\rm N}=0<(\OmFm)_{\infty}=\OmF.  \    
		\lb{SEM<0}  \eneq  
	{	(see Eq.\ \rf{GapC}). }
 We can thus `coordinatize' $\OmFm$, to see changes of various quantities along each FL. That is, when $\ell$ denotes distances from the horizon at $\ell=\ellH$ outwardly, it is convenient firstly to express $\om=\om(\ell,\Psi)$ and $\OmFm=\OmFm(\ell,\Psi)$, and next to `coordinatize' $\om$ and $\OmFm$, instead of $\ell$ (see FIGs.\ \rff{Flux-om}$\sim$\rff{F-WS}; \cite{oka15a}). 

\subsection{Two non-conserved energy fluxes, $\vcSEM$ and $\vcSsd$}  \label{2fluxes}  
Just as the overall energy flux $\vcSE$ corresponds to $c^2 dM/dt$, the other two energy fluxes are necessary to the other two terms  in Eq.\ (\rf{c.mass}). That is, we define the Electromagnetic Poynting flux $\vcSEM$ and the frame-dragging Spin-Down energy flux $\vcSsd$ as follows;  we have, from Eqs.\ (\rf{E-AmFlux}), (\rf{EpBt}), and (\rf{vcBt}); 
	\begin{subequations} \begin{eqnarray}
		\vcSEM=\frac{\al c}{4\pi}(\vcEp\times\vcBt)=\frac{\OmFm I}{2\pi\al c}\vcBp=\OmFm\vcSJ,  \lb{Sem/a} \ \  \\   
		\vcSsd =\frac{\om I}{2\pi\al c}\vcBp =\om \vcSJ,  \lb{Sem/b} \ \   
		\end{eqnarray}  \lb{Sem} \end{subequations}  
(see Eqs.\ (4.13) and (5.7) given in \cite{mac82}), which constitute the overall energy flux $\vcSE$ in Eq.\ (\rf{E-AmFlux}), i.e.,
	\beeq
		\vcSE=\vcSEM+\vcSsd=\OmF\vcSJ.  
		\lb{SEEMsd} \eneq  
Eq.\ (\rf{OmFm}) allows us to express the overall FLAV $\OmF$ in terms of the two non-conserved AVs, $\OmFm$ and $\om$, along each FL;  
\beeq
\OmF=\OmFm+\om.    
\lb{s.i}  \eneq 
This means that $\OmF(\Psi)$ is resolved to the ZAMO-FLAV $\OmFm$ (with `ZAMO-electric potential gradient') and the FDAV $\om$ (with `gravito-electric potential gradient'), and expresses the manner how  frame-dragging couples with unipolar induction, so as to comply with the first law of thermodynamics through the three energy fluxes (see Eqs.\ (\rf{1st-l}) and (\rf{SEEMsd})). 

Note that `distant static observers' may see that Ferraro's law of iso-rotation holds unconditionally, and $\vcSE$ is conserved along each FL, i.e., $\vcnb\cdot\al\vcSE=\OmF\vcnb\cdot\al\vcSJ=0$, and yet they may not notice the existence of the two non-conserved fluxes $\vcSEM$ and $\vcSsd$, whereas the ZAMOs circulating the hole with $\om$ will see that ``the iso-rotation is violated,'' and detect that $\vcSE$ consists of $\vcSEM=\OmFm\vcSJ$ and $\vcSsd=\om\vcSJ$. 
Therefore, the ZAMOs will be correctly aware that the Kerr hole has an exquisite device of frame-dragging to split the `overall' flux $\vcSE=\OmF\vcSJ$ into the two fluxes, thereby observing the first and second laws of thermodynamics strictly. 

Contrary to the `overall' energy flux $\vcSE$ directed always outward, the Poynting flux $\vcSEM$ changes direction on the null surface \SSN, where $\OmFm\ggel 0$ (see FIG.\ \rff{Flux-om}).  This will naturally require a relevant pair of `surface' unipolar induction batteries on \SSN, which must be placed back-to-back to each other, but oppositely directed. In addition,  a particle source related must be established on the null surface in the force-free limit. This means that the force-free condition as well as the freezing-in condition must break down on \SSN\ (see Secs.\ \ref{NullS}, \ref{indMem}).  

The ZAMO-FLAV $\OmFm\ggel 0$ not only expresses the `radiation condition' for the ZAMO-Poynting flux $\vcSEM$, but also indicates the direction of the magnetic sling-shot effect due to magneto-centrifugal force at work in wind theory.  In particular, we remark that the null surface \SSN\ (with $\PN{\OmFm}=\OmF-\omN=0$) is nothing but the magneto-centrifugal divider of the force-free magnetosphere into the two domains; the outer domain $\calDout$ \emph{prograde}-rotating ($\OmFm>0$) and the inner domain $\calDin$ \emph{retrograde}-rotating ($\OmFm<0$). Also, $\OmF=\omN$ means that the whole magnetosphere is frame-dragged into rotation by the hole's rotation in the steady state (see Sec.\ \ref{BC-SN}).  


\subsection{The two light surfaces and two eigenfunctions for $I(\Psi)$ }  \label{I&LSs} 

\subsubsection{The two light surfaces \SOL\ and \SIL}  \label{SoL/SiL}  
The field-line-rotational-velocity (FLRV) is given by $\vF=\OmFm\vp/\al$ (see Eq.\ (\rf{vp/vt})). Then, the ZAMOs will see $\vF=0$ on \SSN\ and $\vF=\pm c$ at the two light surfaces \SOL\ and \SIL\ in $\calDout$ and $\calDin$, respectively  (see FIG.\ \rff{Flux-om}), as follows;
		\begin{subequations} \begin{eqnarray}    
		\OmFmOL=+c(\al/\vp)_{\rm oL}, \  \OmFmIL=-c(\al/\vp)_{\rm iL} ,  \hspace{0.6cm} \lb{Sil/oL} 
		   \end{eqnarray}    
and then we have $\om_{\rm oL}$ and $\om_{\rm iL}$, relative to $\omN$, 
		\begin{eqnarray} 
		\om_{\rm oL} =\omN - c(\al/\vp)_{\rm oL} ,  \  \om_{\rm iL} =\omN + c(\al/\vp)_{\rm iL}. \hspace{0.5cm}
		 \lb{om/iLa}      \end{eqnarray}   \end{subequations}  
 It is because of the counter rotation of $\calDin$ by frame-dragging that the null surface \SSN\ exists between the two light surfaces, \SOL\ and \SIL\ on the \emph{same} FLs. Therefore, we see `$\om_{\rm oL} < \OmF <  \om_{\rm iL}$,' which will correspond to an inequality `$\Omega_{\rm min}<\Omega< \Omega_{\rm max}$' given below equation (15) in \citet{bla22} (see Eqs.\ (\rf{GapC}) \& (\rf{dS>0a})).  	
Also, readers may refer to Eqs.\ (\rf{xoiL}a,b) for the behaviors of \SOL\ and \SIL\ for the slow-rotation limit of $h\to 0$. 

It was already pointed out \citep[the footnote at p.443]{bla77}  that ``The outer light surface corresponds to the conventional pulsar light surface and physical particles must travel radially outwards beyond it. Within the inner light surface, whose existence can be attributed to the dragging of inertial frames and gravitational redshift, particles must travel radially inwards.''   It was thus concluded in \cite{zna77} that ``the particle-production mechanism described in \cite{bla77} must operate between the two light surfaces.'' 
 
We can see furthermore that physical observers will see the electric field reversing direction on the surface $\omN=\OmF$. Inside this surface, they see a   Poynting flux of energy going toward the hole. (When $0<\OmF<\OmH$, i.e., when the hole is losing energy electromagnetically, this surface always exists.) (see the caption of Figure 2 in \cite{bla77}; Eqs.\ (\rf{ThdS/dt}) and (\rf{dS>0a}), FIG.\ \rff{Flux-om}). 	
{The ZAMOs will then understand that {\em a sufficiently strong flux of angular momentum leaving the hole} will be equivalent to {\em a sufficiently strong in-flux of `{negative}' angular momentum leaving this surface \SSN\ inwardly.}} 
We can find a related statement in \citep[p.520]{pun90} that ``the magnetospheric plasma is produced in the spatial region between \SIL\ and \SOL; at the \SIL\ (\SOL), the magnetic field rotates backward (forward) at the speed of light relative the plasma,'' and actually this is consistent with the existence, between \SIL\ and \SOL,\  of the null surface \SSN\ where $\OmFm\ggel 0$. It will be shown in Secs.\ \ref{NullS}, \ref{indMem}, and \ref{m-mdGap} that the breakdown of the force-free condition on the null surface \SSN\ will set up a workplace for unipolar induction and particle production. 
 

\subsubsection{The eigenfunctions $\Iout$ and $\Iin$ }  \label{Iout/in}  

Let us determine the current/angular momentum function $I(\Psi)$ for the two SC and GR domains, by \emph{terminating} the force-free domains on the \emph{resistive} membranes $\Sffinf$ and $\SffH$ near \Sinf\ and \SH, respectively; precisely speaking, on the outer and inner fast-magnetosonic surfaces \SoF\ and \SiF\ (see, e.g.,\ \cite{oka78,ken83, pun90}, \footnote{In the case of treating the whole MHD theory of BH winds, not only the outflow but also the inflow must pass smoothly through three critical surfaces; slow, intermediate, and fast magnetosonic surfaces \cite{web67,mic69, oka78, ken83, pun89, pun90,oka99,oka02,oka03}. In the force-free theory, the last two surfaces reduce to \SOL, \SIL\ and \SoF, \SiF, respectively, although the slow surface is generally neglected.}),  
where \SH$\lo \SffH\lo$\SiF$<$\SIL$<$\SSN$<\,$\SOL$<$ \SiF $\lo\Sffinf\lo$\,\Sinf\ (see Eq.\ (\rf{GapC})).  We have the behavior of $\vcB$ and $\vcEp$ toward \Sinf\ and \SH\ as follows; 
	\[ \vcB^2=\vcBp^2 +\vcBt^2=(2I/\vp\al c)^2+(\Bp\vp)^2/\vp^4   \]
	\begin{eqnarray}
	 \simeq  \left\{
	\begin{array} {ll} 
	 (2\Iout /\vp c)^2 &; \Sffinf, \    \al \to1, \  \vp\to \infty, \\[1mm]
	 (2\Iin/\vp\al c)^2 &; \SffH, \  \al\to 0, \ \om\to \OmH,
		\end{array}  \right.    \lb{Iout/inA} 
	 \end{eqnarray} 
	\newline \\ \newline \\ \newline
	\beeq
	 \vcEp^2= (\OmFm\vp/\al c)^2 \Bp^2  \hspace{4cm} \nonumber \eneq
	\begin{eqnarray}
	 \simeq  \left\{
	\begin{array} {ll} 
	\left( \displaystyle{\frac{\OmF}{\vp c}}\right)^2 (\Bp\vp^2)^2 &; \Sffinf, \    \al\simeq 1, \  \vp\to \infty, \\[2mm]
	\left( \displaystyle{\frac{\OmH-\OmF}{\al\vp c}}\right)^2 (\Bp\vp^2)^2 &; \SffH, \  \al\to 0, \ \om\to \OmH.
		\end{array}  \right.    \lb{Iout/inB}  \end{eqnarray}	
Then, $(\vcB^2-\vcE^2)\to 0$ reduces to the so-called `criticality condition' or, equivalently, the `radiative condition' on $\Sffinf$ and $\SffH$ for $\Iout$ and $\Iin$ (see \cite{zna77,oka92,oka06}); 
	\begin{subequations} \begin{eqnarray}
		\Iout =\displaystyle{ (1/2)\OmF(\Bp\vp^2)_{{\rm ff}\infty}}  &;\ \Sffinf, \lb{Iout/in-a} \\[1mm]
		\Iin =	(1/2) (\OmH-\OmF) (\Bp\vp^2)_{\rm ffH} &;\ \SffH    \lb{Iout/in-b}
	\end{eqnarray} \lb{Iout/in}  \end{subequations}  
(also see Eq.\ (6.6b) in \citep{mac82}). The former $\Iout$ expresses the external resistance of particle acceleration on the resistive membrane $\Sffinf$, and the latter $\Iin$ also does the external resistance of entropy production on another resistive membrane $\SffH$, 
but this is not an internal resistance of a horizon battery (if any) \cite{bla79}. 

For an arrangement of characteristic surfaces, we see 
	\beeq
	\OmH\ggo\omiF>\omiL>\omN=\OmF>\omoL> \omiF \ggo 0 
		\lb{GapC} \eneq  
from the horizon to infinity along each FL,  where $\omiF$ and $\omoF$ are the values of $\om$ on the inner and outer fast-magnetosonic surfaces \SoF\ and \SiF,\ respectively (see [47]).  Note that $\omiF\approx \OmH$ and $\omoF\approx 0$ in the force-free \emph{limit}.  

By Eqs.\ (\rf{Iout/in}a,b), the energy and angular momentum fluxes in Eqs.\ (\rf{TotalFa,b}a,b) possess different forms in the outer and inner domains;    
	\begin{subequations}  	\begin{eqnarray}  
	\calPEout  
		= \frac{1}{c}\int^{\bar{\Psi}}_{\Psi_0} \OmF \Iout  d\Psi   ,  \lb{calPE/out} 	\\ 
	\calPEin  
		= \frac{1}{c}\int^{\bar{\Psi}}_{\Psi_0} \OmF \Iin  d\Psi. 
		\lb{calPE/in} 
		\end{eqnarray} \lb{calPEout-in}  \end{subequations}  
	\begin{subequations}  	\begin{eqnarray}
	\calPJout  
		= \frac{1}{c}\int^{\bar{\Psi}}_{\Psi_0} \Iout d\Psi , \hspace{1.1cm} 	\\ 
	\calPJin 
		= \frac{1}{c}\int^{\bar{\Psi}}_{\Psi_0} \Iin d\Psi . 
		\hspace{1cm} \ \  
		\end{eqnarray} \lb{calPJout-in}  \end{subequations}  
As seen in Sec.\ \ref{2fluxes}, the force-free magnetosphere is divided by the null surface \SSN\, where $\OmFm=0$ (and hence $\vcEp=\vcSEM=0$), but still the overall energy and angular momentum fluxes  always seem to flow outward. Hence, we will impose $\DN{\calPE}=\DN{\calPJ}=0$ (Eq.\ (\rf{DNX})) as well as $\PN{\calPE}=\PN{\calPJ}=0$ (Eq.\ (\rf{SNd})) as the `boundary condition' on the null surface \SSN\ where the force-free condition breaks down, in order to finally determine the eigen-FLAV $\OmF=\omN$ (see Sec.\ \ref{BCagain}). 

Next, the {\em surface} currents $\calI_{{\rm ff}\infty}$ and $\calI_{\rm ffH}$ flowing on the resistive membranes $\Sffinf$ and $\SffH$ with the surface resistivity $\RH=\calRinf=4\pi/c$, respectively, are 	
	\begin{subequations} 
	\begin{eqnarray}
		\calI_{\rm ff{\infty}}=\left(\frac{\Iout}{2\pi \vp}\right)_{\rm ff{\infty}} 
		  =\left(\frac{c}{4\pi} \frac{\OmF\vp}{c} \Bp\right)_{\rm ff{\infty}} =\left(\frac{E_{\rm p}}{\calRinf}\right)_{\rm ff{\infty}}, \nonumber \\[1mm]
		  \lb{calIffinf}  \end{eqnarray}		
		 \begin{eqnarray}  
		 \calI_{\rm ffH}= \left( \frac{\Iin}{2\pi \vp} \right)_{\rm ffH} 
		 =	\left( \frac{c}{4\pi}  \frac{(\OmH-\OmF)\vp}{c} \Bp \right)_{\rm ffH} 
		 =\left( \frac{E_{\rm p}}{\RH} \right)_{\rm ffH}.	 \nonumber \\[1mm]
		   \lb{calIffH} 		
		\end{eqnarray} \lb{calIinfH}  \end{subequations}  
(see FIG.\ \rff{DC-C} and Eqs.\ (\rf{dSHH/dt}) and (\rf{Sffinf-M})). Ohmic dissipation of these two \emph{surface} currents corresponds to particle acceleration and entropy production in each closed-circuit $\calCout$ and $\calCin$ (see Eqs.\ (\rf{ThdS/dt}) and (\rf{Sffinf-M})). 
 
It must be on \emph{this surface} \SSN\ that the influx of negative angular momentum in $\calDin$ (or equivalently the outward flux of positive one) must cancel out the outward flux of positive angular momentum in $\calDout$, 
i.e., $\vcSJout= \vcSJin= -\vcSJinU$.  It turns out  that this condition $\Iout=\Iin$ is to yield the `boundary condition' to finally determine $\omN=\OmF$ for the whole magnetosphere frame-dragged into rotation with $\om=\OmF$ (see Eq.\ (\rf{DN/SN})).

\subsection{A restriction by the second law on the efficiency}  \label{second/restr}  
In order to understand entropy production in the resistive membrane $\SffH$ (i.e.,\ the `stretched' horizon ${\cal H}^S_t$), a general expression (3.99) 
in \cite{tho86} is helpful; 
	\begin{subequations}  \begin{eqnarray} 
		\TTH \frac{dS}{dt}
		=\oint_{{\cal H}^S_t}  \RH \vec{{\cal J}}_{H}^2 dA  
		=\oint_{{\cal H}^S_t} \vec{\EH}\cdot \vec{{\cal J}}_{H} dA \nonumber \hspace{1cm} \\
		= \frac{1}{4\pi} \oint_{{\cal H}^S_t}  ( -\vec{\EH}\times\vec{\BH} )\cdot \vec{n } dA, \hspace{0.5cm} 
		\lb{dSHH/dt}  \end{eqnarray} 
where the Ohm's law holds on $\SffH$, i.e., $\vec{E}_{\rm H}=\RH \vec{{\cal J}}_{\rm H}$ (see Eq.\ (\rf{Sffinf-M}))  with the surface resistivity $\RH=4\pi/c=377$ Ohm (equal to $\calRffH$) and $ \vec{{\cal J}}_{H}\equiv\calI_{\rm ffH}$ is the surface current (see Eq.\  (\rf{calIffH})). Ohm's and Ampere's laws are equivalent to the radiative condition, i.e., $\vec{B}_H=\vec{E}_H\times \vec{n}$, as a result. Thus, the inflow of the Poynting flux gives rise to the Joule heating, leading to the BH's entropy increase. That  is, inserting $\vcSEM$ in the last of expressions (\rf{Sem/a}) into Eq.\ (\rf{dSHH/dt}), we can express the second law in terms of $\OmF$ and $\OmFb$, from Eq.\ (\rf{GTEepsb}),
	\begin{eqnarray} 
			\TTH \frac{dS}{dt} = \int_{{\cal S}_{\rm ffH}} \RH \calI_{\rm ffH}^2 dA  = - \oint_{{\rm S}_{\rm ffH}} \al\vcSEM\cdot d\vcA  \nonumber  \\
			= \frac{1}{c}\int^{\bar{\Psi}}_{\Psi_0}(\OmH-\OmF) I d\Psi =\OmH\calPJin -\calPEin     \nonumber \\ 
			=-(\OmH-\OmFb) \dr{J}{t}>0, 
			  \hspace{0.5cm}   
			\lb{ThdS/dt}  
		\end{eqnarray} \lb{Entro} \end{subequations} 	 
where Eq.\ (\rf{calIffH}) is used for the surface current on $\SffH$. This corresponds to Eq.\ (\rf{second-ib}). Therefore, the influx of \emph{a Poynting flux} \cite{bla77} leads to an increase of the hole's entropy through the Ohmic dissipation of the surface current in Eq.\ (\rf{dSHH/dt}) (see Eq.\  (\rf{H/resistanc}) in Sec.\ \ref{FluxC}).  
 
The loss of the hole's angular momentum due to the surface Lorentz braking torque  also is given in a general form (3.100) in \cite{tho86}, as follows;   
	\begin{subequations} 	\begin{eqnarray}
		\dr{J}{t}=\oint_{{\cal H}^S_t} (\sg_{H} \vec{\EH} + \vec{{\cal J}}_{H}\times\vec{B}_{n})\cdot \vec{\xi} dA  \lb{dJ/dt},     
		\end{eqnarray} \lb{TPM-TD}      
which leads to 		
	\begin{eqnarray}
		\dr{J}{t}  = -\int_{{\cal S}_{\rm ffH}} (\al  \vccalI_{\rm ffH}/c \times \vcBp)\cdot d\vcA  \hspace{2.5cm} \nonumber \ \ \ \ \  \\[1mm]  	  
		= - \int_{{\cal S}_{\rm ffH}} \al \vcSJin \cdot d\vcA= - \frac{1}{c} \int ^{\bar{\Psi}}_{\Psi_0} \Iin (\Psi) d\Psi  \nonumber \ \ \ \ \ \ \ \ \ \\  
		=  \int_{{\cal S}_{\rm ffH}} \al \vcSJinU \cdot d\vcA=  \frac{1}{ c} \int ^{\bar{\Psi}}_{\Psi_0} \IinU (\Psi) d\Psi<0,  	\hspace{0.8cm} 
		\lb{torque/SffH} 
		\end{eqnarray}   \end{subequations} 
where $\IinU=-\Iin$ and $\vcSJinU=-\vcSJin$. Eqs.\ (\rf{torque/SffH}) show that  ``a sufficiently strong flux of angular momentum leaving the hole...'' (as written in \cite{bla77}) takes place by the surface braking torque on $\SffH$, which is equivalent to the inflow of {\em negative} angular momentum due to the ingoing magneto-centrifugal wind in the inner domain $\calDin$ with $\OmFm<0$ (see Eq.\ (\rf{TotFluxJ})).   
		
The hole's ingenious trick of self-extracting its resource from under the horizon will be as follows: 
The second law 
requires an inflow of \emph{a Poynting flux} $\vcSEM=\OmFm\vcSJ$ from \emph{this surface} \SSN. This is followed by an {\em in}-flow of {\em negative} angular momentum, which is equivalent to an {\em out}-flow of {\em positive} angular momentum, leading to $dJ<0$. This couples with frame-dragging to induce an {\em in}-flow of {\em negative} energy, which leads to an outward spin-down energy $\OmH|dJ|$, with a part of it covering the cost of extraction, i.e., the entropy increase $\Th dS=(\OmH-\OmFb)|dJ|$, and with the rest becoming the outgoing overall energy flux $c^2 |dM|=\OmFb|dJ|$ from \emph{this surface} \SSN\ (see Sec.\  \ref{BCagain}). 
 	
As seen in Eqs.\ (\rf{second-ib}) and (\rf{epsGTE1}),  
it is the second law in Eq.\ (\rf{ThdS/dt}) that imposes the following restrictions on $\OmF$, $\OmFb$, $\epsGTE$, $\epsGTEb$, $\vcSE$, and $\calPE$;   
	\begin{subequations}   \begin{eqnarray} 			
		0\lo \OmF\approx \OmFb \lo \OmH, \lb{dS>0a}  \\ 
		0 \lo \epsGTE \approx \epsGTEb \lo 1 \lb{dS>0b},  \\ 
		\vcSE =\OmF\vcSJ \lo \OmH \vcSJ,  	\lb{dS>0c} \\ 
		\calPE=\OmFb \calPJ \lo \OmH\calPJ  \lb{dS>0d}  
	\end{eqnarray} \lb{second}  \end{subequations}  
(see Eqs.\ (4.6) and (4.7) in \cite{bla77}), which ensure that ``when the hole is losing energy electromagnetically, the null surface \SSN\ (with $\om=\OmF$) always exists'' (as discussed in \cite{bla77}). The FLAV $\OmF=\omN$ on \SSN\ will mean that {\em magnetic field lines which thread \SSN\ will be dragged into rotation with $\omN$,} although the final eigenvalue of $\omN$ is not yet determined (see Sec.\  \ref{BCagain}). 
	

\subsection{The energy and angular-momentum densities of the electromagnetic fields} \label{EnegR}	 
For the densities of the field energy and angular momentum, substituting $\vcEp$ from Eq.\ (\rf{EpBt}) into Eqs.\ (2.30a) and (2.31a) in \cite{mac82}, we have 	\begin{subequations} \begin{eqnarray}
		\epsE= \frac{\al\Bp^2}{8\pi} \left[1+\frac{\Bt^2}{\Bp^2} +\frac{\vp^2}{\al^2 c^2}(\OmF^2 -\om^2 )\right] ,  \lb{epsEba} \\ 
		\epsJ=\frac{\vp\vF}{c}\Bp^2 = \frac{\OmFm (\vp\Bp)^2}{\al c}  \lb{epsJbb}  
		\end{eqnarray} \lb{epsEJab}  \end{subequations}  
(see also Eq.\ (2.17a) in \cite{oka92} and Eq (55) in \cite{kom09}), where $\epsE$ and $\epsJ$ are `explicit' functions of $\om$ along each  FL labeled with $\Psi$. We will work out important properties of $\epsE$ and $\epsJ$ in Sec.\  \ref{EnergyD}. 
	
The angular momentum density $\epsJ$ changes sign on the null surface \SSN, and we refer to the zero-angular-momentum-density surface as \SZAM, which accords with \SSN. As the ZAMOs will see, the spin axis of the magnetosphere reverses direction, and hence the inner domain $\calDin$ counter-rotates against the outer domain $\calDout$, and in turn the Poynting flux $\vcSEM$ reverses its direction as well, as seen in $\vcSEM\ggel 0$ for $\OmFm\ggel 0$.  
	
Expression (\rf{epsEba}) for $\epsE$ shows that there will be such a surface $\SepsE$ that divides the inner domain $\calDin$ farther into the two regions by $\epsE(\om,\Psi)\ggel 0$ for $\om \lleg \omepsE$ between \SSN\ and $\SffH$, where $\omepsE$ is a solution of
	\beeq
	\om^2 = \omN^2 +\frac{\al^2 c^2}{\vp^2} \left(1+\frac{\Bt^2}{\Bp^2} \right).  
	\lb{NER} \eneq 
Here, $\al/\vp$ and $\Bt/\Bp$ are thought of as functions of $\om$ and $\Psi$ (see Sec.\ \ref{EnergyD} for some analyses). This obviously indicates $\OmH> \omepsE>\omN=\OmF$ (see Eqs.\ (\rf{nearSinf}) and (\rf{epsE/N})). Therefore, it is the frame-dragging that produces not only the inner domain $\calDin$ of $\OmFm \leq 0$ with \SIL, but also a region of the negative-energy density of $\epsE\leq 0$ in $\OmH\geq\om\geq \omepsE$. We assume $\omepsE\approx \omiL$. 
	
The above result suggests that the `negative-energy region' will surely extend from near \SIL\ to \SH\ in the force-free magnetosphere of the Kerr hole. The existence of the `negative-energy region' may indeed be a necessary condition, but not quite a sufficient condition for electromagnetic extraction of energy. 
What is essential on the first law of thermodynamics is the `inflow of negative angular momentum' (i.e., $dJ<0$, equivalent to outflow of positive one), which is followed by the `inflow of negative energy' (i.e., $\OmH dJ<0$, equivalent to the outflow of positive energy). Therefore, the existence of the `negative energy region' near the horizon itself may not be an exact or direct indicator of energy extraction from a Kerr hole. Important is the evidence that the force-free magnetosphere is divided into the two, prograde-rotating SC and retrograde-rotating GR, domains by the null surface \SSN, which coincidences with the 
 `Zero-Angular-Momentum-Density surface' \SZAM\ where $\epsJ=\OmFm=0$. 

\section{Breakdown of the iso-rotation law, force-free condition, and freezing-in condition} \label{NullS}  
	\setcounter{equation}{0}
	\numberwithin{equation}{section} 
 The direction of the Poynting flux $\vcSEM=\OmFm\vcSJ$ reverses on every FL threading the null surface \SSN, where $\vcEp$ and $\OmFm$ reverse (see Eqs.\ (\rf{vcEp-OmFm}a,b), (\rf{Sem/a})). Therefore, the `natural radiation' at infinity and on the horizon requires energy to flow out from the null surface \SSN\  in opposite ways, i.e., outward and inward (see FIG.\ \rff{Flux-om}). This means that the force-free and freezing-in conditions must break down. But the question is {\em how and where?} This question now seems quite simple. When `$\vcEp=0$' \footnote{We can confirm the vector $\vec{E}$'s reversal of direction, in Figure 2 and its caption in \citet{bla77}, in Fig.\ 3 in \citet{phi83b} and in Fig.\ 38 in \citet{tho86}, as well as in each ordinate of FIGs \rff{Flux-om}, \rff{GapI}, \rff{DC-C}, \rff{F-WS} at the null surface \SSN\ $\vcEp=0$ in this paper. }   
is inserted into Eqs.\ (\rf{ff-a}a,b),  we can check what happens. The ZAMOs will see that following quantities must necessarily vanish on the null surface \SSN\ where $\vcEp=0$; 
	\begin{subequations} \begin{eqnarray} 
			\PN{\vcEp} =\PN{\vcSEM}=\PN{\vre}=\PN{\vF}=\PN{\epsJ} =0,  \hspace{0.3cm} \ \ \ \ \ \ \lb{SNa} \\  
			\PN{\vcj}=\PN{I}  =\PN{\Bt} =\PN{\vcSJ} =\PN{\vcSsd}=\PN{\vcSE}=0, \hspace{0.3cm} \ \ \ \ \ \ \lb{SNb} \\  
			\PN{\vcv}=\PN{\vcj/\vre}=0,  \hspace{1cm}  	\lb{SNc}  \\ 	
			\PN{\calPE}=\PN{\calPJ}=0,  \hspace{1cm}  	\lb{SNd}   	
		\end{eqnarray} \lb{EqSN} \end{subequations}    
where Eqs.\ (\rf{ff-a}a,b), (\rf{bh/rhob}a,b), (\rf{Sem}a,b), (\rf{SEEMsd}), (\rf{vcjpO}), (\rf{charge1}), (\rf{vc-vjA}), (\rf{jtc}a,b), and (\rf{epsJbb}) are used.  We 
denote the value of any function $X(\OmFm,\Psi)$ on \SSN;   
	\beeq
	\PN{X}=X(0,\Psi). 
	\lb{PNX} \eneq   


 The first group in Eq.\ (\rf{SNa}) contains the quantities that reverse direction or change sign across \SSN. The reversal of these quantities originates from the counter-rotation of the inner domain $\calDin$ ($\OmFm<0$). 
The second ones in Eq.\ (\rf{SNb}) stem from the breakdown of force-freeness due to $\PN{\vcEp}=0$, and contain quantities that vanish, but do not reverse direction nor change sign at first sight (see, e.g., FIG.\ \rff{GapI} for $I$). 
Finally, the third one contains $\PN{\vcv}=0$ resulting from the breakdown of freezing-in-ness, and $\PN{\vcj/\vre}=\PN{dI/d\Psi)/\vre}=0$, which, similarly to those in the first group, reverses and changes sign due to the existence of $\vre$. 

The above Constraints will unequivocally rebuild the whole force-free magnetosphere, from a single-pulsar model to a twin-pulsar model, as follows:  
	\benu   
	\item 
	Constraint $\PN{\vcj }=0$ plays a role of a `circuit breaker' as a safety device, to block `acausal' currents on the null surface from a horizon battery (if any) to external resistances such as particle acceleration (cf.\ \cite{bla79,mac82,tho86}).  This instead indicates a necessity of a pair of `surface batteries' placed contiguously (back-to-back) at both sides of  \SSN, yet oppositely directed. Between the batteries is the particle source, in which the voltage drop between the two EMFs will produce pair-particles  (see Sec.\ \ref{indMem}, FIG.\ \rff{DC-C}; \cite{oka15a}). The purpose of breaking down freezing-in-ness on \SSN, $\PN{\vcv}=0$, is also to sever the streamlines there. Although the current $\vcj$ does not reverse direction, the velocity $\vcv$ does, i.e., $\vcv\ggel 0$ for $\OmFm\ggel 0$ (see FIG.\ \rff{F-WS}). 
	\item  
	Since $\vre\vcEp$ vanishes but does not change sign on \SSN, this reacts back on the force-free condition in Eq.\ (\rf{ff-fz}), producing $\PN{\vcj }=\PN{I}=0$ by Eq.\ (\rf{vcjp2}), whereas the change in direction of $\vcEp$ across \SSN\ is taken over the particle velocity $\vcv$ as it is, 
by the freezing-in condition in Eq.\ (\rf{fi-c}).  This is because the axial symmetry $\vcE_{\rm t}=0$ will lead to $\vcvp=\kappa \vcBp$ and  $\kappa=-(1/\vre \al)(dI/d\Psi)=0$ on \SSN\ (see Eq.\ (\rf{kappa})), 
and hence the ZAMOs will see that the particle velocity $\vcv$ behaves like $\OmFm\ggel 0$ across \SSN, contrary to $\vcjp$. Note that, when $\vcEp\propto \OmFm$ across \SSN, this nature must straightly be succeeded to the particle velocity, i.e., $\vcv\propto \OmFm$ as well. 

	\item 
Therefore, there will be no `single' circuit allowed, with such a current crossing \SSN\ due to a `single' battery at any plausible position 
\cite{tho86}. Each electric circuit must be closed within its respective force-free domain ($\calDout$ or $\calDin$), with each EMF ($\calEout$ or $\calEin$) in Eq.\ (\rf{EMF-ab}), and with each eigenvalue $I(\Psi)$ ($\Iout$ or $\Iin$) in Eq.\ (\rf{Iout/in}). 
	
	\item 
Constraint $\PN{\vcv}=0$ means that the particles pair-created in the Gap $\GN$ are `\zam' particles circulating with $\omN=\OmF$ with no other macroscopic motion.  That is, Constraints $\PN{I}=\PN{\vcSJ}=\PN{\vcSE}=0$ mean that no angular momentum nor energy is transported across \SSN, although the FLs are continuous. It is essential to remark that the toroidal field $\Bt$ is a swept-back component of the poloidal component $\vcBp$ due to inertial loadings in the `resistive membranes' $\Sffinf$ and $\SffH$ (see FIG.\ \rff{GapI}).  Thus, $\PN{I}=0$ means that there must be a jump of $I(\Psi)$ from $\Iin$ to $\Iout$, just like on the NS surface (see Eqs. (\rf{I/Pul-M}) and (\rf{OL-I}) and FIG.\ \rff{GapI}).

	\item 
The surface \SSN\ where $\vcv\ggel 0$ and $\epsJ\ggel 0$ for $\OmFm\ggel 0$ in Eq.\ (\rf{epsJbb}) will behave like a watershed in a mountain pass, (i.e.,\ a `plasma-shed') for outflows and inflows of `force-free' particles pair-created by the voltage drop (see Eqs.\ (\rf{EMF-ab}) and (\rf{Dl-V})), and yet both flows are due to the magneto-centrifugal forces at work toward the two opposite directions, inward and outward by $\OmF\ggel 0$. 
As the outer pulsar-type magneto-centrifugal wind flows through \SOL\ in $\calDout$ with $\vF> 0$, the inner anti-pulsar-type wind will pass through \SIL\ in $\calDin$ with $\vF< 0$ (see FIG.\ \rff{F-WS}; Sec.\ \ref{plasma-shed}, \cite{oka92}). Some particle-production mechanism will be at work in the vicinity of \SSN\ between \SOL\ and \SIL\ \citep{zna77}.   

	\item 
	Two vectorial quantities $\vcjp$ and $\vcSJ$ are closely related as the current/angular-momentum function $I(\Psi)$, i.e., the two-sidedness in the force-free domains, and do not reverse direction, despite that $\PN{\vcjp}=\PN{\vcSJ}=0$. This is because an outflow of negative charges means the ingoing current, and an inflow of negative angular momentum means an outflow of positive one (see FIGs.\ \rff{DC-C} and  \rff{GapI}). Despite that the null surface \SSN\ always exists, and yet Constraints $\PN{\vcSJ}=\PN{\vcSE}=0$, the overall energy flow $\vcSE=\OmF\vcSJ$ seem to flow outwards all the way along each open FL, seemingly as if crossing \SSN, despite that the force-free condition `breaks down' on \SSN, i.e., $\PN{I}=0$. 
	
	
	 \enen 
\begin{figure*}
	\begin{center}
	~~~~~~~~~~~~~~~
	\includegraphics[width=12cm, height = 7cm, angle=-0]{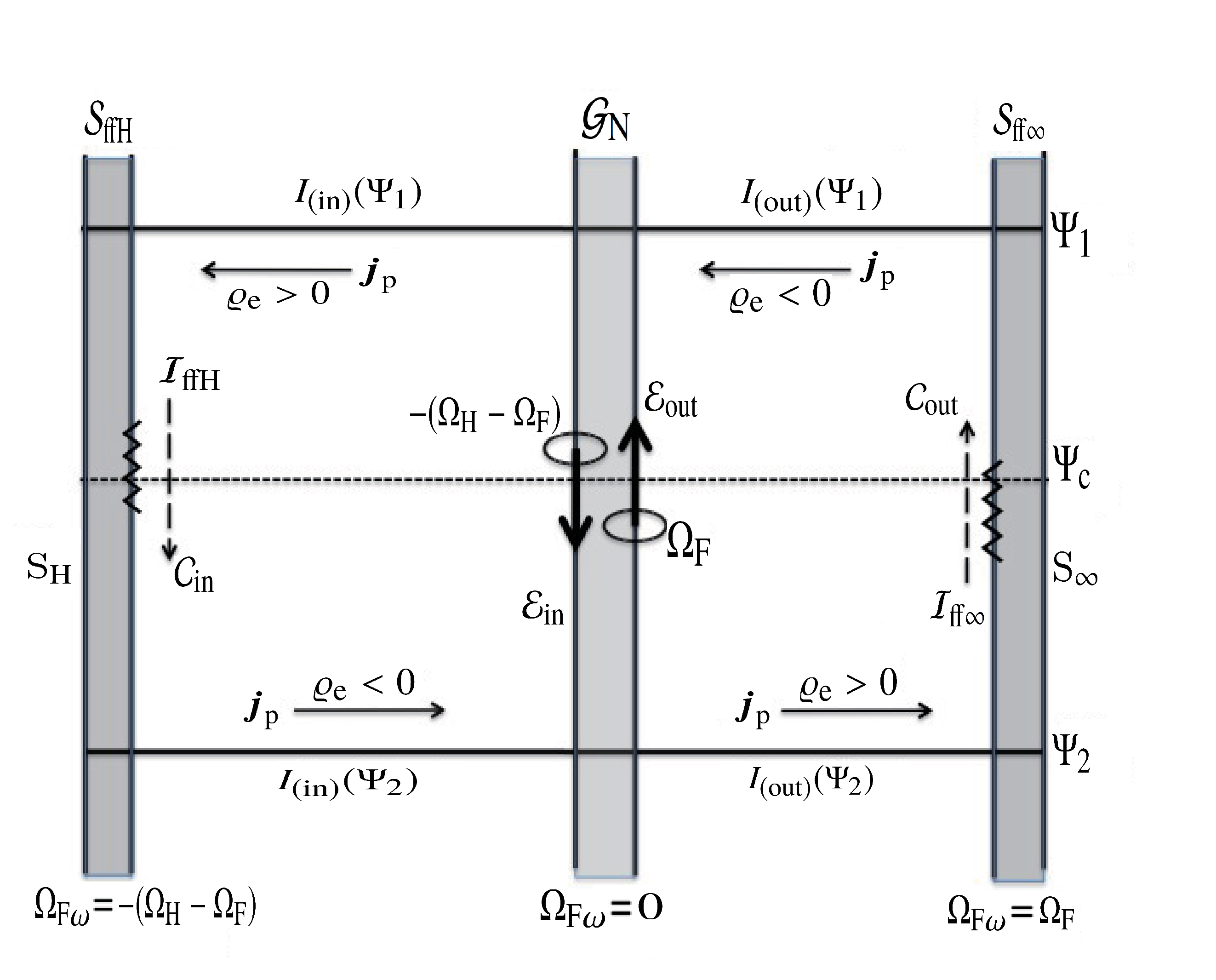}   
	\end{center}
\caption{ 
A schematic figure illustrating a pair of circuits $\calCout$ and $\calCin$, closed respectively in the force-free domains $\calDout$ and $\calDin$, which are separated by the non-force-free Gap $\SG$ with $\PG{\vcj}=\PG{\vcv}=0$ (see Eq.\ (\rf{EqSG}a,b) for the Constraints there).
	There will be the dual unipolar inductors with EMFs, $\calEout$ and $\calEin$, at work with respect to the magnetic spin-axes oppositely directed. The AVs of the magnetic axes are $\OmF$ and $-(\OmH-\OmF)$, respectively, and the former is parallel to the hole's spin axis and the latter is anti-parallel to it. The difference of the two is $\OmH$ (see Eq.\ (\rf{iden-b}); FIGs.\  \rff{GapI}, \rff{F-WS}).  
	Note that $\vcvp=\vcjp/\vre>0$ in the prograde-rotating $\calDout$ and $\vcvp <0$ in the retrograde-rotating $\calDin$, because the magnetic slingshot works outwardly ($\OmFm>0$) in the former and inwardly ($\OmFm>0$) in the latter.  Along the FCSL $\Psio$ where $\vcjp<0$, electrons flow out in $\calDout$, and positrons flow in $\calDin$ (see Eqs.\ (\rf{vcjp2}), (\rf{vcjpPM})).  The opposite is true along the FCSL $\Psit$ where $\vcjp>0$. Positrons flow out in $\calDout$ and electrons flow in $\calDin$.  
	It is a huge voltage drop of $\Dl V\propto \OmH$ (see Eq.\ (\rf{Dl-V})) that will lead to viable particle production of ample plasma particles, which makes happen the development of a dense Gap with the half-width $\Dlom$ (see FIG.\ \rff{GapI}). The particles with $\PG{\vcv}=0$, circulating around the hole's axis with $\omN$, are \zamp s. They are dense enough to pin down magnetic FLs, to fix $\OmF=\omN$ and make the Gap magnetized, thereby enabling the dual batteries to drive currents in each circuit (see FIG.\ \rff{F-WS} and figure 4 in \cite{oka15a}).    
	It is conjectured in the twin-pulsar model (see Sec.\  \ref{TW-P-M}) that the outer half of the Gap in $0\lo\OmFm\lo\Dlom$ plays a role of a `normal' magnetized NS spinning with $\OmF$, while the inner half in $0\ggo\OmFm\ggo -\Dlom$ behaves like an `abnormal' magnetized NS counter-spinning with $-(\OmH-\OmF)$. Relaxation of the RTD (see Sec.\ \ref{TW-P-M}) in-between will lead to widening of the null surface \SSN\ to the Gap $\SG$ with width $\sim 2\Dlom$. 
		}
		\lbf{DC-C}  \end{figure*} 
	
	\newcommand{\SffinfNS}{S_{\rm FF-NS}}

\section{A pair of batteries for the dual circuits	
and particle production occurring between them}  \label{indMem}  
The hole's force-free magnetosphere will be divided on \SSN\ into the SC and GR domains, and the outward Poynting flux $\vcSEMout$ from \SSN\ will be utilized for particle acceleration on $\Sffinf$, while the inward one $\vcSEMin$ will be dissipated for entropy production on $\SffH$ (see Eq.\ (\rf{ThdS/dt})).		
 The inevitable breakdown of the force-free and freezing-in conditions in-between will provide an arena of setting up a pair of batteries and the voltage drop between their EMFs for particle production \cite{oka15a}. The covering surface is referred to as the inductive membrane $\SN$,  
 as opposed to the two {\em resistive} membranes $ \Sffinf$ and $\SffH$ in both ends of the force-free domains (see FIGs.\  \rff{DC-C} and \rff{GapI}).      

	Let us then pick up two such field-current-stream-lines (FCSLs) $\Psio$ and $\Psit$ for the circuits $\calCout$ and $\calCin$ as the two roots of an algebraic equation\  $I(\Psi)=\Iot$, i.e., $I(\Psi_1)=I(\Psi_2)\equiv \Iot$ in the range of $0<\Psio<\Psic<\Psit<\bar{\Psi}$ (see Eq.\ (\rf{c-c-c})), where $(dI/d\Psi)_{\rm c}=0$ and $\vcjp\lleg 0$ for $\Psi\lleg\Psic$ (see FIG.\ \rff{DC-C}). Note that current- and stream-lines are disconnected on \SSN\ between the outer and inner domains along FCSLs $\Psio$ and $\Psit$ to make closed circuits in each domain. The Faraday path integrals of $\vcEp$ in Eq.\ (\rf{EpBt}) or (\rf{bhEp}) along two circuits, $\calCout$ and $\calCin$, yield 
	\begin{subequations} \begin{eqnarray} 
			\calEout= \oint_{{\cal C}_{\rm (out)}} \al\vcEp\cdot d\vcell 
			=-\frac{1}{2\pi c}\int_{\Psi_1}^{\Psi_2} \OmF(\Psi)d \Psi, \hspace{0.5cm} \lb{EMF-out} \hspace{0.3cm}  \\ 
			\calEin=\oint_{{\cal C}_{\rm (in)}} \al\vcEp\cdot d\vcell
			=+\frac{1}{2\pi c}\int_{\Psi_1}^{\Psi_2} (\OmH-\OmF)d \Psi,\hspace{0.5cm}  \lb{EMF-in} \hspace{0.33cm} 
		\end{eqnarray}   \lb{EMF-ab}  \end{subequations}   
	respectively.
There is no contribution to EMFs for the integration along $\Psio$ and $\Psit$ and on the null surface \SSN, because of $\vcEp\cdot d\vcell=\PN{\vcEp}=0$.  The difference between the two EMFs across $\SN$ is  
	\beeq 
	\DN{\calE}  =\calEout- \calEin=-\OmH\Dl\Psi/2\pi c = -\Dl V,
	\lb{Dl-V} \eneq 
	where $\Dl\Psi=\Psit -\Psio$, and the difference in a quantity $X$ across the (infinitely thin) interface \SSN\ is denoted by 
	\beeq 
	\DN{X} =(X)_{\rm N}^{{(\rm out)}}  - (X)_{\rm N}^{{(\rm in)}} . 
	\lb{DNX}  \eneq

The difference of the ZAMO-FLAV $\OmFm$ between \Sinf\ and \SH\ becomes, from Eq.\ (\rf{SEM<0}),
	\begin{subequations}  \begin{eqnarray}
		(\OmFm)_{\infty} -  (\OmFm)_{\rm H}  \lb{iden-c} \\ 
		\hspace{1cm}	=  \OmF  -[- (\OmH-\OmF)]    \lb{iden-b} \\ 
		=  \OmF+ (\OmH-\OmF)=\OmH.   \lb{Iden-a} 
		\end{eqnarray} \lb{distr2}   \end{subequations}  
Although the two FLs $\Psio$ and $\Psit$ thread the null surface \SSN, the pair of unipolar induction batteries for the two circuits $\calCout$ and $\calCin$ are disconnected by the null surface \SSN, but existent there back-to-back (oppositely). It turns out that the `maximum available gravito-potential' $\OmH$ is `concentrated' on \SSN, to produce a voltage drop $\Dl V=(\OmH/2 \pi c)\Dl\Psi$, which will appear not only as the difference in the two EMFs, $\calEout$ and $\calEin$, for the two circuits, but also as the difference $\OmH$ in a pair of magnetic axes spinning with $\OmF$ and $-(\OmH-\OmF)$ of the two domains $\calDout$ and $\calDin$ across the inductive membrane $\SN$.	
Then, the expression (\rf{Dl-V}) for $\Dl V$ across \SSN\ is derivable simply by integrating the identity (\rf{iden-b}) from $\Psio$ to $\Psit$, just as obtaining the difference between the two equipotential lines $\Psio$ and $\Psit$. 

These EMFs for the two DC circuits $\calCout$ and $\calCin$ drive the {\em volume} current $\vcj$ (see Eq.\ (\rf{vc-vjA}) in the force-free domains $\calDout$ and $\calDin$, respectively. The EMFs also drive the \emph{surface} membrane currents $\calI_{\rm ff{\infty}}$ and $\calI_{\rm ffH}$ on $\Sffinf$ and $\SffH$, respectively. 

The outer resistive membrane $\Sffinf$ may also be interpreted as possessing the same surface resistivity $\calRinf=4\pi/c=377$ Ohm as on the inner membrane $\SffH$ above \SH, and Ohm's law holds on $\Sffinf$, i.e., $\calRinf \calI_{\rm ff{\infty}}=(E_{\rm p})_{\rm ff{\infty}}$. Thus, by using Eqs.\ (\rf{Iout/in}a,b) and (\rf{calPE/out},b), we have, similarly to Eq.\ (\rf{ThdS/dt}),
		\begin{subequations} 	\begin{eqnarray}
		 \int_{ {\cal S}_{{\rm ff}\infty } } \calRinf \calI_{\rm ff{\infty}} ^2 dA 
		 =  \int_{ {\cal S}_{{\rm ff}\infty } }  \vccalI_{\rm ff{\infty}}\cdot\vcEp  dA  
		=  \int_{ {\cal S}_{{\rm ff}\infty } } \vcSEM\cdot d\vcA  \nonumber  \\[1mm] 
		 =\frac{1}{2c} \int_{ {\cal S}_{{\rm ff}\infty } } \OmF^2(\Bp\vp^2)_{{\rm ff}\infty} d\Psi =\calPEout ,   \hspace{1cm} 	
		\lb{Sffinf-M}  \end{eqnarray}   
where 
$\vcSEM=\vcSE$ and $\vcSsd=0$ on $\Sffinf$ for $\om\to 0$. On the other hand, from Eq.\ (\rf{ThdS/dt}), one has a resistive membrane $\SffH$ on the horizon,
		\begin{eqnarray}
		 \int_{{\cal S}_{\rm ffH}} \RH \calI_{\rm ffH}^2 dA= \TTH \frac{dS}{dt}  =\OmH\calPJin -\calPEin.  
			\lb{ThdS/dt-b}  
		\end{eqnarray} 
The above two expressions sum up to
		\begin{eqnarray}
		 \int_{{\cal S}_{\rm ffH}} \RH \calI_{\rm ffH}^2 dA + \int_{ {\cal S}_{{\rm ff}\infty } } \RH \calI_{\rm ff{\infty}} ^2 dA  =\OmH\calPJin 
		\end{eqnarray}	\lb{ThdS/dt-c}  \end{subequations} 
(see Eq.\ (\rf{SDenergy})), because $\calPEout=\calPEin=- c^2(dM/dt)$ and $\calPJin=\calPJout= -(dJ/dt)$ hold across the Gap $\GN$ with $\PG{I}=0$ by the `boundary conditions' (see Sec.\ \ref{BC-SN}; Eqs.\ (\rf{DGcalE/a}), (\rf{DN/calPJa})), and hence we have $c^2 dM=\Th dS+ \OmH dJ$. It therefore turns out that the first law of thermodynamics participates directly in Ohmic dissipation of the \emph{surface} currents for entropy production and particle acceleration on the two resistive membranes $\SffH$ and $\Sffinf$ (see Eq.\ (\rf{first-II}a,b)). 


The two EMFs in Eqs.\ (\rf{EMF-ab}a,b) for circuits $\calCout$ and $\calCin$ are also responsible for launching the Poynting energy fluxes in both outward and inward directions, i.e., $\vcSEM\ggel 0$ for $\OmFm\ggel 0$.  Eq.\ (\rf{EMF-out}) for $\calEout$ coincides `almost exactly' with  Eq.\ (\rf{nsEMF}) for a pulsar magnetosphere, because of $\om\ll \omN=\OmF$ and hence $\vcSEMout\approx\vcSE$ in the outer SC domain $\calDout$. 
 
This Ohmic dissipation (in circuit theory) implies that particle acceleration (in wind theory) will take place. 
The rate per unit $\tau$ time at which electromagnetic fields transfer redshifted energy to particles is, by Eq (4.14) \cite{mac82};  
	\begin{eqnarray}
		-\frac{1}{\al} \vcnb\cdot\al\vcSE=\al\vcj\cdot\vcE+(\om/c)(\vcj\times\vcB)\cdot \vcm   \hspace{0.5cm}   \nonumber \\
		=\frac{\OmF\vp}{c}\jvl\Bp  
		=- \frac{\OmF\vp}{c}\frac{\Bp}{2\pi\vp\al}\LPPlDr{\Iout}{\ell}>0, 
		\lb{DivvcSE} \end{eqnarray}  
where Eqs.\ (\rf{jpl/vl}) and (\rf{JcrossB}) are used. This shows that when the current function $I(\ell,\Psi)$ is {\em continuously} decreasing with $\ell$ in the resistive membrane $\Sffinf$ from near \SoF\ to \Sinf, the MHD acceleration will occur (see FIG.\ \rff{GapI}), but the force-free magnetosphere regards the `force-free' domain with $\jvl=0$ formally as extending to the force-free infinity surface $\Sffinf$ where $|\jvl|\gg |\jpl|$. By doing so, the circuit $\calCout$ closes so as not to violate the current-closure condition in the steady state. 

The null surface \SSN\ with Constraints (\rf{EqSN}) seem to be genetically endowed with the discontinuity $\DN{\calE}=-\Dl V$, to widen the surface \SSN\ to a gap $\SG$, thereby constructing a magnetized `\zam' and a charge-neutral Gap between the two force-free domains with $\epsJ\ggel 0$ (see Sec.\  \ref{m-mdGap}). 
Therefore, the voltage drop $\Dl V$ in Eq.\ (\rf{Dl-V}) reveals that the null surface \SSN\ will be a kind of RTD due to the two magnetic rotators, counter-rotating each other, namely, between the  outer and inner domains $\calDout$ and $\calDin$,  although $\OmFm$ and $\vcEp$ likely change sign easily through zero (see Sec.\  \ref{TW-P-M}; cf.\ \cite{lan84}). {  It turns out thus that the maximum available voltage drop $\Dl V=(\OmH/2 \pi c)\Dl\Psi$ will be utilized between the two circuits 
	$\calC_{\rm out}$ and $\calC_{\rm in}$ for particle production (see Sec.\ \ref{TW-P-M}).  }

\newcommand{\vreN}{(\vre)_{\rm N}}
	\begin{figure*}
	\begin{center}
	\includegraphics[width=12cm, height = 7cm, angle=-0]{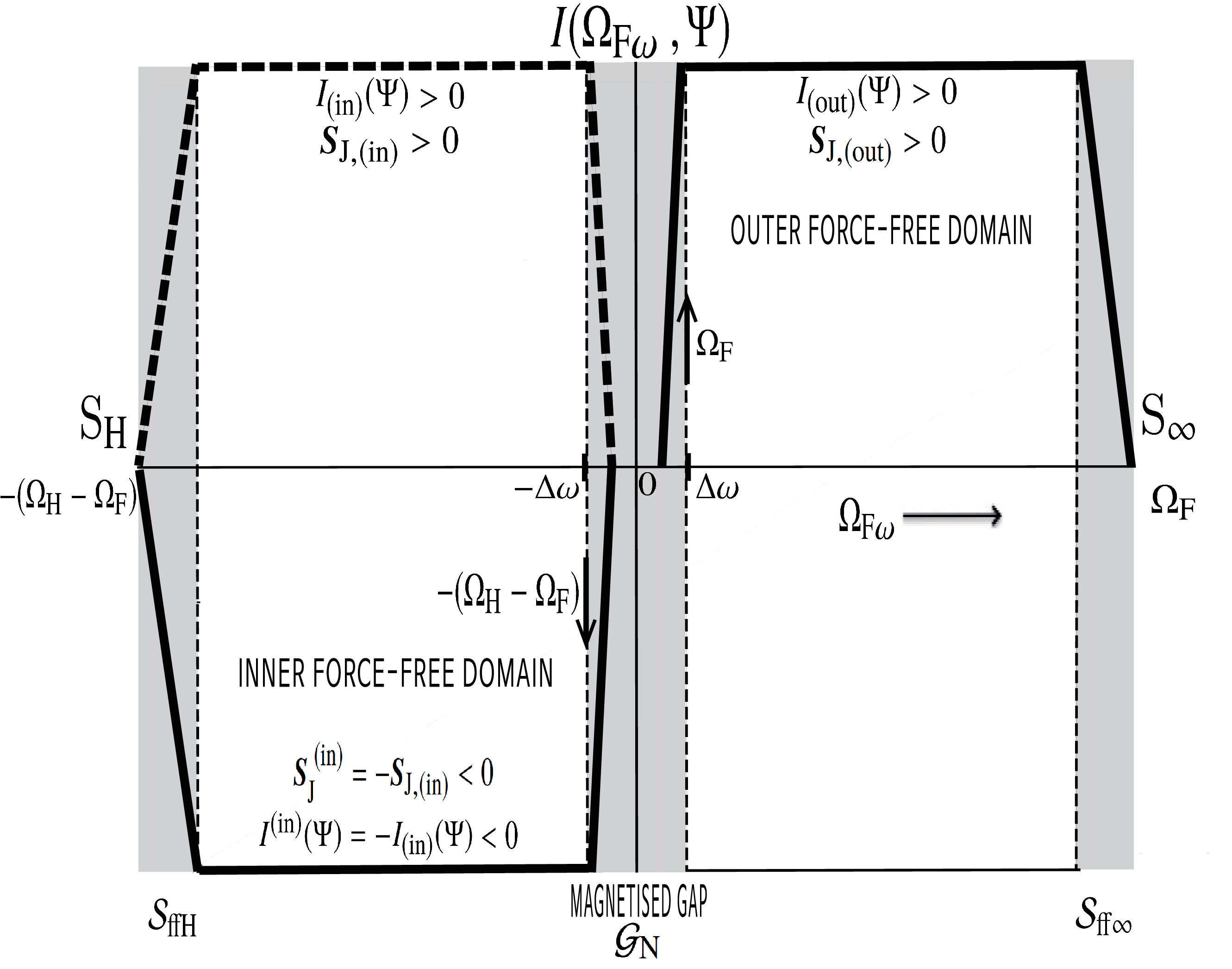}  
	\end{center}
	\caption{
	A plausible behavior of the angular-momentum-flux/current function, $I(\OmFm,\Psi)$ (see Eq.\ (\rf{OL-I})). The abscissa is the `coordinatized' $\OmFm$ along a field line $\Psi=$constant, and the ordinate on the null surface \SSN\ divides the magnetosphere into the two domains of pro- and retro-grade rotation by $\OmFm\ggel 0$.  The breakdown of the force-free and freezing-in conditions on the null surface \SSN\ leads to Constraints $\PG{\vcv}=\PG{\vcj}=0$ in Eqs.\ (\rf{ff-a}), which mean that current- and stream-lines no longer thread $\SG$ (see FIGs.\ \rff{Flux-om}, \rff{DC-C}, \rff{F-WS}). 
	The voltage drop $\Dl V$ between the two EMFs in Eq.\ (\rf{Dl-V}) will induce steady particle production, thereby developing a Gap with $\PG{I}=0$ in a finite zone $|\OmFm|\lo \Dlom$ between the two force-free domains with $I=\Iout(\Psi)$ and $=\Iin(\Psi)$, respectively (see Eqs.\ (\rf{Iout/in}a,b)). 
	When the rate of \emph{positive} angular momentum conveyed outwardly in $\calDout$ is equal to that of \emph{negative} angular momentum conveyed inwardly in $\calDin$, the `zero-angular-momentum state' of the Gap $\PG{\epsJ}=0$ is maintained, i.e., $\DG{I}=\Iout+\IinU=0$. The `boundary condition' $\DG{I}=0$ in Eq.\ (\rf{DN/SN}) yields the eigenfunction $\OmF(\Psi)=\omN$ in Eq.\ (\rf{EigenOmFI}). 
		Just as the watershed at a mountain pass produces two down-streams to both sides by the gravitational force, the {\em plasma}-shed amid the Gap at $\OmFm\approx 0$ will divide pair-produced particles into outflows and inflows $\vcv\ggel 0$ by the magneto-centrifugal force due to $\OmFm\ggel 0$. The Gap filled with the ZAMPs will be well inside between the two light surface \SOL\ and \SIL\ (see Eq.\ (\rf{SG<SoL})). 
		There may be a kind of boundary layers in the vicinity $\OmFm\simeq\pm\Dlom$ where non-force-free ZAM particles pair-created in $\vre\simeq 0$ are changing to force-free charge-separated plasma with $\vre\approx\mp e n^{\pm}$, and $I$ increases (or decreases) to $\Iout$ (or $\IinU$) rather steeply. The non-force-free, matter-dominated Gap, filled with ZAM particles, will ensure the pinning-down of poloidal field lines $\vcBp$ with $\OmF=\omN$, and the pinning-down conversely ensure magnetization of ZAM particles with $\PG{I}=0$ in the Gap within $|\OmFm|\lo \Dlom$. Thus, the whole force-free magnetosphere will be kept in the frame-dragged state by the hole's rotation. 
			}
			\lbf{GapI}  \end{figure*}    

\section{The zero-angular-momentum gap $\GN$ } \label{m-mdGap}  
	\setcounter{equation}{0}
	\numberwithin{equation}{section}
\subsection{A plausible Gap structure with $I(\OmFm,\Psi)$} \label{GapStruc} 
{  We presume that the  RTD  with the voltage drop $\DN{\calE}=-\Dl V$ in the force-free limit will be relaxed as a result of pair-particle production to a ZAM-Gap $\GN$ with a half-width $\Dlom$.} 
For the {\em widened} null Gap $\SG$ in $|\OmFm|\lo\Dlom$, we replace Constraints in Eq.\ (\rf{EqSN}) on \SSN\ as follows;
	\begin{subequations} \begin{eqnarray} 
		\PG{\OmFm}= \PG{\vcEp}=\PG{\vre} = \PG{\vcj}=\PG{\vcv}= \PG{I} \hspace{0.6cm}   \lb{SGa} \\ 
		=\PG{\epsJ}=\PG{\vcSJ}=\PG{\vcSEM}=\PG{\vcSsd}=\PG{\vcSE} \hspace{0.6cm}   \lb{SGb} \\
		=  \PG{\calPJ}= \PG{\calPE} =0.	 \hspace{0.6cm}   \lb{SGc} 
	\end{eqnarray} \lb{EqSG} \end{subequations}    
As before, $\DG{X}$ denotes the difference of $X(\OmFm, \Psi)$ across the Gap width $2\Dlom$ (cf.\ $\DN{X}$ in Eq.\ (\rf{DNX}));  
	\beeq 
	\DG{X} =X(\Dlom,\Psi)-X(-\Dlom,\Psi) \equiv (X)_{{\rm G}{(\rm out)}}  - (X)_{{\rm G}{(\rm in)}} .    
	\lb{DGX}   \eneq   
We think of such a simple form of $I=I(\OmFm,\Psi)$ along a typical open field line in $0\leq\Psi\leq\Psib$ as follows; 
	\begin{eqnarray}
		I(\OmFm,\Psi) =  \left\{
		\begin{array} {ll}
			\to 0 & ;\ \Sffinf\ ( \OmFm\to\OmF), \\ [1mm]
			\Iout & ;\ \calDout\ ( \Dlom \lo \OmFm \lo \OmF),  \\[1mm]   
			0 & ;\ \SG\ ( |\OmFm|\lo \Dlom) , \\[1mm] 
			\Iin &;\ \calDin\ (-\Dlom \ggo \OmFm\ggo-(\OmH-\OmF)),  \\[1mm]
			\to 0 & ;\ \SffH\  (\OmFm\to -(\OmH-\OmF)) 
		\end{array}  \right.    \lb{OL-I}  \end{eqnarray}    
(see FIG.\ \rff{GapI}), where $\Iout$ and $\Iin$ are given by Eqs.\ (\rf{Iout/in}a,b). 
The behavior of $I(\OmFm,\Psi)$ in the outer domain $\calDout$ will be similar to that of the force-free pulsar magnetosphere (see Eq.\ (\rf{I/Pul-M})). 
When the particle source will have to be situated well inside the Gap $\GN$, we have, by Eq.\ (\rf{Sil/oL}),
	\beeq
	|\OmFm| \lo \Dlom \ll c(\al/\vp)_{\rm oL} \approx c(\al/\vp)_{\rm iL}. 
	\lb{SG<SoL} \eneq   
It is not clear yet how helpful or rather indispensable is the above condition, in constructing a reasonable gap model, but we presume that an intense particle production will eventually occur by the voltage drop across the Gap $\GN$, $\Dl V=-(\OmH/2\pi c)\Dl\Psi$, almost independently of the presence of the two light surfaces in wind theory. 

 The Gap $\GN$ under the inductive membrane $\SN$ must be in the {\em zero-angular-momentum} state $\PG{\epsJ}=0$, so that the particles and the field carry no angular momentum nor energy across the Gap, i.e., $\PG{\vcSJ}=\PG{\vcSE}=0$ in Eq.\ (\rf{SGb}), despite that the Gap is threaded by poloidal field lines, i.e., $\PG{\vcBp}\neq 0$.  
Also, the Gap must be {\em magnetized} (i.e., $\PN{\epsE}\neq 0$; see Eqs.\ (\rf{epsEba}), (\rf{epsE/N})), in almost the same sense as a magnetized NS, because the poloidal magnetic field lines (with no toroidal component) {\em threading} the Gap will naturally be {\em pinned down} in the ZAM-particles pair-created and circulating round the hole with $\omN=\OmF$ (see Sec.\  \ref{PPP}).  Therefore, the magnetized ZAM-particles will ensure $\OmF=\omN$ (see Sec.\  \ref{BC-SN}). The non-force-free magnetized ZAM-Gap $\SG$, where $\PG{\vcj}=\PG{\vcv}=0$ but $\PG{\vcBp}\neq 0$, will therefore be formed in the steady-state, with its {\em surfaces} \SgapO\ and \SgapI,  at $\OmFm\approx \pm\Dlom$ (see FIG.\ \rff{GapI}), where $\Dlom\approx|\PN{\partial\om/\partial\ell}| \Dlell$ stands for the Gap half-width (see Eq.\ (\rf{OL-I})), and for $\Dlom\to 0$, $\SG\to$\SSN\ (see FIG.\  4 in \cite{oka15a} for the interplay of microphysics and macrophysics in the magnetized, matter-dominated Gap). Therefore, we conjecture that the voltage drop, $\Dl V=-\DN{\calE}$ on \SSN\ in Eq.\ (\rf{Dl-V}), will produce pair-particles that are copious enough, and the plasma pressure in the steady-state will expand \SSN\ to $\GN$ with a half-width $\Dlom$ in $|\OmFm|\lo \Dlom$  (see Sec.\  \ref{indMem}). 

\subsection{Pinning-down of threading field lines on \zamp s and magnetization of the matter-dominated Gap }  \label{PPP} 
	When we regard the Gap surfaces \SgapO\ and \SgapI\ as being equipped with EMFs $\calEout$ and $\calEin$, respectively, these EMFs will not only drive currents in the respective circuits $\calCout$ and $\calCin$, but also produce a strong voltage drop $\Dl V=-\DG{\calE}$ across the Gap, which will create ZAM-particles copious enough to pin threading FLs down. The ZAM-Gap filled with ZAM-particles will then circulate the hole with $\omN=\OmF$, and the poloidal field lines threading the Gap $\SG$ will surely be pinned down on ZAM-particles with $\OmF=\omN$. Thus, the ZAM-Gap will be in the perfectly magnetized state, with no electric current and no angular momentum flux allowed to cross, i.e., $\PG{\Bt}=\PG{I}=\PG{\vcj}=0$. The  state of inside the ZAM-Gap $\GN$ will be analogous to that of inside the NS  ensuring the boundary condition $\OmF=\OmNS$ for the FLs emanating from the NS surface. 
	
The two circuits $\calCout$ and $\calCin$ are connected  to each other by the FLs $\Psio$ and $\Psit$, and hence the iso-rotation law $\OmF(\Psi)=$ constant  still holds along each FL threading the particle production Gap $\SG$, but current- and stream-lines are severed by breakdown of the force-free and freezing-in conditions in the Gap. { The ZAMOs see that FLs circulate forward (progradely) in the outer domain $\calDout$, but backward (retrogradely) in the inner domain $\calDin$.}  The `boundary condition' of no jump of the angular momentum transport rate across the Gap, i.e., $\DG{I}=\PG{I}=0$, will be maintained, to determine the eigenfunction $\OmF$ (see Sec.\  \ref {BC-SN}). 
	
\subsection{Magneto-centrifugal plasma-shed on the ZAM-surface}  \label{plasma-shed} 
The outer force-free domain $\calDout$ behaves like a pulsar force-free magnetosphere progradely-rotating with $\OmFm>0$, whereas the inner domain $\calDin$ behaves like an anti-pulsar-type magnetosphere retrogradely-rotating with $\OmFm<0$ \citep{oka92}.  
Hence, there must be a particle source. Then, plasma particles pair-created by the voltage drop $\Dl V=- \DG{\calE}$ within the Gap $\SG$ must be ZAM-particles, circulating together with ZAMOs at $\omN=\OmF$, but may not behave as `force-free' particles with negligible inertia within the Gap.  These ZAM-particles with $\PG{\vcv}=0$ will soon become charge-separated, to flow from the Gap outward to the force-free domains as `force-free' particles, with $\vcvp>0$ in $\calDout$ and $\vcvp<0$ in $\calDin$, as the magneto-centrifugal winds (see FIGs.\ \rff{DC-C} and \rff{F-WS}). 
	
	The null surface \SSN$=$\SZAM\  midst the ZAM-Gap $\GN$ will then redefine quite a new general-relativistic type of divider due to magneto-centrifugal force modified by frame-dragging ($\OmFm\ggel 0$), outward and inward ($\vcv\ggel 0$), for particles pair-created in the ZAM-Gap.  That is, {\em this surface} \SZAM\ will be a magneto-centrifugal plasma-shed akin to a gravitational watershed of a mountain pass for heavy rainfalls on Earth. 
This will be quite a natural way to launch `magneto-centrifugal' winds from the ZAM-Gap for both directions toward infinity ($\vcv>0$) and toward the horizon ($\vcv<0$), similarly to the Poynting flux ($\vcSEMout>0$) for particle acceleration on the resistive membrane $\Sffinf$ and the one ($\vcSEMin<0$) for entropy production on another membrane $\SffH$ (see Eqs.\ (\rf{HJ/resistance}), (\rf{HL/resistance})). 
	
	The `spark' models so far used for pair-production discharges in the previous works over the past four decades are based mainly on an extension from a `negligible violation' of the force-free condition (see \cite{bla77,mac82,phi83a,bes92, hir98, son17, hir18b}; also \cite{ruf10,che23}).  It is argued here that the `complete violation' of the force-free condition due to frame-dragging on the null surface \SSN\ leads to a unique gap model for the particle-current sources. It was emphasized already in  \citep{oka15a} that ``the present gap model with a pair of batteries and a strong voltage drop is fundamentally different from any existing models based on pulsar outer-gap models.''  The most significant difference from the previous particle production mechanism in \citep{oka15a} comes mainly from the existence of the counter-rotating inner domain $\calDin$ inside \SSN\ (see Sec.\  \ref{TW-P-M}). 

\newcommand{\dJout}{dJ_{\rm (out)}}  \newcommand{\dJinU}{dJ^{\rm (in)}}  \newcommand{\dJin}{dJ_{\rm (in)}} 	
		
\section{The Eigen-magnetosphere}  \label{BC-SN}  
\setcounter{equation}{0}
\numberwithin{equation}{section}

For a viable force-free magnetosphere, we refer to the condition by which to eventually determine the eigenfunction $\OmF(\Psi)$ as the `boundary condition,' distinguishing from the `criticality condition' for another eigenfunction $I(\Psi)$ in Eqs.\ (\rf{Iout/in}a,b). 
If we had to consider a scenario where the force-free condition does not break down anywhere in the force-free magnetosphere, we would not have figured out a unique procedure on where and how $\OmF$ should be determined for the frame-dragged magnetosphere, except maybe `impedance matching,' which was so far empirically used often, without referring to the `FDAV-eigenvalue' $\omN$ \cite{bla77,mac82}.  
	
\subsection{The `boundary condition' for the eigenfunction $\OmF$} \label{BCagain}   
One of the vital roles of the ZAM-Gap is to anchor the poloidal field $\vcBp$ onto the ZAM-particles pair-created in it, and to accomplish magnetization of the ZAM-Gap, thereby ensuring $\OmF=\omN$ for threading FLs. Accordingly, the ZAM-state of the Gap must always be maintained, in the magnetosphere frame-dragged by the hole into circulation with $\omN$. Nevertheless, the actual position of the Gap and the specific value $\OmF=\omN$ \emph{per se} remain still undetermined. In order to finally determine the eigenfunction $\OmF(\Psi)=\omN$ in terms of $\OmH$, we formulate the `boundary condition,' with Constraints (\rf{EqSN}) or (\rf{EqSG}) appropriately taken into account; in particular, $\PG{I}=\DG{I}=0$ at the place of the ZAM-Gap.  	
This ensures the continuity of overall energy flux as well as angular momentum flowing across the Gap, thereby keeping the ZAM-state, i.e., $\vcSJout+\vcSJinU=0$. 
	
Therefore, we may impose no discontinuity of $I$ across $\SG$ along each FL threading the Gap;
	\begin{subequations} \begin{eqnarray}
			\DG{I}=\Iout(\Psi) -\Iin(\Psi) \quad \quad  \lb{DN/SNa} \\   
			=\Iout(\Psi) +\IinU(\Psi)=0  \lb{DN/SNb}       
		\end{eqnarray}  \lb{DN/SN}  \end{subequations}     
(see FIG.\ \rff{GapI}, Eqs.\ (\rf{Iout/in}a,b)).  Eq.\ (\rf{DN/SNa}) shows that the outward transport rate of \emph{positive} angular momentum leaving \SgapO\ must be equal to that entering \SgapI, although the energy-angular momentum flow actually does not take place inside the ZAM-Gap with $\PG{I}=0$.  Eq.\ (\rf{DN/SNb}) implies equivalently that the outward rate of {\em positive} angular momentum leaving \SgapO\ is offset by the inward rate of {\em negative} angular momentum leaving \SgapI. Now, by Eqs.\ (\rf{E-AmFlux}), we have
	\beeq
	\DG{\vcSE}=\OmF\DG{\vcSJ}=0,
	\lb{DG/SE/SJ}  \eneq  
which `apparently' shows that the overall energy and angular momentum fluxes flow outward continuously across the Gap $\GN$, regardless of $\PG{\vcSE}=\PG{\vcSJ}=0$. 

Likewise, the `boundary condition' in (\rf{DN/SN}) ensures no discontinuity of the power $\calPE$ and  the loss rate of angular momentum $\calPJ$ across the ZAM-Gap, i.e., by Eqs.\ (\rf{TotalFa,b}a,b),
	\begin{subequations}  	\begin{eqnarray}
	\DG{\calPE}=\calPEout- \calPEin   \lb{DGcalE/a}  \\  	
	=\calPEout + \calPEinU =0,   \lb{DGcalE/b}		
	\end{eqnarray} \lb{DGcalEa,b} \end{subequations}	
\begin{subequations} \begin{eqnarray}
	\DG{\calPJ}=\calPJout -\calPJin \quad \quad  \lb{DN/calPJa}  \\ 
	=\calPJout + \calPJinU =0   
\end{eqnarray}  \lb{DN/calPJ}  \end{subequations}     
(see Eqs.\ (\rf{calPEout-in}a,b), (\rf{calPJout-in}a,b)). 

It is in reality the ZAM-state with $\PG{\OmFm}=0$ of the Gap that makes it possible to use expression (\rf{DN/SN}) as the `boundary condition' for $\OmF$ even in the finite value of $\Dlom$ (see Sec.\  \ref{m-mdGap}) and, conversely, the `boundary condition' (\rf{DN/SN}) is necessary to ensure the ZAM-state of the Gap $\GN$.

\newcommand{\zetab}{\bar{\zeta}}	
\subsection{The final eigenfunctions $I(\Psi)$ and $\OmF(\Psi)$ in the force-free magnetosphere}  \label{Feigenv}
From Eqs.\ (\rf{Iout/in}a,b) and (\rf{DN/SN}), we have 
\begin{subequations} \begin{eqnarray}
	\OmF(\Psi)=\omN= \frac{\OmH}{1+\zeta}, 	\hspace{1cm}	\lb{EigenOmFI}      \\  
	I=\Iout=\Iin=- \IinU=\frac{\OmH}{2(1+\zeta)} (\Bp\vp^2)_{\rm ffH}, \hspace{1cm} 	 \lb{EigenOmF}  \\  
	\zeta(\Psi) \equiv (\Bp\vp^2)_{{\rm ff}\infty}/(\Bp\vp^2)_{\rm ffH} \hspace{1cm}   \lb{Zeta}    
\end{eqnarray}  \lb{FL-eigen}   \end{subequations} 
\citep{oka09,oka12a,oka15a}. 
Note that $\PG{\vcBp}\neq 0$, $\PG{\OmF}\neq 0$ and $\DG{\vcBp}=\DG{\OmF}=0$ in our major premises, and also $\PG{I}=\DG{I}=0$ across the Gap $\GN$ in $|\OmFm|\lo \Dlom$ with $\DG{\OmFm}=2\Dlom$ (see FIG.\ \rff{GapI}).  

In this eigenstate, the ZAMOs will see that (i) \SSN$=$\SZAM\ is the magneto-centrifugal plasma-shed, from which the angular momentum and the Poynting fluxes, positive and negative, flow out to both ways toward $\Sffinf$ and $\SffH$,  (ii) their related AVs are given by $(\OmFm)_{\rm out}=\OmF$ in the outer domain $\calDout$, and by $(\OmFm)_{\rm in}=-(\OmH-\OmF)$ in the inner domain $\calDout$, and (iii) the difference $\OmH$ of the two AVs corresponds to the voltage drop $\Dl V$ between a pair of batteries in Eq.\ (\rf{Dl-V}), and this drop will lead to sustainable particle production. 

Constraints $\PG{\vcj}=\PG{\Bt}=\PG{I}=\PG{\vcv}=0$ in Eq.\ (\rf{EqSG}) imply that no transport of angular momentum and energy by the field, the  current, and particles is possible within the ZAM-Gap, i.e., $\PG{\vcSJ}=\PG{\vcSE}=0$. These indicate a disconnection of current- and stream-lines between the two force-free domains, and hence the necessity of the current-particle sources and related EMFs in the Gap.  It will be ensured in Eq.\ (\rf{DN/SN}) that the copious charged ZAM-particles pair-produced in $|\OmFm|\lo\Dlom$ serve to connect and equate both $\Iout$ and $\IinU$ across the Gap $\SG$, despite $\PG{\vcv}=\PG{\vcj}=0$. Also, the \emph{overall} flow of energy-angular momentum is continuous beyond the ZAM-Gap, regardless of $\PG{\calPE}=\PG{\calPJ}=0$ and  $\PG{\OmFb}=0$ from Eqs.\ (\rf{TotalFa,b}a,b)	
as far as the boundary condition $\DG{I}=0$ in Eq.\ (\rf{DN/SN}) is satisfied.  

The eigen-efficiency of extraction is given, from Eq.\ (\rf{EigenOmFI}), by 
\beeq 
\epsGTE=\frac{\OmF}{\OmH}=\frac{1}{1+\zeta}.
\lb{eps} \eneq 
When the plausible field configuration allows us to put $\zeta\approx 1$ and hence $\epsGTE\approx 0.5$, we have, from Eqs.\ (\rf{EigenOmFI}), 
\begin{subequations} \begin{eqnarray}
	\OmF\approx \OmFb \approx \frac12\OmH,  \lb{matching} \\  
	\hspace{1.5cm}	  c^2 |dM| \approx  \Th dS \approx \frac12 \OmH |dJ| . \lb{therfirst-eig} 
\end{eqnarray} \lb{eig/state} \end{subequations}   
The overall value $\OmFb= \bar{\om}_{\rm N}$, together with $\zetab$, in the eigen-state will be given from Eqs.\ (\rf{FL-eigen}a,b,c), and also the average null surface $\bar{\SN}$ and the overall efficiency $\epsGTEb$ will be given from Eq.\ (\rf{eps}). 

 A pair of batteries's EMFs become, for $\zeta \approx 1$, by Eqs.\ (\rf{EMF-ab}), 
	\begin{subequations} \begin{eqnarray} 
			\calEout  =-\frac{\OmH}{2\pi c} \int_{\Psi_1}^{\Psi_2}   \frac{1}{1+\zeta} d\Psi 
			\approx - \frac{\OmH \Dl\Psi}{4\pi c} 	\approx -\frac{\Dl V}{2},   \hspace{0.7cm}
			 \lb{eigEMF-out} 	 \\ 
			\calEin	
			=\frac{\OmH}{2\pi c}\int_{\Psi_1}^{\Psi_2}  \frac{\zeta}{1+\zeta} d\Psi    
			\approx   \frac{\OmH \Dl\Psi}{4\pi c}	 \approx  \frac{\Dl V}{2}.   \hspace{0.5cm}  
			\lb{eigEMF-in} \hspace{0.33cm} 
		\end{eqnarray}   \lb{eigEMF-ab}  \end{subequations}   

\begin{figure*}
\begin{center}
	\includegraphics[width=12cm, height = 7cm, angle=-0]{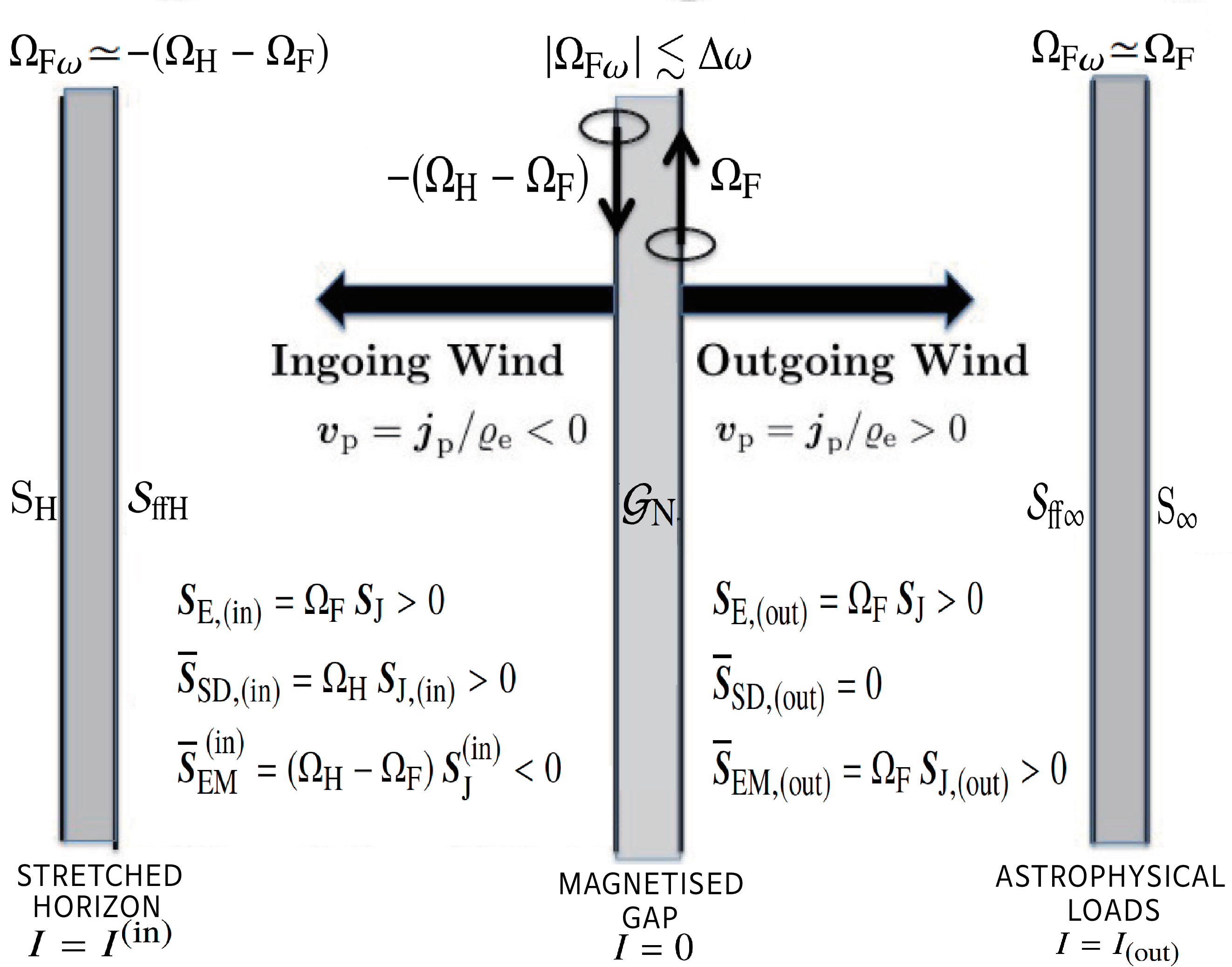}
\end{center}
\caption{A twin-pulsar model with three (two \emph{resistive} and one \emph{inductive}) membranes with $\OLOmFm$ in Eq.\ (\rf{OLOmFm}).  The {\em two} magnetospheres are anti-symmetric to each other with respect to the ZAM-surface \SSN; the outer SC domain $\calDout$ behaves like a normal pulsar-type magnetosphere rotating with the FLAV $(\OLOmFm)_{\rm (out)}=\OmF$, whereas the inner GC domain $\calDin$ does like an anti-pulsar-type magnetosphere \emph{counter}-rotating with the FLAV $(\OLOmFm)^{\rm (in)}= -(\OmH-\OmF)$. 
The inductive membrane $\SN$ covers the magnetized ZAM-Gap in $|\OmFm| \lo\Dlom$ at $\PG{\OLOmFm}=0$, which will be an inevitable product of the widening due to relaxation of RTD (Sec.\  \ref{TW-P-M}).  
The \emph{ingoing} flux of \emph{ negative} energy in $\calDin$ from the $\GN$ is equivalent to the \emph{outgoing} flux of \emph{positive} energy from the hole, i.e., $\OLvcSsdin =\OmH \vcSJin=-\OmH \vcSJinU=- \OLvcSsdinU>0$ (see Eqs.\ (\rf{vcSEMsd}), (\rf{energyH})), just as the \emph{ingoing} flux of \emph{negative} angular momentum is so to the \emph{outgoing} flux of {\em positive} angular momentum.  
	A steady pair-production mechanism due to the voltage drop $\Dl V$ will be at work to supply ZAM-particles, dense enough to anchor threading FLs, thereby ensuring  $\OmF=\omN$. The two batteries supply electricity to `external resistances,' such as Joule heating leading to `particle acceleration' and `entropy production,'  in the resistive membranes $\Sffinf$ and $\SffH$, respectively. The plasma-shed on the ZAM-Gap will easily divide wind particles charge-sepatated into both directions ($\vcv\ggel 0$) by the magnetic sling-shot ($\OmFm\ggel 0$) (see FIG.\ \rff{DC-C}). 
	The ZAMOs 
will see that the ZAM-Gap with $\PG{\epsJ}=0$ will be \emph{frame-dragged} together with the two GR and SC domains into rotation with the AV $\omN=\OmF$. 
}   \lbf{F-WS}  \end{figure*}

\section{A twin-pulsar model with Rotational-tangential discontinuity}	   \label{TW-P-M} 
\setcounter{equation}{0}
\numberwithin{equation}{section}

Path integrals of $\vcEp$ in Eq.\ (\rf{EMF-ab}a,b) along the two closed circuits $\calCout$ and $\calCin$ reveal the sharp potential drop $\Dl V$ between the EMFs for the two circuits in the SC and GR domains \cite{oka15a}.  { This discontinuity (i.e., RTD) comes from the differential rotation of the outer prograde-rotating domain with $\OmF$ and the inner retrograde-rotating one with $-(\OmH-\OmF)$, and distinctly will differ from any ordinary rotational or tangential discontinuities in classical magnetohydrodynamics  (see, e.g.,\ \citet[\S 71]{lan84}).} We attempt to briefly analyze a fundamental feature of this RTD on this surface \SSN\ in the force-free {\em limit}.  Important is the evidence that the above results including the voltage drop $\Dl V$ across \SSN\ come from the \emph{continuous} function of the FDAV $\om$, and all these results also seem  to be obtainable by assuming a \emph{discontinuous} step function $\OLom$ for $\om$;  
\beeq
\OLom= \left\{
\begin{array} {ll}
	0 & ;\ \calDout\ (\OmFm>0), \\[1mm]   
	\omN \equiv \OmF & ;\ \ \ \  \mbox{\rm \SSN}\ \ \ (\OmFm=0), \\[1mm] 
	\OmH  &;\ \calDin\ (\OmFm<0 ).  
\end{array}  \right. \lb{OLom}    \eneq 
Likewise, $\OmFm$, $\vF$, and $\epsJ$ are also replaced by the following step-functions ($\OLOmFm\equiv \OmF-\OLom$); 	
\begin{eqnarray}
\OLOmFm= \left\{
\begin{array} {ll}
	\OmF\equiv (\OLOmFm)_{\rm (out)} & ; {\bf \Uparrow}\ \calDout\ (\OmFm>0),
	\\[1mm] 
	0 \equiv (\OLOmFm)_{\rm (N)}  & ;\ \ \ \ \ \mbox{\rm \SSN}\ (\OmFm=0),  	
	\\[1mm] 
	-(\OmH- \OmF) \equiv (\OLOmFm)^{\rm (in)}  &;\ {\bf \Downarrow}\ \calDin\ (\OmFm<0), 
	\end{array} \right. \nonumber \\
\OLvF=\OLOmFm\vp/\al,  \hspace{5.7cm} \nonumber \\
\OLvarepsJ=\OLvF(\vp\Bp^2 /c), \hspace{5.5cm}
 \lb{OLOmFm}    
 \end{eqnarray} 
where ${\bf \Uparrow}$ and ${\bf \Downarrow}$ show that the $\calDout$ prograde-rotates, while the $\calDin$ retrograde-rotates, respectively (see the two arrows in FIGs.\ \rff{DC-C}, \rff{GapI}, and \rff{F-WS}).  
The differences of $\OLOmFm$ and the EMFs across \SSN\ become, respectively,  
\begin{subequations} \begin{eqnarray}
\DN{\OLOmFm}=(\OLOmFm)_{\rm (out)}- (\OLOmFm)^{\rm (in)}=\OmH,  \lb{DN-OmFm} \\ 
 \DN{\calE}=- \frac{ \DN{\OLOmFm}}{2\pi c}\Dl \Psi=- \Dl V     \lb{Dl-V2} 
\end{eqnarray} \lb{OmF/calE} \end{subequations}  
(see Eqs.\  (\rf{Dl-V}), (\rf{distr2})). That is, there is a  proportional jump in $\DN{\OLOmFm}=\OmH$, leading to a drop between the two EMFs, although there is no jump of the transport rate $I$ of angular momentum across \SZAM, i.e., $\DN{I}=\PN{I}=0$ (see Eqs.\ (\rf{DN/SN}a,b)).  


The related electric field $\OLEp$ and its discontinuity at \SSN\ respectively become, from Eq.\ (\rf{EpBt}),
\begin{subequations}  \begin{eqnarray}
		\OLEp=-  \frac{\OLOmFm}{2\pi\al c}\vcnb\Psi, \hspace{1.0cm}     \lb{OLEp} \\  
		\hspace{0.5cm}	\DN{\OLEp}=-  \frac{\DN{\OLOmFm}}{2\pi c} \LPfrac{\vcnb\Psi}{\al}_{\rm N} 
		= -  \frac{\OmH}{2\pi c} \LPfrac{\vcnb\Psi}{\al}_{\rm N}.  \hspace{1.0cm}	\lb{DNOLEp}  
	\end{eqnarray}  \lb{OLEpDN} \end{subequations}  
The expression for $\OLEp$ naturally reproduces the same results for $\calEout$ and $\calEin$ in Faraday path integrals of $\vcEp$ along the circuits $\calCout$ and $\calCin$, respectively, as given in expressions (\rf{EMF-ab}a,b). Also, the discontinuity of the EMF is given already in Eq.\ (\rf{Dl-V2}). 

The related energy fluxes $\vcSEM$ and $\vcSsd$ are also replaced with the step-functions $\OLvcSEM$ and $\OLvcSsd$, respectively, i.e., 
\beeq		
\OLvcSEM=\OLOmFm \vcSJ, \quad \OLvcSsd =\barom \vcSJ . 
\lb{OLvcSEMsd} \eneq  
There is naturally no discontinuity in the `overall' energy and angular momentum fluxes across the null surface \SSN\ with $\DN{\vcBp}=0$, i.e., similarly to Eq.\ (\rf{DG/SE/SJ}),
\beeq
\DN{\vcSE}=\DN{\OLvcSEM+\OLvcSsd}=\OmF \DN{\vcSJ}=0. 
\lb{DGvcSE} \eneq 
 Likewise, we compute the differences of the Poynting flux $\OLvcSEM$ and the spin-down flux $\OLvcSsd$ across \SSN\ 
from Eqs.\ (\rf{DN-OmFm}) and (\rf{OLvcSEMsd}); 
	\begin{subequations}  \begin{eqnarray}
		\DN{\OLvcSEM}=  \OLvcSEMout - \OLvcSEMinU,  \lb{energyN}  \\ 
		\DN{\OLvcSsd}  = - \OLvcSsdin,     \lb{energyG}  \ \  
	\end{eqnarray}  \lb{vcSEMsd} \end{subequations}  
where $\OLvcSsdout=0$ and $\OLvcSEMout= \vcS_{\rm E,(out)}$, because $\om$ is regarded as negligible in the outer SC domain. Then, by Eq.\ (\rf{DGvcSE}), we have $-\DN{\OLvcSsd}= \DN{\OLvcSEM}$, i.e., 
	\beeq   
	 \OLvcSEMinU+\OLvcSsdin = \OmF \vcSJin=\OLvcSEMout,        \lb{energyH}  
	\eneq    
which after all is equivalent to Eq.\ (\rf{SEEMsd}) (note that $\OLvcSEMinU =-(\OmH-\OmF)\vcSJin <0$).  Integrating Eq.\ (\rf{energyH}) over all open FLs from $\Psi_0$ to $\Psib$ returns to the first law, $c^2 dM=\Th dS+\OmH dJ$, indicating that the equation $c^2 dM=\OmFb dJ$ means nothing but astrophysical loading (see, also, Eqs.\ (\rf{SDenergy})). 

In the pulsar-type force-free magnetosphere, the \emph{conserved} energy flux $\vcSE=\OmF\vcSJ$ alone flows outward from \SNS\ to \Sinf. In contrast, in the hole's force-free magnetosphere, the \emph{conserved} energy flux $\vcSE$ `must' be split into the two \emph{non-conserved} fluxes $\vcSEM$ and $\vcSsd$ in $\calDin$, to comply with the first law of thermodynamics, but these energy fluxes flow along the same equipotential FCSLs, so that the Kerr hole will be unable to discriminate between the sum of $\vcSEM= - (\om-\OmF) \vcSJ$ and $\vcSsd=\om\vcSJ$ and that of $\OLvcSEM=-(\OmH-\OmF)\vcSJ$ and $\OLvcSsd=\OmH\vcSJ$ in the inner GR domain, while $\OLvcSEM=\vcSE$ and $\OLvcSsd= 0$ hold in the outer SC domain (see Eqs.\ (\rf{OLom}), (\rf{OLOmFm}) and (\rf{OLvcSEMsd})). Therefore, the basic properties of the energy fluxes in the curved space with $\om$ and $\OmFm(\om,\Psi)$ will be fully reproduced in the pseudo-flat space with $\OLom$ and $\OLOmFm(\OLom,\Psi)$ \cite{oka15a}.

A kind of inevitable relaxation of RTD will occur due to the pair-particle production by the voltage drop $\Dl V$, thereby leading to widening  the ZAM-surface to a ZAM-Gap $\SG$ with a finite thickness. 
This Gap may be regarded as effectively consisting of two halves of {\em virtual} magnetized NSs consisting of a `non-spherical, shell-like structure,' e.g., the outer one foward-rotating with $(\OmFm)_{\infty}=\OmF\approx (\OmH/2)$ and the inner one backward-rotating with $(\OmFm)_{\rm H}= -(\OmH-\OmF)\approx -(\OmH/2)$. These two halves of the structure are packed together  reversely and threaded by the poloidal field $\PG{\vcBp} \neq 0$ with no toroidal component. They are pinned down in the ZAM-particles pair-produced in the Gap, to ensure $\OmF=\om_{\rm G}$.

The above conjecture is depicted schematically in FIG.\ \rff{F-WS}; the pulsar-type wind is slung outward from the outer magnetized \emph{ virtual} shell-like structure spinning with $(\OLOmFm)_{\rm (out)} =\OmF$ through the outer domain $\calDout$ with the outer \SOL, in which the related Poynting flux $\OLvcSEMout$ is equal to $\OmF \vcSJout>0$, with no frame-dragging spin-down energy flux followed. In contrast, the anti-pulsar-type wind is reversely slung inward from the inner magnetized {\em  virtual} shell-like structure counter-spinning with $(\OLOmFm)^{\rm (in)} =-(\OmH-\OmF)$ through the inner domain $\calDin$ with inner \SIL, 
in which the {\em positive}-valued Poynting flux is directed \emph{inward}, i.e., $\overline{\vcSEMinU} =(\OmH-\OmF) \vcSJinU<0$, 
while the {\em negative}-valued frame-dragging spin-down flux is directed \emph{inward} as well. Hence, $\OLvcSsdin = -\OmH\vcSJinU>0$, which may be understood as an equivalent to an {\em in-}flow of the {\em negative} energy, i.e.,\ $\OLvcSsdinU =\OmH \vcSJinU=- \OLvcSsdin<0$, related to the {\em in}-flow of {\em negative} angular momentum. So, the `overall' energy flux becomes
\beeq 
\vcSEin= \OLvcSEMinU+\OLvcSsdin=-\OmF\vcSJinU= \OmF\vcSJin>0,
\lb{Totalin} \eneq  
which will be equal to $\vcSEout=\OmF\vcSJout$ across the Gap (see Eqs.\ (\rf{energyH}), (\rf{DG/SE/SJ}); Sec.\  \ref{Energetic}). 



\section{The energetics and structure of the twin-pulsar model}  \label{FluxC}   
	\setcounter{equation}{0}
	\numberwithin{equation}{section}

\newcommand{\calPEMin}{\calP_{\rm EM,(in)}}
\subsection{Energetics of the hole's self-extraction of energy}  \label{Energetic}   
A variant of the first law, i.e., $\OmH |dJ|=\Th dS+c^2 |dM|$ indicates that the energy extracted through the spin-down energy flux will be shared at the inductive membrane $\SN$ between the two Poynting fluxes toward the two resistive membranes $\SffH$ and $\Sffinf$ (see Eqs.\ (\rf{first-II})). 
By integrating Eq.\ (\rf{energyH}) over all open field lines from $\Psi_0$ to $\Psib$, we have  
		\beeq
		  \int_{\Psi_0}^{\Psib} \al \OLvcSsdin  \cdot d\vcA 	
		= -  \int_{\Psi_0}^{\Psib} 
		\alpha \OLvcSEMinU \cdot d\vcA  + \int_{\Psi_0}^{\Psib}  
		\alpha \OLvcSEMout \cdot d\vcA ,  
		\lb{SDenergy} \eneq   
which explains that the power $\OmH\calPJ$ self-extracted from the horizon is evenly distributed between entropy production  
in $\SffH$ and particle acceleration in $\Sffinf$ (cf.\ \cite{mac82}, \S 7.3;  \cite{tho86}, Ch.\ IV D).
The two terms of the right-hand side of Eq.\ (\rf{SDenergy}) actually become, by Eqs.\  (\rf{Iout/in}a,b), (\rf{ThdS/dt}), (\rf{torque/SffH}), and (\rf{calPEout-in}a,b), 	respectively;  
	\begin{subequations} \begin{eqnarray}
	\TTH \frac{dS}{dt}	
		=\frac{1}{2 c}   \int_{\Psi_0}^{\Psib} 	  
		(\OmH-\OmF)^2(\Bp\vp^2)_{\rm ffH} d\Psi,	  \hspace{1cm}	 \lb{H/resistanc}    
		\end{eqnarray}  	
which corresponds to $\Th dS=-(\OmH-\OmFb)dJ$ in Eq.\ (\rf{second-ib}), and	
	 \begin{eqnarray}
	-c^2 \dr{M}{t} = \calPEout
		= \frac{1}{2c}   \int_{\Psi_0}^{\Psib}    
		\OmF^2 (\Bp\vp^2)_{{\rm ff} \infty} d\Psi \hspace{1.3cm}  \ \  \lb{L/res/a} \\  
	\quad\quad =\calPEin 
		= \frac{1}{2c}  \int_{\Psi_0}^{\Psib} 	
		\OmF(\OmH-\OmF) (\Bp\vp^2)_{{\rm ffH}} d\Psi \ \hspace{0.9cm}   \lb{L/res/b} 	
	\end{eqnarray}  \lb{HL/resistance}   \end{subequations} 
(see Eq.\ (\rf{Sffinf-M})), where $c^2 dM=\OmFb dJ= -\calPEin dt= -\calPEout dt$ by Eqs.\ (\rf{first-II}a).  
The left hand-side of Eq.\ (\rf{SDenergy}) reduces by Eqs.\ (\rf{calPJout-in}a,b) to 
	\begin{subequations} \begin{eqnarray}
		- \OmH\dr{J}{t} 	\hspace{6cm}	\nonumber  \\
		=\OmH \calPJin = \frac{\OmH}{2c} \int_{\Psi_0}^{\Psib} 	
		(\OmH-\OmF) (\Bp\vp^2)_{{\rm ffH}} d\Psi, \ \hspace{0.3cm} \lb{J/res/b} 	\\ 
		=\OmH\calPJout= \frac{\OmH}{2c}  \int_{\Psi_0}^{\Psib} 
		\OmF (\Bp\vp^2)_{{\rm ff} \infty} d\Psi \hspace{1.3cm} \lb{J/res/a}   
		\end{eqnarray}  \lb{HJ/resistance}   \end{subequations} 
(see Eq.\ (27) in \cite{bla22}). 
Summing up Eqs.\ (\rf{H/resistanc}) and (\rf{L/res/a}) or (\rf{L/res/b}) and using the `boundary condition' $\Iout=\Iin$ yield $-\OmH (dJ/dt)=\OmH\calPJ$. Also, Eqs.\ (\rf{HL/resistance}) and (\rf{HJ/resistance}) show $\DG{\calPE}=\DG{\calPJ}=0$, in spite of the RTD of the EMFs $\DG{\calE}=-\Dl V$ existent  in the Gap $\GN$ (see Eq.\ (\rf{Dl-V})). 

It therefore seems in one interpretation that the {\em positive} angular momentum and energy `self-extracted' by the surface magnetic torque through $\SffH$ from the hole will be  transported outwardly in the inner domain $\calDin$ and then beyond the ZAM-Gap with $\PG{\vcSJ}=\PG{\vcSE}=0$,     
to the outer domain $\calDout$ with astrophysical loads in $\Sffinf$ (see FIG.\ \rff{GapI}), despite that the angular momentum flux cannot pass through the ZAM-state in the Gap with $\PG{\epsJ}=\PG{\OmFm}=\PG{I}=0$. 

The point is that ZAM-particles created inside the Gap $\GN$ are spinning with $\omN=\OmF$ dragged by the hole's rotation, literally with no angular momentum. The particles will effortlessly flow out of the Gap, flung both outwards or inwards from the surfaces \SgapO\ or \SgapI\ on the `plasma-shed,' with positive or negative angular momenta by the respective magneto-centrifugal forces, thereby keeping the ZAM-state of the Gap. 
This corresponds to the situation where the outgoing Poynting flux $\vcSEM> 0$ is related to the outer EMF $\calEout$, whereas the ingoing Poynting flux $\vcSEM<0$ is related to the inner EMF $\calEin$.  Then, the distant observers may think as if the spin-down energy `self-extracted' through the resistive horizon membrane $\SffH$ were shared between the out- and in-going Poynting fluxes reaching the two resistive membranes $\Sffinf$ and $\SffH$, respectively, to dissipate in particle acceleration and entropy generation as seen in Eq.\ (\rf{SDenergy}). 

The above depiction of energy-sharing on the ZAM-Gap will be another interpretation of the extraction process. This is in a sense a reversal of the same phenomenon that the hole can be an acceptor of {\em negative} angular momentum only for self-extraction, thereby spinning down, eventually and equivalently, to extract {\em positive} angular momentum and energy from the hole. 



It seems that the Kerr hole makes very skillful use of the properties of the inner GR domain $\calDin$ counter-rotating to pass {\em negative} angular momentum inward from the ZAM-Gap for the sake of defending the first and second laws, thereby enabling self-extraction of {\em positive} energy successfully. 


\subsection{The stream equation for a twin-pulsar magnetosphere }	 \label{streamEq}  
The \emph{stream} equation in the force-free limit determines the FCSL structure in terms of the FLAV $\OmF(\Psi)$ (and also ZAMO-FLAV $\OmFm$) and the current/angular-momentum function $I(\Psi)$: 
	\begin{eqnarray} 
		\nabla\cdot\left\{ \frac{\al}{\vp^2} \left[1-\frac{\OmFm^2\vp^2}{\al^2 c^2} \right] \nabla\Psi \right\} \hspace{1cm}	\nonumber \\
		\hspace{0.5cm}	+ \frac{\OmFm}{\al c^2}\dr{\OmF}{\Psi}(\nabla\Psi)^2  +\frac{16\pi^2}{\al\vp^2 c^2} I\dr{I}{\Psi}=0   
		\lb{stream/MT} \end{eqnarray} 	
(see Eq.\ (6.4) in \cite{mac82}), which contains the two conserved quantities $\OmF(\Psi)$ and $I(\Psi)$; with regard to the former, not only the two light surfaces \SOL\ and \SIL, but also the null surface \SSN\ between them, are contained rather explicit way by $\vF=\OmFm\vp/\al=\pm c$ and $\vF=\OmFm=0$ (see Sec.\ \ref{I&LSs}; Eqs.\ (\rf{vp/vt})).  On the latter, the breakdown of the force-free condition appears in a complicated form of severance of both current- and stream-lines, $\vcj=\vcv=0$, or $I=\OmFm=0$ on the null surface \SSN.
All these complications are thereby to allow the particle-current sources to be inserted, by particle production due to the voltage drop between a pair of batteries under the null surface \SSN\ (Secs.\ \ref{NullS}, \ref{indMem}). 



When the null surface \SSN\ develops into a gap $\GN$ with any finite width where $I(\ell,\Psi)=0$ (see, e.g.,\ FIG.\ \rff{GapI}), then the \emph{stream} equation (\rf{stream/MT}) will relevantly be modified. 
We conjecture now that the poloidal magnetic field $\vcBp$ with no toroidal component $\Bt=I(\ell, \Psi)=0$ will be robust enough to thread the particle-production Gap $\GN$ due to the voltage drop $\Dl V$, with the ZAM-state of the Gap maintained to keep circulation with $\omN$ around the hole. Probably, this situation will be compatible with the solution of the \emph{stream} equation  $\nabla\cdot ((\al/\vp^2) \nabla\Psi)=0$ for the particle production Gap within $|\OmFm|\leq \Dl \om$ where $\OmFm=I=0$. 

\section{Discussion and conclusions}   \label{Dis-Con}
	\setcounter{equation}{0}
	\numberwithin{equation}{section}
		
\subsection{Astrophysical roles of frame-dragging in energy extraction}  \label{role-FD}  
The frame-dragging effect plays an indispensable role in reforming the pulsar force-free magnetosphere into the specification suitable for the Kerr hole's so as, in particular, to be adaptive to the first and second laws of thermodynamics and to include the current-particle sources. The observance of the two laws demands breakdown of the force-free condition.  But the issue, ``{\em where does the breakdown take place in the force-free magnetosphere?,}'' seems to have remained almost untouched for more than four decades since \cite{bla77}, probably because the astrophysical roles of frame-dragging remain so far ill-understood (see, e.g., \cite{bla02}).  This may explain that the topic of energy extraction from Kerr holes still is a big challenge even today \cite{tho17}, although several hints toward understanding of the process of the breakdown were given in some references (e.g.,\ \cite{bla77,mac82}). 

The ZAMOs circulating with $\om$ round the hole will in fact be certain that the Kerr hole can self-extract its extractable energy, if and only if frame-dragging is correctly taken into account. The coupling of the FDAV $\om$ with the FLAV $\OmF$ begins with the ZAMO-FLAV, $\OmFm=\OmF-\om$. Then, they will see for $0<\OmF<\OmH$ that the coupling necessarily leads to nesting the inner domain $\calDin$ counter-rotating ($\OmFm<0$) inside the outer domain $\calDout$ with the null surface \SSN\ (or the Gap $\GN$) between them. The magnetic sling-shot effect works inwardly in the former domain, and outwardly in the latter, and hence the ZAMOs will see {\em a sufficiently strong flux of \emph{negative} angular momentum leaving the null surface \SSN} ($\vcSJinU=-\vcSJin <0$), 
and this does not contradict {\em a Poynting flux going towards the hole} ($\vcSEMin<0$) (as discussed in \cite{bla77}). 
They will understand that the overall energy flux $\vcSE=\OmF\vcSJin$ always flows outward. 
It is on the null surface \SSN$=$\SZAM\ that the spin-axis of the hole's force-free magnetosphere changes from positive in $\calDout$ to negative in $\calDin$, and the emitted Poynting flux changes direction from outward to inward, and hence the \emph{complete} violation of the freezing-in-ness and force-freeness \emph{must} take place. 

The force-free magnetosphere is `magneto-centrifugally' divided by \emph{this surface} \SZAM$=$\SSN\ into the SC and GR domains, spinning reversely each other.  The inner domain $\calDin$ with \emph{negative} angular momentum density $\epsJ<0$ may be referred to as the `effective ergosphere' \citep{oka92,oka06}, because FCSLs there are {\em negative-angular-momentum orbits} themselves and hence correspond to {\em negative-energy orbits} in the ergosphere in the Penrose process \cite{pen69}.  
When a property of  $\vcSJin=-\vcSJinU>0$ is fully used, the ZAMOs will see $\OLvcSEMinU=(\OmH-\OmF)\vcSJinU<0$ for entropy production on $\SffH$, but they see $\OLvcSsdin=\OmH\vcSJin>0$ and hence $\vcSEin=\vcSEMout=\OmF\vcSJin>0$ for particle acceleration on $\Sffinf$. The counter-rotating inner domain thus nested inside \SZAM\ will be designed so that the electrodynamic process of self-extraction of energy can surely obey the thermodynamic laws.

The null surface \SSN\ is the \emph{key} surface where the eigen-FLAV $\OmF$ and the eigen-FDAV $\omN$ can simuletaneously be determined uniquely, thereby dragging    
the force-free magnetosphere into circulation round the hole with the FDAV $\omN=\OmF$, and hence enabling the Kerr hole to self-extract its rotational energy. It was recently pointed out in \cite{cos21} that \emph{what are dragged by the Kerr hole are the ZAMOs and the compass of inertia.}  The force-free magnetospheres circulating with $\omN=\OmF$ round the Kerr hole may be included among them as well. 
 Thanks to frame-dragging, ``a physical observer will see not only a Poynting flux of energy from \SSN\ entering the hole,'' (as suggested in \cite{bla77}) 
 but he will also see another Poynting flux from \SSN\ outward, to transform into kinetic energy through the particle acceleration zone in the outer resistive membrane $\Sffinf$, and then to evolve eventually into a high-energy gamma ray jet \citep{oka15b}. In any case, the force-free theory, inclusive of the `complete violation' of its force-freeness, will be a consistent, unique theory for extracting energy from Kerr holes.  These ingenious actions of frame-dragging may seem to be due to a wire-puller maneuvering behind the scenes, and we may so far never have seen them as its real astrophysical effects.  The ZAMOs will not think of these, rather modest and peculiar actions of frame-dragging, as `spooky.' 
 

\subsection{Concluding remarks}  

Kerr holes will construct an exquisite nested structure, which consists of a counter-rotating inner GR domain $\calDin$ and the outer SC domain $\calDout$, with the particle-producing ZAM-Gap $\GN$ between them, thereby enabling to self-extract their reducible energy, probably free from violating any physical law. 
We emphasize that we consider magnetic field lines to be not only threading the ZAM-Gap but also be pinned down in it, which circulates round the hole with $\omN=\OmF$ dragged by the hole rotation.  The resultant twin-pulsar model will be a natural outcome from the unification of pulsar electrodynamics and BH thermodynamics with the help of frame-dragging.  The secret of the physics of the magnetized ZAM-Gap,  its structure, particle production in it, etc.,\ are still waiting to be unveiled as the Gaps power the ultimate high-energy central engines in the universe.  

If observed large-scale high-energy $\gamma$-ray jets from AGNs are really originated from quite near the event horizon of the central super-massive BHs, it appears  plausible that these jets are a magnificent manifestation of the trinity of general relativity, thermodynamics, and electrodynamics (GTE);  precisely speaking,  frame-dragging, the first and second laws, and unipolar induction. The heart of the black hole's central engine may lie in the Gap $\GN$  between the two, outer and inner, light surfaces just above the horizon, and the embryo of a jet will be born in the Gap $\GN$ under the inductive membrane $\SN$ in the range of $2^{1/3}\rH \lo r_{\rm N} \lo 1.6433\rH$ quite near the horizon (see, e.g., \cite{oka15b}: Appendix Sec.\ \ref{SNshapAp} for an approximate topology of \emph{this surface} \SSN). 
A further illumination on the BH Gap physics is needed for confirming this postulate.

It will nevertheless be the dragging of inertial frames that creates a synergy effect between thermodynamics and electrodynamics for the purpose of energy self-extraction through the hole's force-free magnetosphere. Some remaining issues out of the scope of this paper may include: 
When the gravito-electric potential gradient of the hole $\OmH$ and the strength of magnetic field $\vcBp$ threading and pinned down in the Gap are given by  observations, for instance, {how efficiently does the voltage drop due to the EMFs, $\Dl V=\DG{\calE}$, contribute to particle production?}  The pinning-down of the poloidal field $\vcBp$ by the ZAM-particles pair-produced in the Gap will result in a complete magnetization of the plasma to ensure that field lines possess $\OmF=\omN$. If so, how much density of ZAM-particles is needed for pinning-down and magnetization to ensure $\OmF=\omN$? How large is the Gap width $\Dlom=|\partial\om/\partial\ell| \Dlell$, to provide neutral plasma particles with $\vre\approx 0$, being charge-separated to both the outflow and inflow in the force-free domains? 
\footnote{The referee of \cite{oka92} asked one of the present authors to comment on the question of causality then arising about the BZ process.  This paper now contains the final detailed response to his request and has been recently submitted to the same Journal. However, two members of the Journal's Editorial Board have agreed that the paper is unsuitable for publication in that Journal.  That journal's Scientific editor's Comments are as follows:  ``This Journal publishes only direct astrophysical applications of physics. Your paper deals with fundamental physics, and aims to clarify results in General Relativity. I recommend you to submit the paper to a journal publishing in fundamental physics, such as Phys Rev.'' }  

\appendix
	\section{The $3+1$ formulation for the modified BZ process}  \label{ED3/1F}  
\setcounter{equation}{0}
\numberwithin{equation}{section}

\subsection{Fundamental equations and conditions}  \label{basics}  
The absolute space around a Kerr hole with mass $M$ and angular momentum per unit mass $a=J/Mc$ is described in Boyer-Lindquist coordinates: 
\begin{subequations}
	\begin{eqnarray}
		ds^2=(\rho^2/\Delta) dr^2+\rho^2 d\theta^2 +\vp^2 d\phi^2, \\[1mm]	
		\rho^2\equiv r^2+a^2\cos^2 \theta,\ \Delta \equiv r^2-2GMr/c^2 +a^2,\\[1mm]	
		\Sigma^2\equiv (r^2+a^2)^2-a^2\Dl\sin^2\theta, \ \  \vp=(\Sigma/\rho)\sin\theta, \\[1mm]	
		\al=\rho\Dl^{1/2}/\Sigma , \ \  \om=2aGMr/c\Sigma^2,  \lb{alom}  
	\end{eqnarray} \lb{Kmetric} \end{subequations}  
(see \cite{mac82,oka92}), where $\al$ is the lapse function/redshift factor and $\om$ is the FDAV. The \emph{two} parameters $\al$ and $\om$ are reminiscent of the no-hair theorem in specifying the Kerr spacetime. They are given as unique functions of $\vp$ and $z$ in the Boyer-Lindquist coordinates, and $0\leq\al\leq 1$ and $\OmH\geq\om\geq 0$. Note that for $\al\to 0$, $\om\to\OmH=$constant on \SH\ by the zeroth law of thermodynamics. 

When we introduce curvilinear orthogonal coordinates $(\ell,\Psi)$ in the poloidal plane, where $\ell$ stands for the distances measured along each FL $\Psi=$ \,constant, then we express, e.g.,\ $\om=\om(\ell,\Psi)$. Just as $\al$ was `coordinatized' in the stretched horizon \citep{mac82,tho86}, we `coordinatize' $\om$ along FLs in the whole magnetosphere \citep{oka15a}. The ZAMO-measured FLAV $\OmFm$ as well is  `coordinatized' (see, e.g., FIG.\ \rff{Flux-om}). 

We revisit basic expressions for the poloidal and toroidal components of $\vcB$, $\vcE$, the charge density $\vre$, the particle velocity $\vcv$, and the FL rotational velocity (or FLRV) $\vF$ in the steady axisymmetric state \citep{mac82,tho86,oka92,oka15a}.  
For the electric field $\vcE$ in a curved spacetime, we use Eq.\ (\rf{vcEp-A}), which is the kick-off equation to make frame-dragging couple with unipolar induction, by utilizing both the freezing-in and force-free conditions (Eq.\ \rf{ff-a}a,b). 
When inertial forces are regarded as negligible in Eq.\ (\rf{ff-fz}), Eq.\ (\rf{fi-c}) 
implies that `force-free' magnetic FLs are freezing-in particles and yet dragged around by the motion of `massless' particles (i.e., the velocity $\vcv$). The combination of those two opposite conditions then creates a kind of extreme physical state \citep{oka06} 
where current-field-streamlines are equipotentials (see Eq.\ (\rf{vc-vjA})), and no particle acceleration takes place in the force-free domains. This circumstance creates a good chemistry between electrodynamics and thermodynamics. 

The theory cannot however be viable unless the two conserved quantities are determined by the criticality-boundary condition formulated by the breakdown of the above conditions (see Sec.\  \ref{NullS}). The flows of angular-momentum and energy, particles, and currents are described by {\em flux}, {\em wind}, and {\em circuit} theories, respectively, not to mention that these theories must be consistent with each other. 

\subsection{The electric current }  \label{coupling}  
We decompose the magnetic field $\vcB=\vcnb\times\vcA$ as
\begin{subequations} \begin{eqnarray}
		\vcBp=-(\uvt \times\nabla \Psi)/2\pi\vp,  \lb{bhBp}\\  
		\vcBt=- (2I/ \vp\al c)\uvt,  \lb{bhBt}   
	\end{eqnarray} \lb{Bpt} \end{subequations}  
where the `current function' is denoted with $I=I(\ell,\Psi)=I(\OmFm,\Psi)$ as it is generally defined.  From Eq. (2.17c) in \cite{mac82} for $\vcj$, we have 
\beeq 
\vcj=\frac{c}{4\pi\al} \left[\vcnb\times\al\vcB+ \frac{1}{c}(\vcE\cdot\vcnb\om)\vcm \right],
\lb{vcjP}  \eneq   
where $\vcm=\vp\uvt$ is a Killing vector,  and then for $\vcjp$, 
\beeq 
\vcjp= 
\frac{\uvt\times\vcnb I}{2\pi\vp\al}. 
\lb{vcjpO}   \eneq   
Introducing the two orthogonal unit vectors $\uvp$ and $\uvn$ in the poloidal plane, i.e., 
$\uvp=\vcBp/|\vcBp|$ and $\uvn=-\vcnb\Psi/|\vcnb\Psi|$, and $\uvn\times\uvp=\uvt$, we have for the current function $I=I(\ell,\Psi)$,
\beeq
\vcnb I=\LPPlDr{I}{\ell} \uvp - 2\pi\vp\Bp\LPPlDr{I}{\Psi}\uvn
\lb{vcnbI} \eneq 
and hence, 
we express the electric current $\vcjp$ as 
\begin{eqnarray}
	\vcjp=\jpl\uvp+\jvl\uvn, \hspace{3cm} \nonumber \\[1mm]
	\jpl= -\frac{\Bp}{\al}\LPPlDr{I}{\Psi} ,\ \ \jvl=-\frac{1}{2\pi\al} \LPPlDr{I}{\ell}  
	\lb{jpl/vl} \end{eqnarray}  
(see \cite{oka99}).  In addition, for $\jt$, we have, from Eq.\ (\rf{vcjP}),
\beeq
\jt= \frac{\vp c}{8\pi^2 \al}   
\left[- \vcnb\cdot \LPfrac{\al\vcnb\Psi} {\vp^2}  + \frac{2\pi}{c} \vcE\cdot\vcnb\om  \right]. 
\lb{vcjtO} \eneq 
When $(\vcBp\cdot\vcnb)X=\Bp(\partial X/\partial\ell)=0$ for an arbitrary function $X$, we have $X=X(\Psi)$, and then we can say that $X$ is {\em conserved} along each FL. For example, $I=I(\Psi)$ in the `force-free' domains (see Eq.\ (\rf{SJa})). 		
We presume that each current line given by $I(\ell,\Psi)=$ constant must close itself as in {\em circuit} theory; each current line starts flowing  from one terminal of a unipolar induction battery and in the end returns to the other terminal in the steady-state, after supplying power to the acceleration zone with $\jvl>0$ (the current-closure condition).

\newcommand{\Nbe}{n^{(-)}}    \newcommand{\Nbp}{n^{(+)}} 

\subsection{The velocity $\vcv$ of `force-free' particles}  \label{velocityP}  
Combining the two conditions (\rf{ff-a}a,b), we have
\beeq
\vcv=\vcj/\vre 
\lb{charge1} \eneq 
\citep{bla77}. When we denote the number densities of electrons and positrons by $\Nbe$ and $\Nbp$, the charge density is given by $\vre=e(\Nbp - \Nbe)$. Here, Eq.\  (\rf{charge1}) implies that the `force-free' plasma must be charge-separated, i.e., $\vre=-e\Nbe$ or $+e\Nbp$, and that the role of `massless' or `inertia-free' particles is just to carry charges, exerting no dynamical effect. 

The `force-free' domains of no particle acceleration must be terminated by restoration of particle inertia for particles to accelerate, thereby determining the eigenfunction $I(\Psi)$ (see Eqs.\ (\rf{Iout/in}a,b)). This requires a change of the {\em volume} currents, parallel to the poloidal field $\vcBp$ ($\jvl =0$) in the force-free domain, into the {\em surface} currents, perpendicular to $\vcBp$ ($\jpl \approx 0$) on the terminating surfaces of the outer and inner force-free domains $\Sffinf$ and $\SffH$ (see Sec.\ \ref{Iout/in}).  Moreover, the breakdown of the freezing-in and force-free conditions on \emph{this surface} \SSN\ imposes $\vcv=\vcj=0$, because $\vcv>0$ far outside and $\vcv<0$ near the horizon and $\vcj$ does not change direction but must vanish. Also, $\vre$ must certainly change its sign at the place of breakdown (see Sec.\  \ref{NullS}).  This implies that the breakdown on \emph{this surface} \SSN\ must locate the sources of particles and currents there.

\subsection{The field angular momentum flux/the current function $I(\Psi)$} \label{FAgMF} 
The inner product of Eq.\ (\rf{ff-fz}) and $\vcm=\vp\uvt$ yields, using Eq.\ (\rf{vcjpO}), 
\begin{eqnarray}
0=\vcm\cdot\left[ \vre\vcEp+\frac{\vcj}{c}\times \vcB\right]=\frac{\vp}{c}(\vcjp\times\vcBp)_{\rm t} =\frac{\vp}{c}\jvl\Bp  \nonumber   \\ 
=-\frac{(\vcBp\cdot\vcnb) I }{2\pi\al c} =- \frac{1}{\al}\vcnb\cdot \left(\frac{I\vcBp}{2\pi c}\right)=- \frac{1}{\al}\vcnb\cdot \al\vcSJ, 	
\lb{SJa} \hspace{0.4cm}    \end{eqnarray} 
where $\vcSJ$ is given by the second of Eqs.\ (\rf{E-AmFlux}). It turns out that the {\em field} angular momentum $-\al\vp\Bt=(2/c)I(\Psi)$ is conserved along each FL. 
From Eqs.\ (\rf{bhBp}) and (\rf{jpl/vl}), we have $\jvl=0$ and then 
\beeq
\vcjp =-(1/\al)(dI/d\Psi)\vcBp.  
\lb{vcjp2} \eneq 
As $\vcm\cdot(\vcjp\times\vcBp)=0$ shows, it is precisely the `torque-free condition' included in the force-free condition that leads to $(\vcBp\cdot\vcnb)I=0$, i.e., $I=I(\Psi)$. It turns out that $I=I(\Psi)$ expresses not only the `current function' but also the `angular momentum flux per unit magnetic flux tube' in the force-free domains. Here, $I(\Psi)$ and $\OmF(\Psi)$ are both {\em two-sided}, and current lines are coincident with the corresponding field-streamlines. Note that the {\em two-sidedness} of $I(\Psi)$ holds only in the force-free domains, and the $\ell$- or $\OmFm$-dependence of $I$ must be restored, i.e., $I=I(\ell,\Psi)=I(\OmFm,\Psi)$, in the resistive and inductive membranes (i.e., decreasing in the former and vanishes in the latter; see Eq.\ (\rf{OL-I})).

\subsection{Two potential gradients $\OmF(\Psi)$ and $\OmFm(\ell,\Psi)$}  \label{D-OmF}  
The coupling of frame-dragging and unipolar induction in BH electrodynamics begins with Eq.\ (\rf{vcEp-A}). Inserting relations  $\vcB=\vcBp+\Bt\uvt$ and $\vcv=\vcvp+\vtt\uvt$ into Eq.\ (\rf{fi-c}) yields 
\[ \vcE=-\vcv/c\times\vcB= -\vcvp/c\times\vcBp+\frac{\uvt}{c}\times(\vcvp\Bt-\vtt\vcBp),  \]
and, by axial symmetry, $\vcE_{\rm t}=-\vcvp/c\times\vcBp=0$, and hence
\beeq
\vcvp=\kp\vcBp,
\lb{vp-Bp} \eneq  
where $\kp$ is a scalar function (see Eq.\ (\rf{kappa})). Then, we have 
\beeq
\vcEp= -\frac{(\vtt-\kp\Bt)}{2\pi\vp c}\vcnb\Psi .
\lb{vcEp-b} \eneq  
Equating two Eqs.\ (\rf{vcEp-A}) and (\rf{vcEp-b}) for $\vcEp$ yields 
\beeq
\vcnb A_0 = -K\vcnb\Psi,  \quad K \equiv -\frac{\al(\vtt-\kp\Bt) -\om\vp}{2\pi\vp c}, 
\lb{vcEp-A0} \eneq   
and taking the curl of $\vcnb A_0$, we get
\[ 
0= \vcnb\times \vcnb A_0= - \vcnb\times (K\vcnb\Psi) =-\vcnb K\times \vcnb\Psi=2\pi\vp\uvt  (\vcBp\cdot\vcnb K),  \]
which indicates that $K$ is a function of $\Psi$ only, and hence 
\beeq 
K=-\dr{A_0}{\Psi}\equiv \frac{\OmF(\Psi)}{2\pi c}.   
\lb{K}  \eneq  
Equating this $K$ to the one in Eqs.\ (\rf{vcEp-A0}) yields the FLAV $\vF$ in Eq.\ (\rf{vp/vt}) later. From Eqs.\ (\rf{vcEp-b})$\sim$(\rf{K}), we get 
\begin{subequations} \begin{eqnarray}
	\vcEp =- \frac{\OmFm}{2\pi\al c}\nabla\Psi = \frac{\vF}{ c} \Bp \uvn , \lb{bhEp} \\  
	\vre= -\frac{1}{8\pi^2 c} \nabla\cdot \left( \frac{\OmFm}{\al}\vcnb\Psi  \right), \lb{bh/rhoa} 
\end{eqnarray} \lb{bh/rhob}   \end{subequations}	
where $\vcEp$ is already given in Eq.\ (\rf{EpBt}). Note that it is the freezing-in condition that ensures Ferraro's law of iso-rotation for FLs in the steady axisymmetric state, i.e., $\OmF(\Psi)=$constant, but the ZAMOs see that the iso-rotation law for $\OmFm$ is violated by the FD effect, as shown by the $\ell$-dependence of $\OmFm=\OmFm(\ell,\Psi)$. The importance of \emph{this surface} \SSN\, where $\OmFm=\vcEp=0$, resulting from the violation was already pointed out in \cite{bla77}. 

The ZAMO-measured particle-velocity $\vcv$ and the FLAV $\vF$ are summarized as follows;  
\begin{subequations} \begin{eqnarray}
	\vcv=\kappa\vcB+\vF \uvt, \lb{vcv} \\  
	\vcvp=\kappa \Bp, \quad \vvt=\kappa\Bt +\vF, \lb{vcvp/t} \\ 
	\vF=\OmFm\vp /\al, \lb{vp/vt} \\  
	\kp=-(1/\vre\al) (dI/d\Psi).  \lb{kappa}  
\end{eqnarray}   \end{subequations}   
Because $\vF$ stands for the physical velocity of FLs relative to the ZAMOs, $\vcEp$ seen by the ZAMO is entirely induced by the motion of the magnetic FLs, i.e., $\vcEp =-(\vF\uvt/c)\times \vcBp$ \citep{mac82}. It is the `$\al\om$ mechanism' \citep{oka92}, with which one can define the inner light surface \SIL\ by $\vF=-c$ and \emph{this surface} \SSN\ by $\vF=0$, in addition to the outer light surface \SOL\ by $\vF=c$. 
We decompose the Lorentz force $(\vre\vcE+\vcj/c\times\vcB)$ as
\begin{eqnarray} 
\vre\vcE+\frac{1}{c} \vcj\times\vcB =\frac{1}{c}\left[ -\jvl \Bt\uvp+\jvl \Bp\uvt   \right.    \nonumber \\ 
\left.  +  \left(\jpl\Bt-\jt \Bp +\frac{\OmFm\vp}{\al}\Bp\vre\right)\uvn \right]  =0
\lb{JcrossB} \end{eqnarray}  
(see \cite{oka99}). 		 
The force-free and torque-free conditions are given simply by $\jvl\propto -(\partial I /\partial\ell)=0$, i.e., $I=I(\Psi)=$ constant along each FL (see Eqs.\ (\rf{SJa}) and (\rf{energya})). The $\uvn$-component yields
\begin{subequations} \begin{eqnarray}  
	\jt=\vre\vvt=(\OmFm\vp/\al) \vre + (1/\al^2 \vp c)(dI^2/d\Psi) , \ \ ~~ \ \lb{jta}  \\ 
	= -\frac{\OmFm\vp}{8\pi^2 \al c}\vcnb\cdot \left(\frac{\OmFm}{\al}\vcnb\Psi\right) +\frac{1}{\vp\al^2 c}\dr{I^2}{\Psi} , \ \ \  ~~ \ \lb{jtb}   
\end{eqnarray} \lb{jtc}  \end{subequations} 
which accord with the result from $\jt=\vre\vvt$ in Eq.\ (\rf{charge1}), utilizing $\vre$ in (\rf{bh/rhoa}), $\vvt$ in (\rf{vcvp/t}) and $\vF$ in (\rf{vp/vt}).  By Eqs.\ (\rf{vcjtO}) and (\rf{bhEp}), we  also have
\beeq 
\jt = - \frac{\vp c}{8\pi^2\al} \left[\vcnb\cdot\left(\frac{\al}{\vp^2}\vcnb\Psi\right)+\frac{\OmFm}{\al c^2}(\vcnb\Psi\cdot\vcnb)\om \right]     
\lb{jt-Farad}  \eneq    
(see Eqs. (2.17c), (5.6b) in \cite{mac82}). Equating two expressions (\rf{jtb}) and (\rf{jt-Farad}) for $\jt$ leads to the stream equation\ (\rf{stream/MT}).   

Putting relations among $\vcv$, $\vcj$, and $\vcB$ together from Eqs.\ (\rf{charge1}), (\rf{vcjp2}), and (\rf{jta}), we have  
	\beeq 
	\vcv=\frac{\vcj}{\vre}= -\frac{1}{\vre\al}\dr{I}{\Psi}\vcB+\frac{\OmFm\vp }{\al}  \uvt  , 
	\lb{vc-vjA} \eneq 
which indicates that FCSLs are equipotentials in the force-free domains, but when FLs are continuous, i.e., $\PN{\vcBp}\neq 0$,  stream- and current-lines are severed on the null surface \SSN, by $\PN{\vcj}=\PN{\vcv}=0$. 

In passing, we clarify an important constraint imposed by the `current closure condition' in the steady axisymmetric state: i.e., no net gain nor loss of charges over any closed surface threaded by current lines in the force-free domains. For a closed surface from the first open FL $\Psi=\Psiz$ to the last open FL $\Psi=\bar{\Psi}$ in the poloidal plane, we have 
\beeq
\oint \al\vcj\cdot d\vcA\propto I(\Psib)-I(\Psiz)=0,\ \ 
I(\Psib)=I(\Psiz)=0,  
\lb{c-c-c}  \eneq	
when there is no line current at $\Psi=\Psiz$ nor at $\Psi=\Psib$. This requires that function $I(\Psi)$ has at least one extremum at $\Psi=\Psic$ where $(dI/d\Psi)_{\rm c}=0$ (see figure 2 in \citep{okam06} for one example of $I(\Psi)$), and hence 
\beeq
\vcjp=\vre \vcvp \left\{
\begin{array} {ll}
>0; \ \ \Psio<\Psi< \Psic , \\[1mm]
=0;  \ \ \Psi=\Psic, \\[1mm]
<0; \ \  \Psic<\Psi<\Psit, 
\end{array}  \right. \lb{vcjpPM}    \eneq 
where $\Psiz<\Psio<\Psic<\Psit<\Psib$  (see FIG.\ \rff{DC-C}).

\subsection{The `conserved' and `non-conserved' energy fluxes}  \label{TEF} 
Multiplying Eq.\ (\rf{SJa}) with $\OmF$, we have 
\beeq
0=- \frac{\OmF}{\al}\vcnb\cdot \al\vcSJ =- \frac{1}{\al}\vcnb\cdot \al\OmF\vcSJ,  \lb{energya} 
\eneq 
which indeed reproduces Eq.\ (\rf{E-AmFlux}) for the $\vcSE=\OmF\vcSJ$ relation. However, this procedure of derivation does not yield the non-conserved fluxes $\vcSEM+\vcSsd$ between  $\vcSE$ and $\OmF\vcSJ$, although one can quickly obtain {\em a Poynting flux} in \cite{bla77}, which accords with $\vcSEM$, from Eqs.\ (\rf{bhEp}) and (\rf{bhBt}). By replacing $\OmF$ with $\OmFm + \om$ with the use of the identity (\rf{s.i}), one obtains Eq.\ (\rf{SEEMsd}), 
which shows that frame-dragging splits the overall conserved flux $\vcSE$ into the two non-conserved fluxes $\vcSEM$ and $\vcSsd$ [see Eq.\ (\rf{Sem}a,b)]. 


\subsection{The densities of the electromagnetic energy and angular momentum in the force-free domains}  \label{EnergyD} %

From expressions (\rf{epsEJab}a,b), we clarify important properties of $\epsE$ and $\epsJ$. Toward the surface at infinity \Sinf, where $\al\approx 1$, $\om\approx 0$, $\Bt^2\gg\Bp^2$, and $\vp^2\gg \vp^2_{\rm oL}$, we have, from expressions (\rf{epsEJab}a,b),
\beeq
\epsE\approx \frac{1}{8\pi} \left(\frac{\OmF^2 \vp^2}{c^2}\Bp^2+\Bt^2\right)_{{\rm ff}\infty}, \ \ 
\epsJ \approx \frac{\OmF(\vp\Bp)_{{\rm ff}\infty}^2}{c},  
\lb{nearSinf} \eneq	
which will be dissipated for particle acceleration in the resistive membrane $\Sffinf$ (see, e.g.,\ Eq.\ (\rf{DivvcSE}) and FIG.\ \rff{GapI}). 

Near the null surface \SSN,\ where $\OmFm\approx 0$ and $\vp\Bt\propto I\approx 0$, 
\beeq
\epsE \approx \left(\frac{\al \Bp^2}{8\pi}\right)_{\rm N}, \ \ \epsJ \approx 0,
\lb{epsE/N} \eneq  
and the region under \SSN\ (i.e.,\ the ZAM-Gap $\GN$; see Sec.\ \ref{m-mdGap}) will be matter-dominated by charged particles pair-created by the voltage drop, thereby breaking down the force-free condition (see Secs.\ \ref{NullS}, \ref{indMem}). 



Near $\SffH$, where $(\Bp\vp^2)_{\rm ffH}/(\Bp\vp^2) \approx 1$, and hence $\Bt^2/\Bp^2\approx (2\Iin /\vp c\al\Bp)^2 \approx ((\OmH-\OmF)\vp/(\al c))^2$ by Eq.\ (\rf{Iout/in-b}),  we have 
		$$ \epsE \approx -\left(\frac{\Bp^2}{8\pi \al} \frac{2\OmF\OmH\vp^2}{c^2} \right)_{\rm ffH} \left[\left(1-\frac{\OmF}{\OmH}\right) -\left( \frac{c^2\al^2}			{2\OmF\OmH\vp^2}\right)_{\rm ffH} \right]  $$ 
		\beeq
		\approx - \left(1-\frac{\OmF}{\OmH}\right)\left(\frac{\Bp^2}{8\pi \al} \frac{2\OmF\OmH\vp^2}{c^2} \right)_{\rm ffH}   <0   
		\lb{epsE/ffH} \eneq  
  toward the resistive horizon membrane $\SffH$ for $\al\to 0$. Also, for the density of angular momentum near the horizon, we obtain
\beeq
\epsJ=-(\OmH-\OmF) \LPfrac{ (\vp\Bp)^2}{\al c}_{\rm ffH}<0.   
\lb{epsJffH}  \eneq  
\section{The place and shape of the null surface \SSN }  \label{SNshapAp} 

It is the final eigenvalue $\OmF(\Psi)$ that determines not only the efficiency $\epsGTE(\Psi)$ of energy extraction, but the place and shape of the null surface \SSN, which hides a magnetized ZAM-Gap $\GN$ underneath it in the force-free limit. Some basic properties of the structure of force-free eigen-magnetospheres had been already clarified in some details in Sec.\  7 of \cite{oka92} and Sec.\  2 in \cite{oka09} (see figure 3 in \cite{oka06} for a schematic shape of the magnetosphere;  also, see \cite{oka09, oka12a} for the monopolar {\em exact} solution in the slow-rotation limit). So,  for the FDAV $\om$, we deduce
\beeq 
\frac{\om}{\OmH}= 
\frac{(1+h^2)^2 x}{(x^2+h^2)^2-h^2(x-1)(x- h^2)\sin^2\theta,  }=\frac{1}{1+\zeta} ,
\lb{om-N} \eneq  
from Eq.\ (\rf{alom}), where $x\equiv r/\rH$ and $\OmH=c^3 h/2GM$. When we use $\omN=\OmH/(1+\zeta(\Psi))$ from Eq.\ (\rf{EigenOmFI}), the expression of $\xN=\xN(\theta)$ for the shape of \SSN\ reduces to an algebraic equation, i.e.,
\begin{eqnarray}  
F_{\rm N}(x,\theta, \zeta;h)=(x^2+h^2)(x^2+h^2\cos^2\theta)   \nonumber   \hspace{1cm} \\ \hspace{1cm} 
-(1+h^2)[(1+h^2\cos^2\theta)+(1+h^2)\zeta]x=0  . \quad 
\lb{FNul} \end{eqnarray}    
When $\zeta(\Psi)\simeq 1$, it is useful to define a `mid-surface' \SSM\ with $\omM=0.5\OmH$, and to examine topological features of \SSM, by using
\begin{eqnarray}
F_{\rm M}(x,\theta; h)=(x^2+h^2)(x^2+h^2\cos^2\theta)   \nonumber   \hspace{1cm} \\ \hspace{1cm} 
-(1+h^2)[2+h^2(1+\cos^2\theta)]x=0 .
\lb{FM} \end{eqnarray}	
In addition, we introduce the static-limit surface as the surface limiting the ergosphere from $g_{tt}=-(\Delta-a^2\sin^2\theta)/\rho^2=0$, 
\beeq
F_{\rm E}(x,\theta; h)=(x-1)(x- h^2)- h^2\sin^2\theta,
\lb{FE}  \eneq  
and its solution is expressed as
\beeq
\xE(\theta,h)=\frac{1}{2} \left((1+h^2) +\sqrt{(1-h^2)^2+4h^2\sin^2\theta} \right).
\lb{xE/sol} \eneq  
From Eqs.\ (\rf{FM}) and (\rf{xE/sol}), for $h\ll 1$, we have 
\begin{subequations}   \begin{eqnarray}
	\xE = 1+h^2\sin^2\theta, \lb{xE} \hspace{2cm}  \\  
	\xM = 2^{1/3} \left[1+\frac{h^2}{6} \left( 2(2- 2^{1/3})+(2^{1/3}-1)\sin^2\theta \right) \right], \hspace{1cm}   \lb{xM}  
\end{eqnarray}  \lb{xEM}  \end{subequations}  
while by Eqs. (7.7a,b) in \cite{oka92}, the two light surfaces, \SOL\ and \SIL, become 
\begin{subequations}   \begin{eqnarray}  
	\xoL=\frac{2}{h}\left(1-\frac{\sin\theta}{4} \right), \\  
	\xiL=1+\frac{h^2}{4}\sin^2\theta.   
\end{eqnarray}  \lb{xoiL}  \end{subequations}  
Now, $\xiL<\xE<\xM<\xoL$, and, for $h\to 0$, it turns out that both of $\xiL$ and $\xE$ $\to 1$ and $\xoL\to \infty$, but also $\xM\to 2^{1/3}=1.2599$. Therefore, when $\zeta\simeq 1$ (and hence \SSN$\simeq$\SSM), \SSN\ interestingly keeps a position of $\xN\to 2^{1/3}$ above the horizon between $\xiL=\xE=1$ and $\xoL\to\infty$ (i.e., \SE$\leftarrow$\SIL$<$\SSN$<$\SOL$\to$\Sinf), even for $h\to 0$.  

There is a certain surface \SSMc, which contacts with \SE\ from the outside at the equator, i.e., $\xM=\xE$. This occurs when $\hc=\sqrt{\sqrt{2}-1}=0.6436$, and then $\xM=1.3960$ at $\theta=0$ and $\xE=\xM=1+\hc^2=\sqrt{2}$ at $\theta=\pi/2$ (see Figs 1 and 2 in \cite{oka09}).  

For the extreme-Kerr state, with $h\to 1$, Eq.\ (\rf{FM}) reduces to 
\beeq
F_{\rm M}(x,\theta; 1)=(x^2+1)(x^2+\cos^2\theta)-2(3+\cos^2\theta)x=0,
\lb{FM1} \eneq	
which yields $\xM=1.6085$ for $\theta=0$ at the pole and $\xM=1.6344$, while by $F_{\rm E}(x,\pi/2; 1)=0$, we have $\xE=2$ at the equator (see figure 3 in \cite{oka92}). 

When $\zeta\simeq 1$, from the above analysis, one can read such interesting features at $\theta=\pi/2$ for $0\leq h\leq 1$ that 
\beeq
\begin{array} {ll}
1 \leq\xE(h) 
\leq 2, &  \mbox{for \SE}, \\ 
2^{1/3}=1.2599\leq\xN(h) 
\leq 1.6433, & \mbox{for $\SN$}, 
\end{array}  \lb{rE/Neq}    \eneq 
and that $\xN \ggel \xE$ for $h\lleg \hc=(2^{1/2}-1)^{1/2}=0.6436$. This shows that, for $1\geq h\geq \hc$, the equatorial portion of the null surface \SSN\ lies within the ergosphere \SE, while, for $h<\hc$, the whole ergosphere \SE\ lies within the null surface. It turns out that the ergosphere changes from a spherical shape at $h=0$ to a spheroidal one at $h=1$, while when $\zeta\simeq 1$ the null surface keeps an almost spherical shape from $h=0$ to $h=1$. 
In any case, it appears that mechanical properties in the ergosphere have no direct connection with electrodynamic properties of the null surface \SSN\ and the inner domain $\calDin$. 

\begin{acknowledgments}
	The authors thank Professor Kip Thorne for strong encouragement to continue this research. I. O.\ is grateful to T.\ Uchida for discussion and  to O. Kaburaki for the joint works, which helped significantly deepen his understanding of thermodynamics.  Y. S.\ thanks National Astronomical Observatory of Japan for kindness and support during his visits in 2017 and 2018.
\end{acknowledgments}


\begin{thebibliography}{3}%
\makeatletter
\providecommand \@ifxundefined [1]{%
 \@ifx{#1\undefined}
}%
\providecommand \@ifnum [1]{%
 \ifnum #1\expandafter \@firstoftwo
 \else \expandafter \@secondoftwo
 \fi
}%
\providecommand \@ifx [1]{%
 \ifx #1\expandafter \@firstoftwo
 \else \expandafter \@secondoftwo
 \fi
}%
\providecommand \natexlab [1]{#1}%
\providecommand \enquote  [1]{``#1''}%
\providecommand \bibnamefont  [1]{#1}%
\providecommand \bibfnamefont [1]{#1}%
\providecommand \citenamefont [1]{#1}%
\providecommand \href@noop [0]{\@secondoftwo}%
\providecommand \href [0]{\begingroup \@sanitize@url \@href}%
\providecommand \@href[1]{\@@startlink{#1}\@@href}%
\providecommand \@@href[1]{\endgroup#1\@@endlink}%
\providecommand \@sanitize@url [0]{\catcode `\\12\catcode `\$12\catcode
  `\&12\catcode `\#12\catcode `\^12\catcode `\_12\catcode `\%12\relax}%
\providecommand \@@startlink[1]{}%
\providecommand \@@endlink[0]{}%
\providecommand \url  [0]{\begingroup\@sanitize@url \@url }%
\providecommand \@url [1]{\endgroup\@href {#1}{\urlprefix }}%
\providecommand \urlprefix  [0]{URL }%
\providecommand \Eprint [0]{\href }%
\providecommand \doibase [0]{https://doi.org/}%
\providecommand \selectlanguage [0]{\@gobble}%
\providecommand \bibinfo  [0]{\@secondoftwo}%
\providecommand \bibfield  [0]{\@secondoftwo}%
\providecommand \translation [1]{[#1]}%
\providecommand \BibitemOpen [0]{}%
\providecommand \bibitemStop [0]{}%
\providecommand \bibitemNoStop [0]{.\EOS\space}%
\providecommand \EOS [0]{\spacefactor3000\relax}%
\providecommand \BibitemShut  [1]{\csname bibitem#1\endcsname}%
\let\auto@bib@innerbib\@empty
\bibitem [{Note1()}]{Note1}%
  \BibitemOpen
  \bibinfo {note} {In the case of treating the whole MHD theory of BH winds,
  not only the outflow but also the inflow must pass smoothly through three
  critical surfaces; slow, intermediate, and fast magnetosonic surfaces \cite
  {web67,mic69, oka78, ken83, pun89, pun90,oka99,oka02,oka03}. In the
  force-free theory, the last two surfaces reduce to S$_{\protect \rm oL}$,
  S$_{\protect \rm iL}$\ and S$_{\protect \rm oF}$, S$_{\protect \rm iF}$,
  respectively, although the slow surface is generally neglected.}\BibitemShut
  {Stop}%
\bibitem [{Note2()}]{Note2}%
  \BibitemOpen
  \bibinfo {note} {We can confirm the vector $\protect \mathaccentV
  {vec}2AE{E}$'s reversal of direction, in Figure 2 and its caption in \protect
  \citet {bla77}, in Fig.\ 3 in \protect \citet {phi83b} and in Fig.\ 38 in
  \protect \citet {tho86}, as well as in each ordinate of FIGs \ref
  {ff:Flux-om}, \ref {ff:GapI}, \ref {ff:DC-C}, \ref {ff:F-WS} at the null
  surface S$_{\protect \rm N}$\ $\unhbox \voidb@x \hbox {\protect \boldmath
  $E$}_{\protect \rm p}=0$ in this paper.}\BibitemShut {Stop}%
\bibitem [{Note3()}]{Note3}%
  \BibitemOpen
  \bibinfo {note} {The referee of \cite {oka92} asked one of the present
  authors to comment on the question of causality then arising about the BZ
  process. This paper now contains the final detailed response to his request
  and has been recently submitted to the same Journal. However, two members of
  the Journal's Editorial Board have agreed that the paper is unsuitable for
  publication in that Journal. That journal's Scientific editor's Comments are
  as follows: ``This Journal publishes only direct astrophysical applications
  of physics. Your paper deals with fundamental physics, and aims to clarify
  results in General Relativity. I recommend you to submit the paper to a
  journal publishing in fundamental physics, such as Phys Rev.''}\BibitemShut
  {NoStop}%
\end{thebibliography}%


\begin{thebibliography}{99}
	%
	
	\bibitem[\protect\citeauthoryear{Blandford \& Znajek}{1977}]{bla77}	
		Blandford R. D.,  Znajek R. L., 1977, MNRAS, 179, 433
	\bibitem[\protect\citeauthoryear{Thorne \& Blandford}{2017}]{tho17} 	
		Thorne K. S., Blandford, R. 2017,  Modern Classical Physics,  Princeton University Press	 
	\bibitem[\protect\citeauthoryear{Macdonald \& Thorne}{1982}]{mac82} 	
		Macdonald D. A, Thorne K., 1982, 	MNRAS, 198, 345 	 
	\bibitem[\protect\citeauthoryear{Thorne et.al.}{1986}]{tho86} 	
		Thorne K. S., Price R. H., Macdonald D. A. 1986,  Black Holes: The Membrane Paradigm, Yale University Press, New Haven   	 
	\bibitem[\protect\citeauthoryear{Znajek}{1977}]{zna77}   
		Znajek R. L., 1977, MNRAS, 179, 457 
	\bibitem[\protect\citeauthoryear{Znajek}{1978}]{zna78}   
		Znajek R. L., 1978, MNRAS, 185, 833 	
	\bibitem[\protect\citeauthoryear{Phinney}{1983a}]{phi83a} 
		Phinney S., 1983,  in Proc. Torino Workshop on Astrophysical Jets, ed. A. Ferrari \& A. Pacholczyk, Reidel, Dordrecht	
	\bibitem[\protect\citeauthoryear{Blandford}{1979}]{bla79}   
		Blandford R. D., 1979, ``Accretion disc and black hole electrodynamics", in ``Active Galactic Nuclei", eds  C. Hazard and S. Mitton. p.241, 			Cambridge University Press, Cambridge   
	\bibitem[\protect\citeauthoryear{Phinney}{1983b}]{phi83b} 
		Phinney S., 1983, A theory of radio sources, Ph.D. thesis, Univ. of Cambridge	
	\bibitem[\protect\citeauthoryear{O78}{}]{oka78}		
		Okamoto I.,  1978, MNRAS, 167, 457 			
	\bibitem[\protect\citeauthoryear{Kennek et.a.}{1983}]{ken83}    	
		Kennel C. F., Fujimura F. S., Okamoto I., 1983, Geophys. Ap. Fluid Dyn., 26, 147  
	
	\bibitem[Punsly \& Coroniti(1989)]{pun89}		
		Punsly B., Coroniti F. V., 1989, Phys. Rev. D, 40, 3834
	\bibitem[\protect\citeauthoryear{Punsly \& Coroniti}{1990}]{pun90}	
		Punsly B., Coroniti F. V., 1990, ApJ, 350, 518	
	\bibitem[\protect\citeauthoryear{OK90}{1990}]{OK90} 
		Okamoto I., Kaburaki O.,  1990, MNRAS, 247, 244 		
	\bibitem[\protect\citeauthoryear{OK91}{1991}]{OK91} 
		Okamoto I., Kaburaki O.,  1991,  MNRAS, 250, 300 
	\bibitem[\protect\citeauthoryear{KO91}{1991}]{KO91} 
		Kaburaki O., Okamoto I.,  1991, Phys. Rev. D, 43, 340	
	\bibitem[\protect\citeauthoryear{Blandford}{2002}]{bla02}   
		Blandford R. D., 2002, ``To the Lighthouse", in Lighthouse of the Universe, eds Gilfanov, M.\ et al. (Springer: Berlin)
	\bibitem[\protect\citeauthoryear{Blandford \& Globus}{2022}]{bla22}	
		Blandford R. D., Globus, N., 2022, MNRAS, 514, 5141
	\bibitem[\protect\citeauthoryear{O92}{1992}]{oka92} 		
		Okamoto I.,  1992, MNRAS, 254, 192  	
		
	\bibitem[\protect\citeauthoryear{Landau et al.}{1984}]{lan84}	
		Landau L. D., Lifshitz E. M., Pitaevskii L. P., 1984, Electrodynamics of Continuous Media, 2nd edition, Butterworth-Heinemann, Oxford	 
	\bibitem[\protect\citeauthoryear{O15a}{2015a}]{oka15a}	   
		Okamoto I.,  2015a,  PASJ, 67, 89
	\bibitem[\protect\citeauthoryear{O15b}{}]{oka15b}		
		Okamoto I.,  2015b, ``Frame dragging, unipolar induction and jet source,'' 
		in Proceedings of the Texas Symposium on Relativistic Astrophysics, 2015.12.13--18, Geneva, Switzerland  
	
	\bibitem[\protect\citeauthoryear{Goldreich \& Julian}{1969}]{gj69} 
		Goldreich P., Julian W. H., 1969, ApJ,  157, 869  
	\bibitem[\protect\citeauthoryear{O74}{1974}]{oka74}		
		Okamoto I.,  1974, MNRAS, 167, 457   
	\bibitem[\protect\citeauthoryear{O12b}{2012}]{oka12b}  
		Okamoto I., ``How does unipolar induction work in Kerr black holes?,'' Proceedings of the Ginzburg Conference on physics, Lebedev Physical 			Institute, Moscow, Russia, May 28 - June 2, 2012 
	\bibitem[\protect\citeauthoryear{OS06}{2006}]{okam06}  
		Okamoto I., Sigalo B. F.,  2006, PASJ, 58, 987  			
	\bibitem[\protect\citeauthoryear{Basu \& Lynden-Bell}{1990}]{bas90}	
		Basu, B., Lynden-Bell, D., 1990, Q.\ Jl.\ R.\ astr.\ Soc., 31, 359	
	\bibitem[\protect\citeauthoryear{O06}{}]{oka06}	
		Okamoto I.,  2006, PASJ, 58, 1047 		 
	\bibitem[\protect\citeauthoryear{OK93}{1993}]{OK93} 
		Okamoto I., Kaburaki O., 1993, MNRAS, 225, 539 	  
	\bibitem[Cveti\v{c} et al.(2018)]{cve18}	
		Cveti\v{c} M., Gibbons G.W., L\"{u} H., Pope C.N., 2018, Phys. Rev. D 98, 106015	
	\bibitem[\protect\citeauthoryear{O09}{}]{oka09}	
		Okamoto I.,  2009, PASJ, 61, 971
	\bibitem[Komissarov(2009)]{kom09}	
		Komissarov S. S., 2009, J.\ Korean Phys.\ Soc., 54, 2503  		
	
	
			
	
	\bibitem[\protect\citeauthoryear{Beskin et al.}{1992}]{bes92}  
		Beskin V. S., Istomin Ya. N.,  Par'ev  V. I., 1992, Sov. Astron., 36(6), 642
	\bibitem[\protect\citeauthoryear{Hirotani \& Okamoto}{1998}]{hir98}   
		Hirotani K., Okamoto I., 1998, ApJ, 497, 563
	\bibitem[\protect\citeauthoryear{Song, et al.}{2017}]{son17}	
		Song Y., Pu H.-Y., Hirotani K., Matsushita S., Kong A. K. H., Chang H.-K., 2017, MNRAS, 471, L135 
	\bibitem[\protect\citeauthoryear{Hirotani et al.}{2018}]{hir18b}	
		Hirotani, K., Pu, H.-Y., Outmani, S., Huang, H., Kim, D., Song, Y., Matsushita, S., Kong, A. K. H., ApJ, 867, 120	
	\bibitem[\protect\citeauthoryear{Ruffini, et al.}{2010}]{ruf10}  
		Ruffini, R., Vereshchagin, G., Xue, She-Schen, 2010, Physics Reports, 487, Issues 1- 4, 1-140
	\bibitem[\protect\citeauthoryear{H.\ Chen \& F.\ Fiuza}{2023}]{che23}
		Chen, H.,  Fiuza, F., 2023, Phys.\ Plasmas 30, 020601


	\bibitem[\protect\citeauthoryear{O12a}{}]{oka12a} 	%
		Okamoto I.,  2012, PASJ, 64, 50 		
		
		
		
		
	\bibitem[\protect\citeauthoryear{Weber \& Davis}{1967}]{web67}  
		Weber E.\ J., Davis L.\ J., 1967, ApJ, 148, 217		
	\bibitem[\protect\citeauthoryear{}{1969}]{mic69}	 
		Michel F.\ C.,1969, ApJ, 158, 727	
	\bibitem[\protect\citeauthoryear{O99}{}]{oka99}		
		Okamoto I.,  1999,   MNRAS, 254, 192 
	\bibitem[\protect\citeauthoryear{O02}{}]{oka02}	
		Okamoto I.,  2002, ApJ, 573, L31		
	\bibitem[\protect\citeauthoryear{O03}{}]{oka03}	
		Okamoto I.,  2003, ApJ, 589, 671   		
	\bibitem[\protect\citeauthoryear{Penrose 1969}{}]{pen69}  
		Penrose, R.,  1969, Nuovo. Cim., 1, 252
	\bibitem[\protect\citeauthoryear{Costa \& Nat\'{a}rio}{2021}]{cos21}  
		Costa, L.F.O., Nat\'{a}rio, J., Universe, 2021, 7, 388. 

		
		
		
	\end{thebibliography}
\end{document}